\documentclass[longauth]{aa}  
\usepackage{graphicx}
\usepackage{natbib}
\usepackage{subcaption}
\usepackage{hyperref}
\usepackage{rotating}
\usepackage{xcolor}
\usepackage{amsmath}
\usepackage{CJKutf8}
\usepackage{siunitx}
\usepackage{threeparttable}
\usepackage{comment}
\usepackage{flushend} 

\usepackage{txfonts}
\newcommand{\ngist}{\textsc{nGIST}}
\newcommand{\gist}{\textsc{GIST}}
\newcommand{\ppxf}{\textsc{pPXF~}}

\DeclareMathOperator{\arcsinh}{arcsinh}

\begin{document} 

   \title{The GECKOS Survey: Identifying kinematic sub-structures in edge-on galaxies}
\titlerunning{GECKOS: Identifying kinematic sub-structures}
\authorrunning{Fraser-McKelvie et al.}

   \author{A. Fraser-McKelvie,\inst{1,2}   
          J. van de Sande,\inst{3,2}
          D.~A. Gadotti,\inst{4}
          E. Emsellem,\inst{1}
          T. Brown,\inst{5}
          D.~B. Fisher,\inst{6,2}
          M. Martig,\inst{7}
          M. Bureau,\inst{8}
          O. Gerhard,\inst{9}
          A.~J. Battisti,\inst{10,11,2}
          J. Bland-Hawthorn,\inst{12,2}
          A. Boecker, \inst{13}
          B. Catinella,\inst{10,2}
          F. Combes,\inst{14,15}
          L. Cortese,\inst{10,2}
          S.~M. Croom,\inst{12,2}
          T.~A. Davis,\inst{16}
          J. Falc\'on-Barroso,\inst{17,18}
          F. Fragkoudi,\inst{19}
          K.~C. Freeman,\inst{11,2}
          M.~R. Hayden,\inst{20,2}
          R. McDermid,\inst{21,2}
          B. Mazzilli Ciraulo,\inst{5,2} 
          J.~T. Mendel,\inst{11,2}
          F. Pinna,\inst{17,18}
          A. Poci,\inst{8}
          T.~H. Rutherford,\inst{1,12,2}
          C. de S\'a-Freitas\inst{22},
          L.~A. Silva-Lima,\inst{23}
          L.~M. Valenzuela,  \inst{24}
          G. van de Ven,\inst{13}
        \begin{CJK*}{UTF8}{gbsn} Z. Wang (王梓先)\end{CJK*},\inst{25,2},
          \and A.~B. Watts.\inst{10,2} 
}

\institute{European Southern Observatory, Karl-Schwarzschild-Stra{\ss}e 2, Garching, 85748, Germany\\
    \email{a.fraser-mckelvie@eso.org}   
\and
ARC Centre of Excellence for All Sky Astrophysics in Three Dimensions (ASTRO-3D)
\and 
School of Physics, University of New South Wales, NSW, 2052, Australia
\and
Centre for Extragalactic Astronomy, Department of Physics, Durham University, South Road, Durham DH1 3LE, UK
\and
National Research Council of Canada, Herzberg Astronomy and Astrophysics Research Centre, 5071 W. Saanich Rd. Victoria, BC, V9E 2E7, Canada
\and
Centre for Astrophysics and Supercomputing, Swinburne University of Technology, PO Box 218, Hawthorn, VIC 3122, Australia
\and
Astrophysics Research Institute, Liverpool John Moores University, 146 Brownlow Hill, Liverpool L3 5RF, UK
\and 
Sub-department of Astrophysics, Department of Physics, University of Oxford, Denys Wilkinson Building, Keble Road, Oxford OX1 3RH
\and
Max-Planck-Institut f\"ur Extraterrestrische Physik, Gie{\ss}enbachstra{\ss}e 1, 85748 Garching, Germany
\and
International Centre for Radio Astronomy Research (ICRAR), The University of Western Australia, M468, 35 Stirling Highway, Crawley, WA 6009, Australia
\and
Research School of Astronomy and Astrophysics, Australian National University, Cotter Road, Weston Creek, ACT 2611, Australia
\and 
Sydney Institute for Astronomy, School of Physics, A28, The University of Sydney, NSW, 2006, Australia
\and 
Department of Astrophysics, University of Vienna, T\"urkenschanzstra{\ss}e 17, 1180 Vienna, Austria
\and
Observatoire de Paris, LERMA, CNRS, PSL University, Sorbonne University, 75014 Paris, France
\and
Coll\`ege de France, 11 Pl. Marcelin Berthelot, 75231 Paris, France
\and
Cardiff Hub for Astrophysics Research \& Technology, School of Physics \& Astronomy, Cardiff University, Queens Buildings, Cardiff CF24 3AA, UK
\and
Instituto de Astrof\'isica de Canarias, Calle V\'ia L\'actea s/n, E-38205 La Laguna, Tenerife, Spain
\and
Departamento de Astrof\'isica, Universidad de La Laguna, Av. del Astrof\'isico Francisco S\'anchez s/n, E-38206, La Laguna, Tenerife, Spain
\and 
Institute for Computational Cosmology, Department of Physics, Durham University, South Road, Durham DH1 3LE, UK 
\and
Homer L. Dodge Department of Physics \& Astronomy, University of Oklahoma, 440 W. Brooks St., Norman, OK 73019, USA
\and
Research Centre for Astronomy, Astrophysics, and Astrophotonics, Department of Physics and Astronomy, Macquarie University, NSW 2109, Australia
\and
European Southern Observatory, Alonso de C\'ordova 3107, Vitacura, Regi\'on Metropolitana
\and
Universidade Cidade de São Paulo/Universidade Cruzeiro do Sul, Rua Galv\~ao Bueno 868, S\~ao Paulo-SP, 01506-000, Brazil
\and
 Universit\"ats-Sternwarte, Fakult\"at f\"ur Physik,  
Ludwig-Maximilians-Universit\"at M\"unchen, Scheinerstr. 1, 81679 M\"unchen,  
Germany
\and
Department of Physics and Astronomy, University of Utah, Salt Lake City, UT 84112, USA\\
             }


 
  \abstract
{The vertical evolution of galactic discs is governed by the sub-structures within them. Several of these features, including bulges and kinematically-distinct discs, are best studied in edge-on galaxies, as the viewing angle allows for easier separation of component light.
In this work, we examine the diversity of kinematic sub-structure present in the first 12 galaxies observed from the GECKOS survey, a VLT/MUSE large programme providing a systematic study of 36 edge-on, Milky Way-mass disc galaxies.  Employing the \ngist\ analysis pipeline, we derive the mean luminosity-weighted line-of-sight stellar velocity ($V_{\star}$), velocity dispersion ($\sigma_{\star}$), skew ($h_{3}$), and kurtosis ($h_{4}$) for the sample, and examine 2D maps and 1D line profiles. 
Common clear kinematic signatures are observed: all galaxies display $h_{3}$ - $V_{\star}$ sign mismatches 
in outer disc regions consistent with a (quasi-)axisymmetric, rotating disc of stars.
Scrutinising visual morphologies, the majority of this sample (8/12) are found to possess boxy-peanut bulges and host the corresponding kinematic structure predicted for stellar bars viewed in projection. Inferences are made on the bar viewing angle with respect to the line of sight from the strength of these kinematic indicators, finding one galaxy whose bar is close to side-on with respect to the observer, and two that are close to end-on. 
Four galaxies exhibit strong evidence for the presence of nuclear discs, including central $h_{3}$-$V_{\star}$ profile anti-correlations, `croissant'-shaped central depressions in $\sigma_{\star}$ maps, strong gradients in $h_{3}$, and positive $h_{4}$ plateaus over the expected nuclear disc extent. The strength of the $h_{3}$ feature corresponds to the size of the nuclear disc, measured from the $h_{3}$ turnover radius, taking into account geometric effects. 
We can explain the features within the kinematic maps of the four unbarred galaxies via disc structure(s) alone. We do not find any need to invoke the existence of dispersion-dominated bulges in any of the sample galaxies. Obtaining the specialised data products for this paper and the broader GECKOS survey required significant development of existing integral field spectroscopic (IFS) analysis tools.
Therefore, we also present the \ngist\ pipeline: a modern, sophisticated, and easy-to-use pipeline for the analysis of galaxy IFS data, and the key tool employed by the GECKOS survey for producing value-added data products.
We conclude that the variety of kinematic sub-structures seen in GECKOS galaxies requires a contemporary view of galaxy morphology, expanding on the traditional view of galaxy structure, and uniting the kinematic complexity observed in the Milky Way with the extragalactic. }


   \maketitle
%


\section{Introduction}
Most galaxies are multi-component systems, consisting of multiple photometrically, chemically, and kinematically distinct stellar structures. In the local Universe, the majority of Milky Way-mass galaxies comprise a rotationally-supported disc \citep[e.g.][]{guo2020}, a bulge\footnote{The term `bulge' has become something of a provocative descriptor for the over-concentration of light situated in the central regions of many galaxies. Here we use the term simply in a morphological sense to describe the light that bulges out of the disc plane.}
\citep[e.g.][]{gadotti2009}, a stellar bar \citep[e.g.][]{erwin2018}, and a diffuse stellar halo \citep[e.g.][]{helmi2020}.

The Milky Way itself is largely considered to be a typical disc galaxy \citep[e.g.][]{bland-hawthorn2016}. In addition to the structural components listed above, the Milky Way's disc may be separated into two chemically distinct discs, whose stellar populations can be divided into young, metal-rich and [$\alpha$/Fe]-poor, and old, metal-poor and [$\alpha$/Fe]-rich \citep[e.g.][]{gilmore1983, haywood2013, hayden2015}. Observations of external galaxies have hinted at similar chemical structures \citep[e.g.][]{pinna2019a, pinna2019b, scott2021}.
The kinematic thick disc of the Milky Way has also been found to host stars with a larger velocity dispersion and slower net rotation than stars in the thin disc \citep[e.g.][]{chiba2000, soubiran2003, girard2006}. Similar vertical dispersion gradients are also seen in external galaxies  \citep[e.g.][]{yoachim2008, comeron2019, pinna2019a, pinna2019b, kasparova2020, martig2021, bhattacharya2023}. 
Lastly, the Milky Way also hosts a bar, and a boxy-peanut (BP) bulge \citep[e.g.][]{kent1991, wegg2013, ness2016}. As such,  the integrated kinematic measurements along a line of sight in an extragalactic Milky Way-like galaxy will include contributions from at least four different kinematic sub-structures near the centre.

The diversity of galactic kinematic sub-structure is most conspicuous in the central regions, where many structures may contribute to the overall surface brightness distribution. Each sub-structure hosts stars on specific orbits, though the diversity of central structure (including the relative contributions of various stellar orbital families) is not well quantified, and lacks comprehensive kinematic descriptors to categorise central structure, despite efforts in this space \citep[e.g.][]{fabricius2012, mendez-abreu2014}. In fact, it seems that the more depth with which one studies galactic sub-structures, the more complicated they become \citep[e.g.][]{gadotti2020}. 

Edge-on galaxies represent a unique opportunity to study galactic central components, as the light that `bulges' out of the disc is readily observable for these geometries. One such central structure for which much work has been devoted is the BP bulge \citep[e.g.][]{burbidge1959, lutticke2000}, widely believed to be the out-of-plane projection of a buckled stellar bar \citep[e.g.][]{desouza1987, kuijken1995, bureau1999}. Present in $\sim80\%$ of Milky Way-mass galaxies \citep{erwin2017}, this structure is a natural consequence of disc galaxy evolution \citep[e.g.][]{combes1981, debattista2006, valluri2016, fragkoudi2017a, parul2020, tahmasebzadeh2024}. 
Detecting the presence of a BP bulge and bar can be difficult from imaging alone, as these structures leave distinct photometric imprints for only a limited range of bar viewing angles. Complicating matters, they can visually resemble a ellipsoidal `classical'
bulge in projection when viewed end-on \citep[e.g.][]{combes1990, laurikainen2017}. For this reason, the addition of kinematic information can aid in identifying BP bulges and bars within their host galaxies. 

Significant progress in quantifying the motions of stars within BP bulges has been made, both in- and off-plane. Several studies have applied stellar kinematic indicators to classify the behaviour of the stars \citep[e.g.][]{chung2004, bureau2005, mendez-abreu2008, mendez-abreu2014}, and dynamically, the bar orbits are well-understood and modelled.
Barred galaxies contain stars on more elongated orbits: chiefly the $x_{1}$ orbits coincident with the canonical bar shape, and the $x_{2}$ orbits, which are smaller, generally more circular, and oriented perpendicular to the $x_{1}$ orbits \citep[e.g.][]{athanassoula1992, sellwood1993}. The angle at which the bar is viewed by the observer (i.e.\ more side-on or more end-on) determines both i) the relative contribution of each bar orbit family to a given line-of-sight velocity distribution (LOSVD) and ii) the shape of the LOSVD of that orbit (as bar orbits are not axisymmetric), hence the bar viewing angle has a measurable effect on the stellar kinematics observed.

\citet{chung2004} and \citet{bureau2005} investigated the kinematics of BP bulge galaxies and compiled a set of kinematic criteria from long-slit observations that described the various structures present in the mid-plane.
They describe rotation curve `humps', and velocity dispersion peaks and plateaus. They also see a complex relation between the mean velocity, ($V_{\star}$), and skewness of the LOSVD, ($h_{3}$), which anti-correlate, correlate, and anti-correlate again in the nuclear disc, bar, and outer disc regions respectively.
Taken together, these models and observations provide a solid foundation to begin investigating the kinematic sub-structure behaviour, although only in one dimension (i.e.\ radially along the midplane).

The advent of large-scale integral field spectroscopic (IFS) observations, including highly spatially-resolved
observations, has allowed studies off the plane of galactic discs and prompted observations of kinematic structures in the form of 2D moment maps \citep[e.g.][]{dezeeuw2002,guerou2016, pinna2019a, poci2019, martig2021}.
Motivated by the influx of such data, studies such as \citet{iannuzzi2015} and \citet{li2018} provided a set of 2D kinematic predictors of various central structures, including BP and classical bulges. 
The off-plane structure of the kinematic high-order moments is coherent, and 1D long-slit data insufficient to fully describe the kinematic trends seen in BP bulge galaxies. 
The more complex AURIGA cosmological zoom-in simulations \citep{grand2017} confirm that off-plane kinematic structure is a regular feature in disc galaxies \citep{pinna2024}, with BP bulges contributing to distinct kinematic signatures in the inner regions of galaxies \citep{fragkoudi2020}.

Much can also be learnt about stellar kinematic distributions from simpler analytic galactic models. \citet{wang2024} generated mock IFS data cubes of the Milky Way by forward-modelling the chemodynamical model of \citet{sharma2021}. Based on star particles distributed in a simple disc structure (i.e.\ no bulge, halo, or nuclear disc structure), outputs were provided in the form of spatially-binned 2D maps of stellar velocity moments, allowing for a direct comparison to observations. In particular, once analysed by industry-standard spectral fitting packages, output stellar velocity dispersion maps displayed high dispersion values in the central region and up to several kpc off the plane; much higher than input from the mock IFS cube.  
Importantly, this dispersion peak was caused only by the disc structure; there was no central spheroid included in these models.

Some galactic features are difficult to model.
The small scales of some central galactic features such as nuclear discs \citep{gadotti2020} have precluded them from being resolved in cosmological or even zoom simulations. Numerical heating of stellar particles also wreaks havoc on the size and shape of small, thin discs, even at high resolutions \citep[e.g.][]{ludlow2019, anta2023}.
Furthermore, most mock observations from simulations tend not to include typical observational features such as dust lanes, although some advances have been made in this area \citep[e.g.][]{barrientosacevedo2023}.
Given the prevalence and obvious complexity of galactic kinematic sub-structures coupled with both simulation predictions and limitations, a census of kinematic sub-structures in a representative sample of Milky Way-mass galaxies will allow us to make progress towards a better interpretation and framework of all central disc structures.

In this paper, we categorise the diversity of kinematic sub-structures in the first 12 galaxies observed as part of the GECKOS\footnote{Generalising Edge-on galaxies and their Chemical bimodalities, Kinematics, and Outflows out to Solar environments} survey. We aim to define a range of kinematic sub-structures and link them to the assembly and evolutionary history of the host galaxies. 
In Section~\ref{survey_desc} we describe the GECKOS survey (including data reduction and analysis) and ancillary data. In Section~\ref{results} 
we identify BP bulges from imaging and present 1D and 2D stellar kinematics. In Section~\ref{discussion} we interpret the results within a framework of galactic central structure assembly.
We also introduce the \ngist \footnote{https://github.com/geckos-survey/ngist} package in Appendix~\ref{ngist_appendix}: a modern, freely-available IFS data analysis package, flexible enough to analyse any galaxy IFS data. Throughout this paper, we use $\Lambda$CDM cosmology, with $\Omega_{\rm{m}}=0.30$, $\Omega_{\rm{\Lambda}}=0.70$, and $H_{0} = 70$ km s$^{-1}$ Mpc$^{-1}$, and a \citet{chabrier2003} initial mass function. 

\section{Data}
\label{survey_desc}

\begin{figure*}
\centering
\begin{subfigure}{0.99\textwidth}
\includegraphics[width=\textwidth]{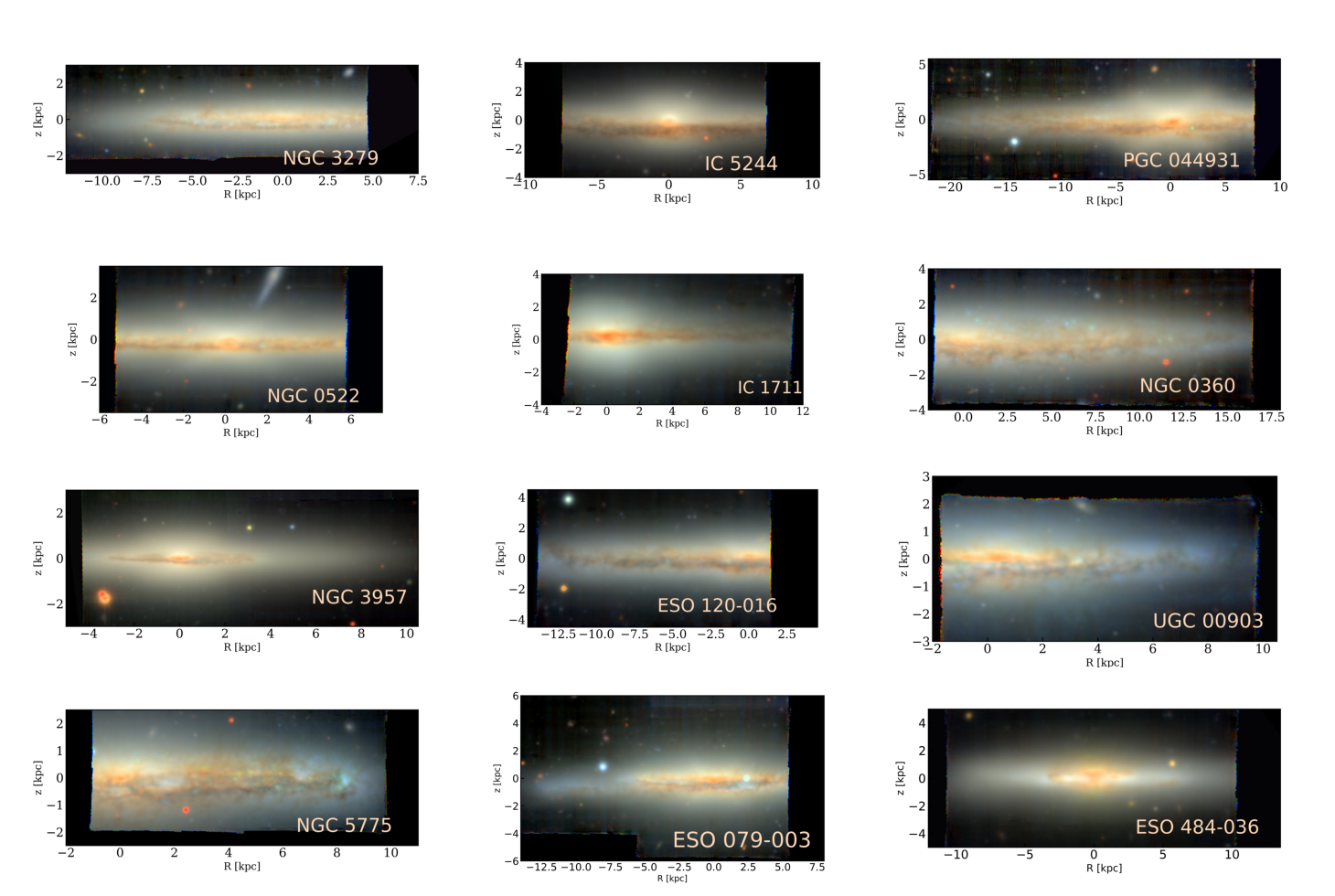}
\end{subfigure}
\caption{$g^{\prime}ri$ colour images of the sample galaxies reconstructed from the MUSE data cubes of the first 12 galaxies observed as part of the GECKOS survey. In each case, the mosaicked cube has been rotated and aligned such that the photometric major axis is horizontal, and if the dust lane is offset from it, it is always on the lower side of the image. For galaxies not exactly edge-on this means that we look at them from above (i.e.\ the lower half is the near half) if the dust lane is assumed to be in the midplane of the galaxy. The GECKOS pointing strategy was such that the entire (visible) bulge was always included within the MUSE field of view, and at least one side of the galactic disc. Here we show only regions where the signal-to-noise ratio is greater than 2.5; the MUSE field of view often extends further vertically.} 
\label{Fig:cube_images}
\end{figure*}

Our sample consists of the first 12 galaxies from the GECKOS survey, an ESO/VLT large programme awarded 317 hours on MUSE \citep{vandesande2024}. In Section \ref{subsec:sample}-\ref{subsec:data_reduction} we summarise the sample selection, observing strategy, and data reduction respectively, but a full description will be presented in van de Sande et al.\ (in prep.). We also introduce the updated data analysis pipeline package \ngist\ in Section \ref{data_analysis} and Appendix \ref{ngist_appendix}.

\subsection{GECKOS sample}
\label{subsec:sample}

GECKOS aims to investigate the relative importance of the internal and external physical processes that drive the evolution of disc galaxies. The GECKOS MUSE survey consists of 36 edge-on disc galaxies, covering out to a surface brightness of $\mu_{V}=23.5~ \rm{mag}~\rm{arcsec}^{-2}$, similar to that at the Sun's position in the Galactic disc when viewed face-on \citep{melchior2007}. Targets were selected within a distance range of $10 < D~[\textrm{Mpc}] < 70$, 25 of which were sourced from the S4G survey \citep{sheth2010} and the remaining 11 from HyperLeda \citep{makarov2014}. Seven of these galaxies that meet our sample selection criteria possess MUSE archival observations. 

Galaxies with a stellar mass within $\pm0.3$ dex of the Milky Way \citep[$5\times10^{10}\rm{M}_{\odot}$;][]{bland-hawthorn2016} that are edge-on (or highly inclined) are targeted. Furthermore, GECKOS aims to include a diversity of morpho-kinematic and star-formation properties by selecting galaxies within a 2 dex range of mid-IR-derived star-formation rates \citep[SFRs;][]{cutri2013, leroy2021}. The GECKOS targets are evenly divided into three bins of SFR, comprising a sub-main sequence, a main sequence, and a super-main sequence sample. This spread in SFR not only maximises the variety of possible assembly histories, but also increases the probability of detecting outflows, present in $\sim30 - 50\%$ of main sequence galaxies \citep{stuber2021} and ubiquitous in starbursts \citep[e.g.][]{veilleux2020}. 

Galaxies with clear signs of ongoing or recent mergers are avoided, as are those near the centre of clusters where external gas-removal processes may strongly impact disc evolution \citep[e.g.][]{cortese2021}. The final GECKOS sample contains BP bulges, non-BP bulges, and some galaxies with no visible bulge, with the hope being that given the diversity in observables, as wide a range in assembly histories as possible is covered. 

In this paper, we use all galaxies whose observations were completed by December 2023, for either the entire galaxy or the central pointing. This subset of galaxies forms the core of the GECKOS first internal data release (iDR1) and includes fully reduced and mosaicked data cubes for the galaxies ESO 484-036, NGC 5775, ESO 079-003, UGC 00903, ESO 120-016, NGC 3279, NGC 0360, PGC 044931, IC 1711, and NGC 3957, and the central pointing only of NGC 0522 and IC 5244. While not a statistical nor representative sample of the local Universe, this sub-sample contains galaxies from all three bins of star formation rate, and is hence a representative sample of GECKOS galaxies. 

\subsection{GECKOS MUSE observing strategy}
\label{subsec:obs_strategy}

GECKOS galaxies are observed using VLT/MUSE \citep{bacon2010} in wide-field mode (field of view 1$^{\prime}\times1^{\prime}$, $0\farcs2 \times 0\farcs2$ pixel sampling), and the nominal wavelength coverage of 4800 – 9300 \AA~ with 1.25 \AA~ pixel$^{-1}$ wavelength sampling ($R\sim3000$). The GECKOS tiling strategy is based on the science requirement to measure the stellar kinematics and stellar populations out to the 23.5 $V$ mag arcsec$^{-2}$ isophote. 

Depending on the distance to each galaxy, one MUSE pointing is required for the most distant galaxies, and up to six radially-tiled pointings for the nearest objects. For the super-main sequence sample, the MUSE coverage extends to a height of 10~kpc, based on the extent of the outflow in M82 \citep{shopbell1998}. For the six most highly star-forming galaxies, we therefore add a pointing off the plane of the disc near the centre. 

All galaxies are observed in natural seeing conditions. However, to ensure a spatial resolution of $<200$ pc as set by the science requirements, depending on the distance to the target, we requested the observations to be carried out in the turbulence category $<30\%$ -- $<85\%$, or a seeing threshold of 0\farcs7--1\farcs3 \citep{martinez2010}. Exposure times were calculated using the ESO MUSE exposure time calculator based on a S/N requirement of S/N$>40~\AA^{-1}$ for a spatial bin with a maximum size of $200$ pc.

Observations commenced in October 2022 and are carried out in service mode, split into 1-hour observation blocks (OBs). Within each OB, four $\sim9-10$ min object (O) and two $\sim2-3$ min sky observations (S) are executed in an OSOOSO sequence. Within each 1-hour OB, four 90-degree rotations combined with small offset dithers were used to optimise data homogeneity in the final cube. Between pointings, a 3$^{\prime\prime}$ overlap was adopted to create a smooth and continuous final mosaicked data cube.

\subsection{Data reduction}
\label{subsec:data_reduction}

The dedicated \textsc{python} package \textsc{pymusepipe} v2.23.4 \footnote{https://github.com/emsellem/pymusepipe} was utilised to calibrate the science exposures and create mosaicked data cubes. \textsc{pymusepipe} is a nearly fully automated pipeline, described in detail in \citet{emsellem2022}. It serves as a data organiser and wrapper for the sequential execution of the ESO Recipe Execution Tool \citep[\textsc{esorex};][]{esorex2015} recipes, and also provides additional utilities for improved alignment, flux and sky calibration, and mosaicking. \textsc{pymusepipe} is built around the MUSE Data Reduction Pipeline (DRP) that was developed to remove instrumental signatures \citep{weilbacher2020}\footnote{https://www.eso.org/sci/software/pipelines/muse/}. A full description of the GECKOS data reduction workflow will be presented in van de Sande et al.\ (in prep.), and we note that a similar setup was used in \citet{watts2024}.

Three-colour $g^{\prime}ri$ images reconstructed from the MUSE cubes of the 12 galaxies used in this paper are shown in Figure~\ref{Fig:cube_images}. We note here that while the MUSE nominal wavelength range of $\sim$4800--9300 \AA~ covers the SDSS $r$ and $i$ bands well, the $g$ band is truncated, and so we refer to it as $g^{\prime}$. The reconstructed $g^{\prime}$ band image follows the SDSS $g$-band filter response curve for $\sim$4800--5500 \AA.  

\subsection{Data analysis}
\label{data_analysis}
There exist many pipelines for the analysis of IFS data, several of which wrap existing well-established spectral fitting routines, for example, the data analysis pipelines of MaNGA \citep{westfall2019} and PHANGS \citep{emsellem2022}, \textsc{Pipe3D} \citep{sanchez2016} and the \gist\ pipeline \citep{bittner2019}. 
The future of extragalactic, spatially-resolved spectroscopy will involve pushing current instrumentation to its limits, requiring highly customisable analysis software packages. The GECKOS Survey alone necessitates the development of techniques to deal with low S/N, and highly sky-dominated regions to extract science from the galactic outer disc regions. Additionally, large co-added mosaics of several pointings of nearby galaxies are becoming a common technique to resolve the small-scale physics driving galaxy evolution. Such mosaics are large, and computationally expensive to run through existing, non-optimised software.  
Here, we present \ngist, an upgraded version of the \gist~ pipeline, with added features, improved usability, and retained functionality and core principles. Described more fully in Appendix~\ref{ngist_appendix}, \ngist\ is publicly available, documented, and maintained via a GitHub repository. %

\ngist\ is flexible, and allows for user specification of over 40 different variables via a compact and human-readable configuration file. 
A full description of all input parameters used for GECKOS iDR1 for each \ngist\ module will appear in van de Sande et al.\ (in prep.) and the \ngist\ documentation\footnote{https://geckos-survey.github.io/gist-documentation/}. Here, we describe the data products produced and employed in this work.

For this work, we employ the \ngist\ version 7.2.1 to output 2D maps of stellar kinematics Voronoi binned to S/N=100 pixel$^{-1}$ (or 80 \AA$^{-1}$, calculated over 4800--7000 \AA). We chose this wavelength range so that the S/N estimate includes a minimal number of sky lines. We chose this binning S/N to ensure the highest-quality kinematic maps possible, whilst noting that the science goals mostly concern the central regions of the galaxies for this work, where S/N is typically very high.
Spaxels with a S/N$<5$ were masked along with bright stars and any other contaminating foreground objects. We tested the effects on the derived stellar kinematics from the wavelength range employed. We present the results in Appendix~\ref{wavelength_dependence}, and determine that for our science goal of understanding the kinematics of structures at the centres of galaxies, whilst still recovering $h_{3}$ and $h_{4}$ in the outskirts of low-dispersion discs, a wavelength range of 4800--8900 \AA~yielded the best results compared to small ranges centered on individual features such as H$\beta$ and Mg b in the blue and the Calcium II triplet (CaT) in the red.
All spectroscopic analyses are run over the wavelength range 4800--8900 \AA, and a Milky Way foreground dust extinction was calculated and corrected for, using the \citet{cardelli1989} dust extinction model. 

The \ngist\ stellar kinematics (\texttt{KIN}) module employs the \textsc{python} implementation of the penalised Pixel Fitting (\textsc{pPXF}) routine of \citet{cappellari2004} and \citet{cappellari2017} and the X-shooter stellar library DR3 \citep{verro2022}. We chose the X-shooter library for its extended wavelength coverage and excellent spectral resolution about the CaT, and following the recommendations of previous works that stars are generally preferred over SSPs for stellar kinematic determinations \citep[e.g.][]{vandesande2017, belfiore2019}.  
Motivated by previous IFS works including SAURON \citep{emsellem2004}, ATLAS$^{\rm{3D}}$ \citep{cappellari2011}, SAMI \citep{vandesande2017}, MaNGA \citep{belfiore2019,westfall2019}, and PHANGS \citep{emsellem2022}, to improve the quality of the derived
kinematics, we fit a 23$^{\rm{rd}}$ order additive Legendre polynomial to provide a closer match between data and the spectral templates. This polynomial order was guided by the detailed analysis of the effect of polynomial order as a function of wavelength range by \citet{vandesande2017}. We additionally fit a 1$^{\rm{st}}$ order multiplicative polynomial, to mitigate any small variation in continuum shape from imperfect sky subtraction and dust attenuation.
We mask spectral regions affected by sky emission and nebular gas emission lines. Initial guesses for velocities were taken from the NASA Extragalactic Database, and an initial guess for the stellar velocity dispersion of 100 km s$^{-1}$. Example spectral fits can be seen in Figure~\ref{Fig:ogist_ngist_spectra}.

We use a \textsc{pPXF} penalizing bias optimised for our specific science case. The default penalization is too high, in particular in the outer discs where the velocity dispersion is close to, or below the instrumental resolution. In those regions, the default or \texttt{auto} bias resulted in the high-order moments being penalised to zero, whereas a run without any penalizing bias showed a significant detection of $h_3$ and $h_4$. 

Following the recommendations outlined in the \textsc{pPXF} code documentation, we determined the optimized bias value by performing a large ensemble of Monte Carlo simulations, testing how well the LOSVD parameters are recovered as a function of velocity dispersion. The ideal bias is defined as one that reduces the scatter in the velocity dispersion, $h_3$ and $h_4$, without creating a systematic offset in the velocity and velocity dispersion when the dispersion is larger than the spectrum's velocity scale (41 km s$^{-1}$). In the tests, we used a MILES SSP with an age of 5.01 Gyr, [M/H] = 0.0, and a wavelength range of 4800--7000\AA.

Based on the examples in \citet{emsellem2004}, \citet{cappellari2011}, and \citet{vandesande2017}, we derived a simple analytic expression for the ideal penalizing bias for MUSE spectra as a function of S/N, implemented by invoking the \texttt{`muse\_snr\_prefit'} option for the \texttt{BIAS} keyword in \ngist. We find:

\begin{equation}
bias_{\,(S/N)}  = 0.01585 \times (S/N)^{0.54640} - 0.016879.
\label{eq:bias_sn}
\end{equation}

\noindent To account for changes in the wavelength range and a number of masked spectral pixels, we then adopt a function similar to the \textsc{pPXF} \texttt{auto} bias:

\begin{equation}
BIAS_{\,(N_{\lambda pixels})}= bias_{~(S/N)} \times \sqrt{500/N_{\lambda pixels}} ~/~ 0.505,
\label{eq:bias_pixels}
\end{equation}
where $N_{\lambda pixels}$ is the number of unmasked spectral pixels used in the \ppxf fit.
\noindent For an $S/N=100$ and wavelength range of 4800--5500 \AA, 4800--7000 \AA, and 4800--8900 \AA, typical values for the penalizing bias are {0.27}, 0.18, and {0.14}, respectively. For our study, we use the S/N of each individual Voronoi bin to calculate the optimal bias.

\subsection{Imaging}
 For comparison to kinematic indicators, we examined the visual morphologies of the galaxies. $r$-band The Dark Energy Camera Legacy Survey (DECaLS) images \citep{dey2019} are available for the whole sample, which were downloaded as 10-arcmin cutouts. Mid-IR imaging suffers less from dust obscuration, and can therefore help identify structures that would otherwise have been hidden by dust. Spitzer imaging is available for NGC 3957, NGC 0522, IC 1711, NGC 0360, UGC 00903, NGC 3279, and NGC 5775, and cutouts of 3.6-$\mu$m imaging were obtained for these galaxies from the NASA IPAC Infrared Science Archive, shown in Appendix~\ref{spitzer_imaging}. 

\section{Results}
\label{results}

\begin{figure*}
\centering
\begin{subfigure}[t]{0.3\textwidth}
\centering
\includegraphics[width=\textwidth]{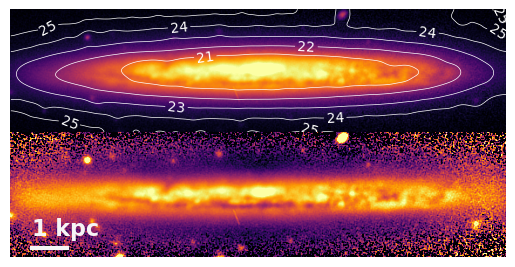} 
\caption{NGC 3279} 
\end{subfigure}
\begin{subfigure}[t]{0.3\textwidth}
\centering
\includegraphics[width=\textwidth]{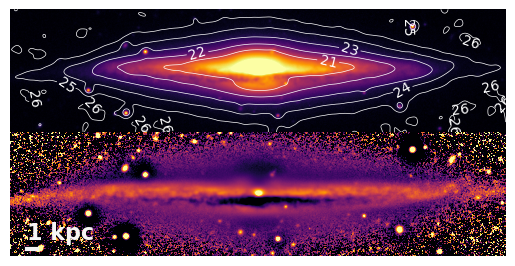} 
\caption{IC 5244$^{\star}$}
\end{subfigure}
\begin{subfigure}[t]{0.3\textwidth}
\centering
\includegraphics[width=\textwidth]{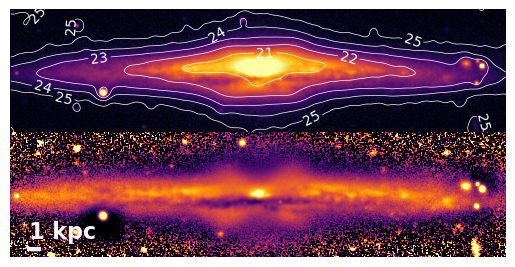} 
\caption{PGC 044931$^{\star}$} 
 \end{subfigure}

\begin{subfigure}[t]{0.3\textwidth}
\centering
\includegraphics[width=\textwidth]{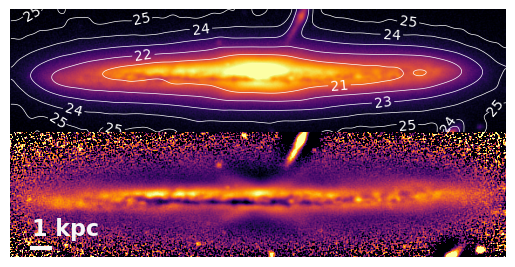} 
\caption{NGC 0522$^{\star}$} 
\end{subfigure}
\begin{subfigure}[t]{0.3\textwidth}
\centering
\includegraphics[width=\textwidth]{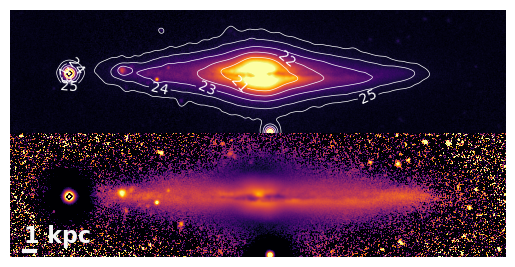} 
\caption{IC 1711$^{\star}$} 
\end{subfigure}
\begin{subfigure}[t]{0.3\textwidth}
\centering
\includegraphics[width=\textwidth]{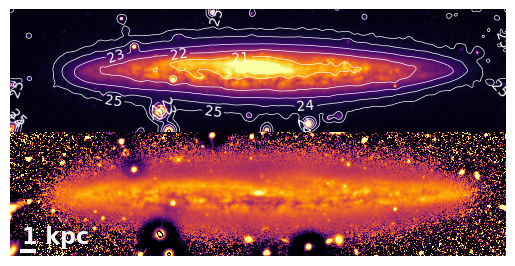} 
\caption{NGC 0360}
 \end{subfigure} 

\begin{subfigure}[t]{0.3\textwidth}
\centering
\includegraphics[width=\textwidth]{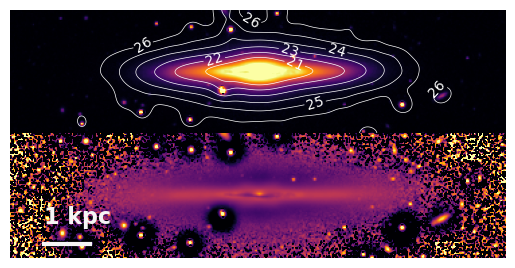} 
\caption{NGC 3957$^{\star}$} 
\end{subfigure}
\begin{subfigure}[t]{0.3\textwidth}
\centering
\includegraphics[width=\textwidth]{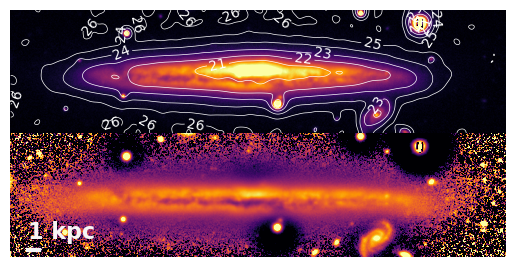} 
\caption{ESO 120-016$^{\star}$} 
\end{subfigure}
\begin{subfigure}[t]{0.3\textwidth}
\centering
\includegraphics[width=\textwidth]{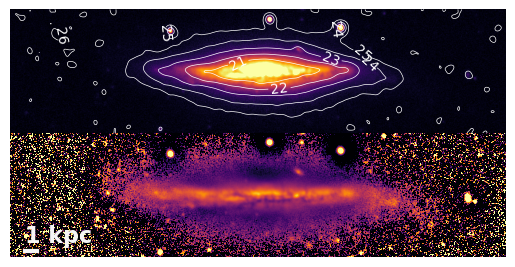} 
\caption{UGC 00903} 
 \end{subfigure} 

 \begin{subfigure}[t]{0.3\textwidth}
\centering
\includegraphics[width=\textwidth]{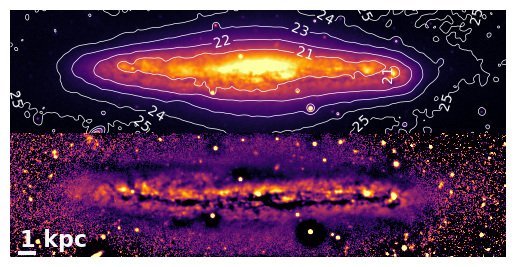} 
\caption{NGC 5775}
\end{subfigure}
\begin{subfigure}[t]{0.3\textwidth}
\centering
\includegraphics[width=\textwidth]{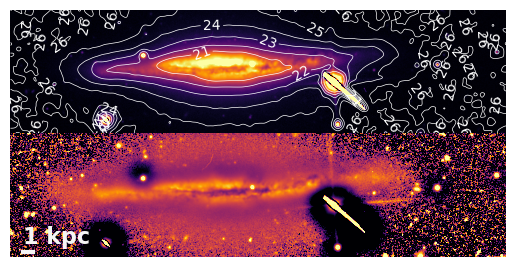} 
\caption{ESO 079-003$^{\star}$} 
\end{subfigure}
\begin{subfigure}[t]{0.3\textwidth}
\centering
\includegraphics[width=\textwidth]{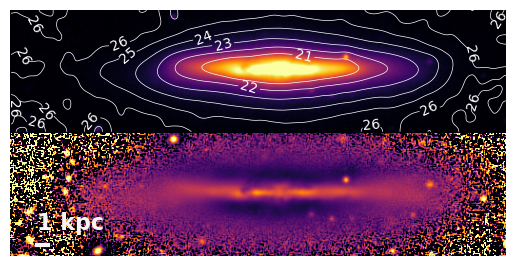} 
\caption{ESO 484-036$^{\star}$} 
 \end{subfigure} 

 \caption{$r$-band DECaLS image (top) scaled with an inverse hyperbolic sine ($\arcsinh$) scaling with surface brightness contours overlaid in white (units of $r$-band magnitude arcsec$^{-2}$), and unsharp masked image created with a Gaussian smoothing kernel of width $\sigma=7-20$ pixels (bottom). Galaxy names marked with a $\star$ are considered to host a boxy-peanut central structure in this work. The orientation of galaxies here, and throughout, is the same as in Fig~\ref{Fig:cube_images}.}
 \label{Fig:unsharp}
\end{figure*}

\subsection{Off-plane structure from unsharp masked imaging}
We first wish to classify the visual morphology of the off-plane structures in the GECKOS galaxies. To do so, we examine unsharp-masked DECaLS DR9 $r$-band images of each galaxy in the sample and 3.6$-\mu$m Spitzer imaging where available. To supplement our imaging data, we also refer to historical morphological classifications from the literature.

Unsharp masking is a common technique whereby a smoothed version of an image is divided by the unsmoothed version to highlight `sharp' high-spatial-frequency features. It has been shown to work well for identifying X-shaped BP bulge features \citep[e.g.][]{aronica2003,bureau2006,laurikainen2017}. We chose DECaLS imaging for this task, as it has good spatial resolution and an average full width at half maximum of 1$^{\prime\prime}$.2 in the $r$ band \citep{dey2019}. In addition, this imaging is available for the entire GECKOS iDR1 sample. 
We note that the three-colour images derived from the MUSE cubes themselves could also be suitable for this purpose, though given that the whole galaxy is not included in the MUSE field of view, it can be a little more difficult to elucide central structures. 
We first smooth a DECaLS $r$-band cutout image of each galaxy with a 2D circular Gaussian kernel with $\sigma=7-20$ pixels depending on galaxy distance (corresponding to $\sim2-5^{\prime\prime}$), and then divide the original image by this smoothed image. 

The results of the unsharp masking are shown in Figure~\ref{Fig:unsharp}. Immediately, we identify two obvious X-shaped bulges from the DECaLS imaging: PGC 044931 and NGC 0522. In addition, faint peanut-shaped structure is seen in IC 5244, IC 1711, ESO 079-003, ESO 484-036, and possibly ESO 120-016. NGC 3279 and UGC 00903 do not display any BP bulge structure, and while spiral arms are visible in NGC 0360 and NGC 5775, no vertically extended bulge features are seen in these galaxies, either. 

IR imaging can also help to reveal BP bulge structures as it suffers less from dust obscuration, and Spitzer 3.6-$\mu$m imaging is shown where available in Figure~\ref{Fig:spitzer}. We also perform unsharp masking on the 3.6-$\mu$m imaging as a comparison. We note that a faint boxy or peanut shape is visible in NGC 3957, NGC 0522, and IC 1711, and clear spiral arms for NGC 0360 and NGC 5775.

From the combined DECaLS and Spitzer images (where available), we reach the consensus that NGC 3957, PGC 044931, NGC 0522, IC 1711, IC 5244, ESO 079-003, and ESO 484--036 contain BP bulge structures. For ESO 120-016, we find tentative evidence for a boxy shape.

\subsection{Visual morphologies from the literature}

\begin{table*}
\centering                        
 \begin{threeparttable}
\caption{Summary of galaxy morphological classifications from the literature.} 
\label{table:classifications}    
\begin{tabular}{l c c c c}       
\hline\hline   
  & \multicolumn{4}{c}{\textbf{Bulge Type}} \\
Galaxy & This work\tnote{1,2} &\citet{lutticke2000}\tnote{3} & \citet{buta2015}\tnote{2} & \citet{bureau1999} \\   
\hline                   
   NGC 3279 & No obv.\ bulge & Elliptical & No BP & -- \\     
   IC 5244 & BP & Close to box-shaped & -- & -- \\
   PGC 044931 & BP & -- & -- & BP \\
   NGC 0522 & BP & Box-shaped & BP & -- \\
   IC 1711 & BP & Elliptical & BP & -- \\ 
   NGC 0360 & No obv.\ bulge & Close to box-shaped & -- \\
   NGC 3957 & BP & Close to box-shaped & -- & BP  \\
   ESO 120-016 & BP & Elliptical & -- & -- \\
   UGC 00903 & No obv.\ bulge & -- & No BP & --\\
   NGC 5775 & No obv.\ bulge & Box-shaped & No BP & -- \\
   ESO 079-003 & BP & Close to box-shaped & -- \\
   ESO 484-036 & BP & -- & -- & -- \\
 
\hline                                   
\end{tabular}
 \begin{tablenotes}
    \item [1] DECaLS imaging; pixel scale 1.7$^{\prime\prime}$ pixel$^{-1}$.
       \item [2] Spitzer 3.6-$\mu$m imaging; pixel scale 0.75$^{\prime\prime}$ pixel$^{-1}$.
   \item [3] Digitised Sky Survey imaging; pixel scale 1.7$^{\prime\prime}$ pixel$^{-1}$.
\tablefoot{Classifications from this work come from the unsharp masking of DECaLS imaging. A `--' denotes that a galaxy was not included in a particular sample.}
 \end{tablenotes}
 \end{threeparttable}
\end{table*}

There have been several previous surveys that classified central structure in large samples of galaxies from imaging of varying spatial resolution, including \citet{lutticke2000} and \citet{buta2015}. The results of literature classifications for the galaxies in this work are shown in Table~\ref{table:classifications}, along with the classifications from unsharp masking described above.
Neither ESO 484-036 nor PGC 044931 were classified by either of these works, though PGC 044931 has been studied extensively and classified as boxy/peanut by works such as \citet{bureau1999} and \citet{chung2004}. The classification of NGC 3957 by \citet{bureau1999} differs from \citet{lutticke2000}, where the former classifies it as an ellipsoidal bulge, while the latter calls it boxy. In this current work, we see a clear X-shape in the Spitzer imaging, confirming the unsharp masking results of \citet{bureau2006}. \citet{lutticke2000} also classify NGC 0360 as `close to box-shaped', while we see no evidence of this in the DECaLS or Spitzer imaging. Indeed, central, tightly-wound spiral arms are clearly visible in the 3.6-$\mu$m imaging, which may have been confused for a box shape in optical and lower-resolution imaging. Given the poor spatial resolution of DSS imaging, we take the classifications from this current work, \cite{buta2015}, and \citet{bureau1999} as our primary references.
 
\subsection{Stellar kinematic maps and radial profiles}
\label{results_maps}
In Figures~\ref{Fig:vel_maps1}--\ref{Fig:vel_maps12}, we present the derived stellar line-of-sight luminosity-weighted mean velocity ($V_{\star}$), velocity dispersion ($\sigma_{\star}$), and high-order moments skew ($h_{3}$) and kurtosis ($h_{4}$) for the GECKOS galaxies, derived over the wavelength range 4800--8900 \AA.
We also tested just the Calcium II triplet (CaT) range \citep[8450--8735 \AA, similar to][]{mcdermid2002}, but found that while the redder wavelength range resulted in more spatially coherent structure in the high-order moment maps, this short wavelength range often failed to produce a cohesive $h_{3}$ and $h_{4}$ signal in the outer, lower dispersion regions of some of the galactic discs.
For a further discussion of wavelength-dependent kinematic structures, see Appendix~\ref{wavelength_dependence}. 
In Figures~\ref{Fig:vel_maps1}--\ref{Fig:vel_maps12} each galaxy is rotated such that the photometric major axis lies horizontal and the dust lane (if offset), is always to the lower side of the image. We note that for some galaxies, the MUSE mosaic covers only one side of the disc. The GECKOS pointing strategy was such that the entirety of the bulge was included in the MUSE pointings, however. 
At first glance, all galaxies appear to host regularly rotating discs. However, upon more careful examination, the diversity in structure in the kinematic maps is considerable. 

For a more straightforward comparison to previous literature, Figures~\ref{Fig:linemaps1} and \ref{Fig:linemaps2} depict the 1D radial profiles extracted from the corresponding 2D maps. These profiles were extracted from slits of width 1 kpc between $-0.3<\rm{z}<0.7$, which we found to be wide enough to encompass central kinematic structure of interest, whilst minimising the contribution of dust lane regions, keeping in mind that the dust lanes are always offset to the lower side of the images for which we place the slits on. 
Furthermore, rather than presenting $h_{3}$ directly against $V_\star$, we present all 1D kinematic quantities as a function of galactocentric radius. 
To facilitate interpretation, for 1D profiles we compare the radial gradients of $V_{\star}$ and $h_{3}$, where if both gradients are positive or negative, the quantities are correlated, but if one is positive and the other is negative, the two quantities are anti-correlated.
For the 2D maps, we discuss the \textit{sign} of the $V_{\star}$ and $h_{3}$ regions (i.e.\ a positive value of $V_{\star}$ and $h_{3}$ have a matching sign).

To aid in the recognition of kinematic structure, we have flipped the direction of the profiles of ESO 120-016, NGC 0360, NGC 0522, ESO 484-036, IC 1711, and IC 5244 such that the receding side of the resultant rotation curve is on the right for all galaxies. 
Below, we describe the trends seen in the 2D kinematic maps and 1D radial profiles, and in Section~\ref{discussion} we interpret these results in terms of galactic sub-structures.

\subsubsection{Velocity, $V_{\star}$, maps and profiles}
\label{sect:velocity}
\begin{figure*}[h]
\centering
\begin{subfigure}[t]{0.97\textwidth}
\centering
\includegraphics[trim={0cm 0cm 0cm 0cm},clip, width=\textwidth]{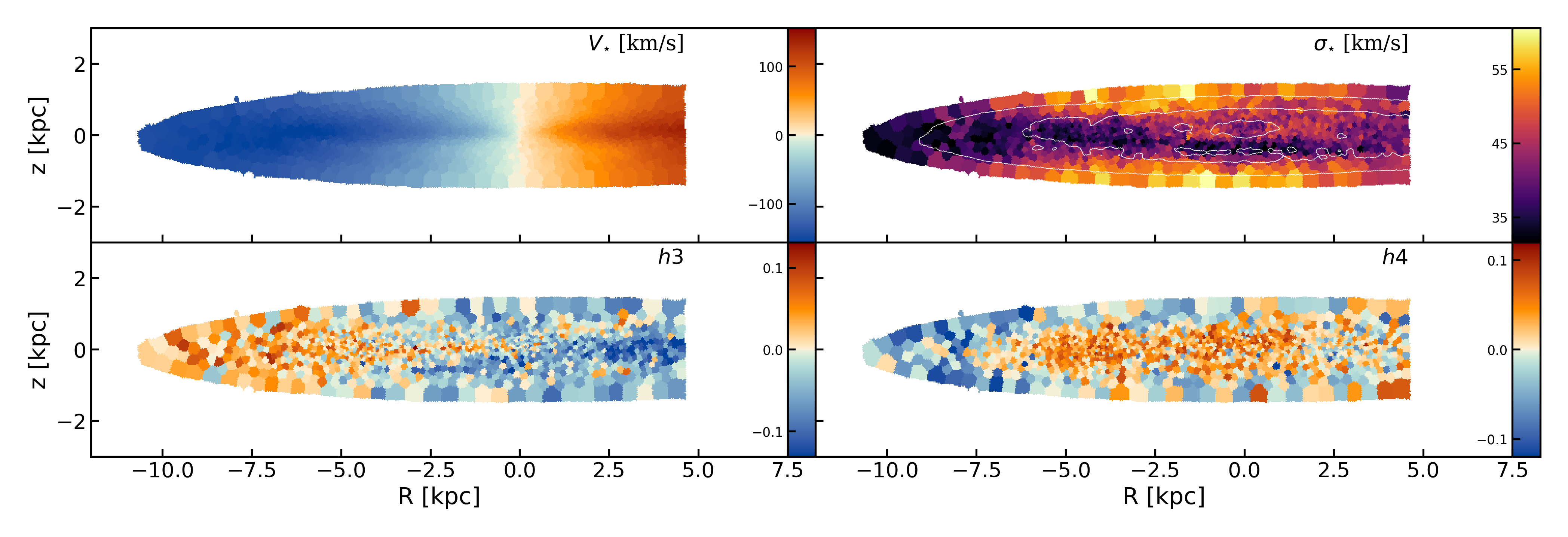} 
\end{subfigure}
 \caption{Stellar luminosity-weighted line-of-sight velocity, $V_\star$, (top left), velocity dispersion, $\sigma_{\star}$, (top right, with $r$-band surface brightness contours overlaid in white), $h_{3}$ (bottom left) and $h_{4}$ (bottom right) maps derived with \ngist\ and binned to S/N=100 for the galaxy \textbf{NGC 3279}, a galaxy with no obvious visual bulge structure. Galaxies are shown such that the dust lane (and photometric major axis) are oriented horizontally, and such that the dust lane (if offset) is to the lower side of the image. Holes in the maps are regions where bright stars have been masked. Flux contours are overlaid on the $\sigma_\star$ maps to highlight the dust lane effects, but not for the $V_\star$, $h_{3}$ and $h_{4}$ maps, to aid in central structure recognition.}
 \label{Fig:vel_maps1}
 \end{figure*}

\begin{figure*}[h]
\centering
\begin{subfigure}[t]{0.97\textwidth}
\centering
\includegraphics[trim={0cm 0cm 0cm 0cm},clip, width=\textwidth]{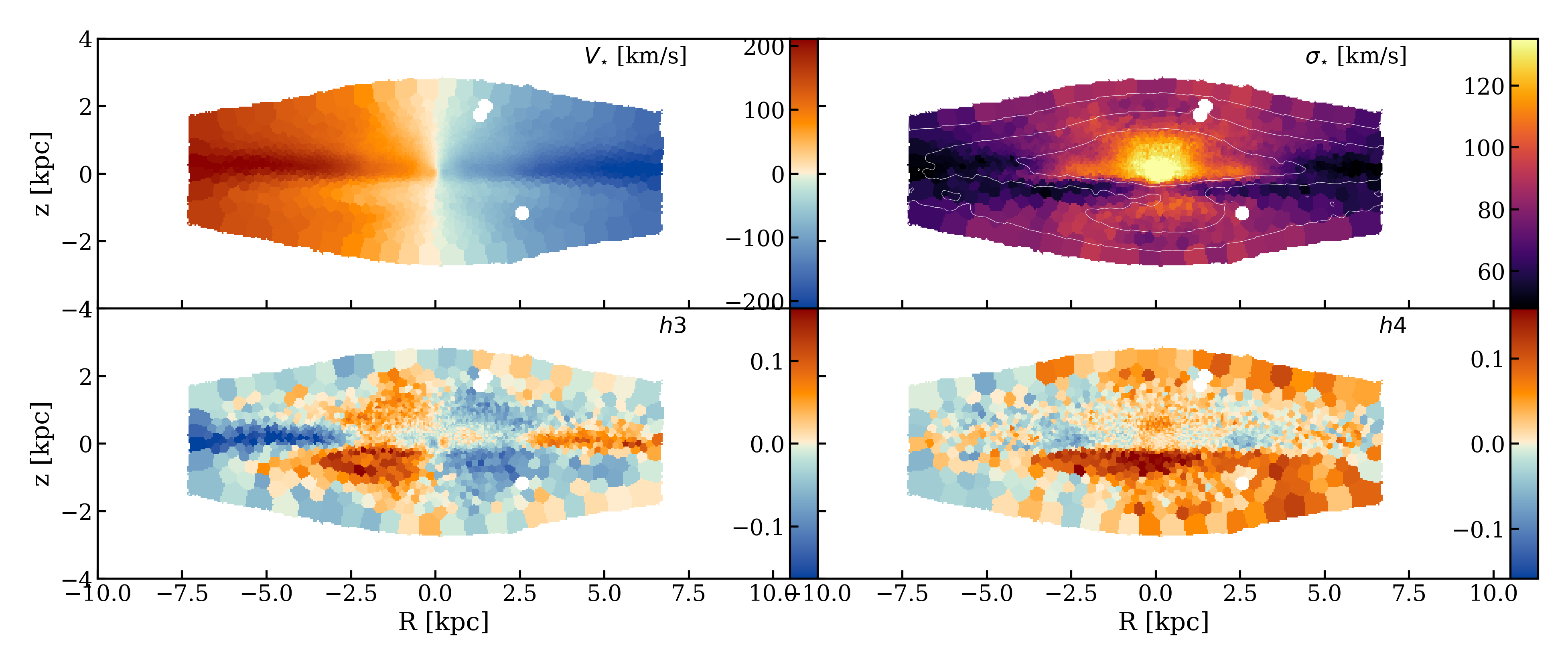} 
\end{subfigure}
 \caption{As for Figure~\ref{Fig:vel_maps1}, but for \textbf{IC 5244} - BP bulge.}
 \label{Fig:vel_maps2}
 \end{figure*}

\begin{figure*}[h]
\centering
\begin{subfigure}[t]{0.97\textwidth}
\centering
\includegraphics[trim={0cm 0cm 0cm 0cm},clip,width=\textwidth]{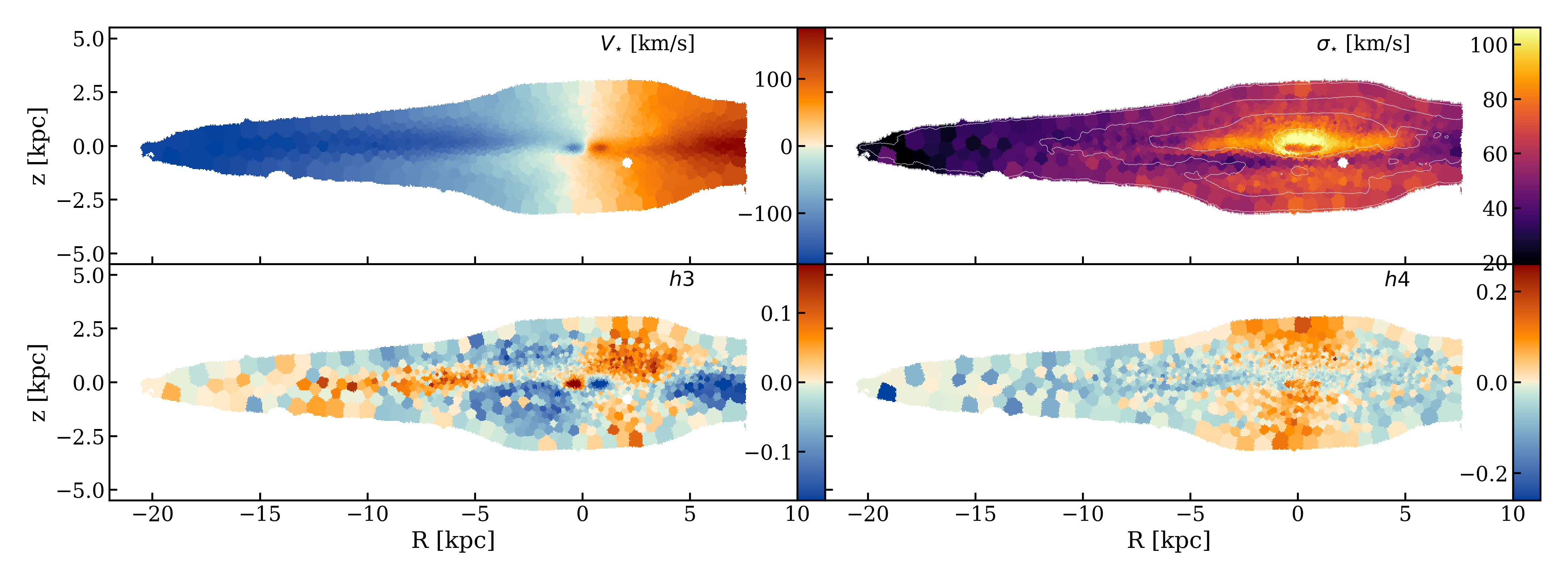} 
\end{subfigure}
 \caption{As for Figure~\ref{Fig:vel_maps1}, but for \textbf{PGC 044931} - BP bulge.}
  \label{Fig:vel_maps3}
 \end{figure*}

\begin{figure*}[h]
\centering
\begin{subfigure}[t]{0.97\textwidth}
\centering
\includegraphics[trim={0cm 0cm 0cm 0cm},clip,width=\textwidth]{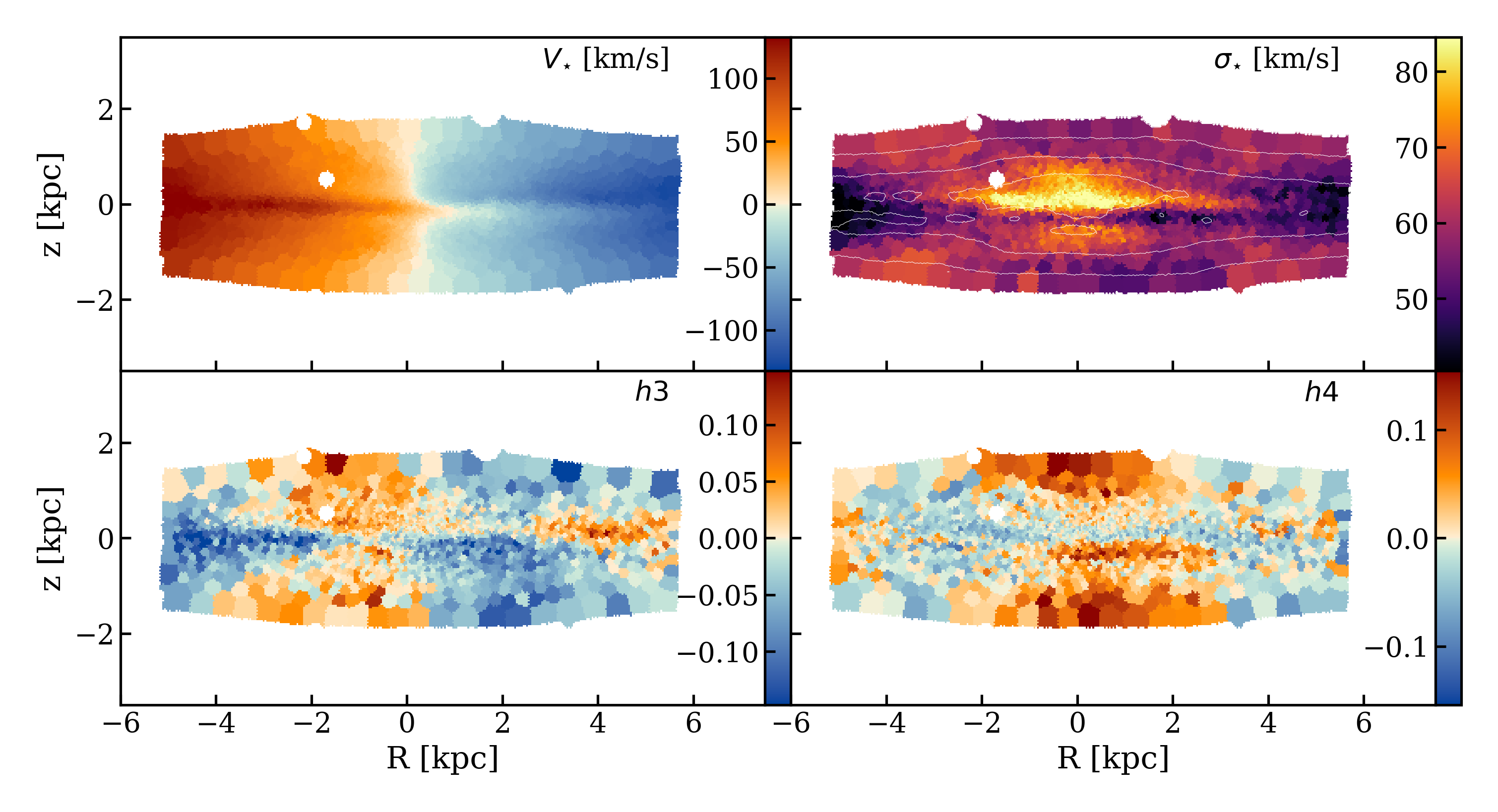} 
\end{subfigure}
 \caption{As for Figure~\ref{Fig:vel_maps1}, but for \textbf{NGC 0522} - BP bulge.}
  \label{Fig:vel_maps4}
 \end{figure*}

\begin{figure*}[h]
\centering
\begin{subfigure}[t]{0.97\textwidth}
\centering
\includegraphics[trim={0cm 0cm 0cm 0cm},clip,width=\textwidth]{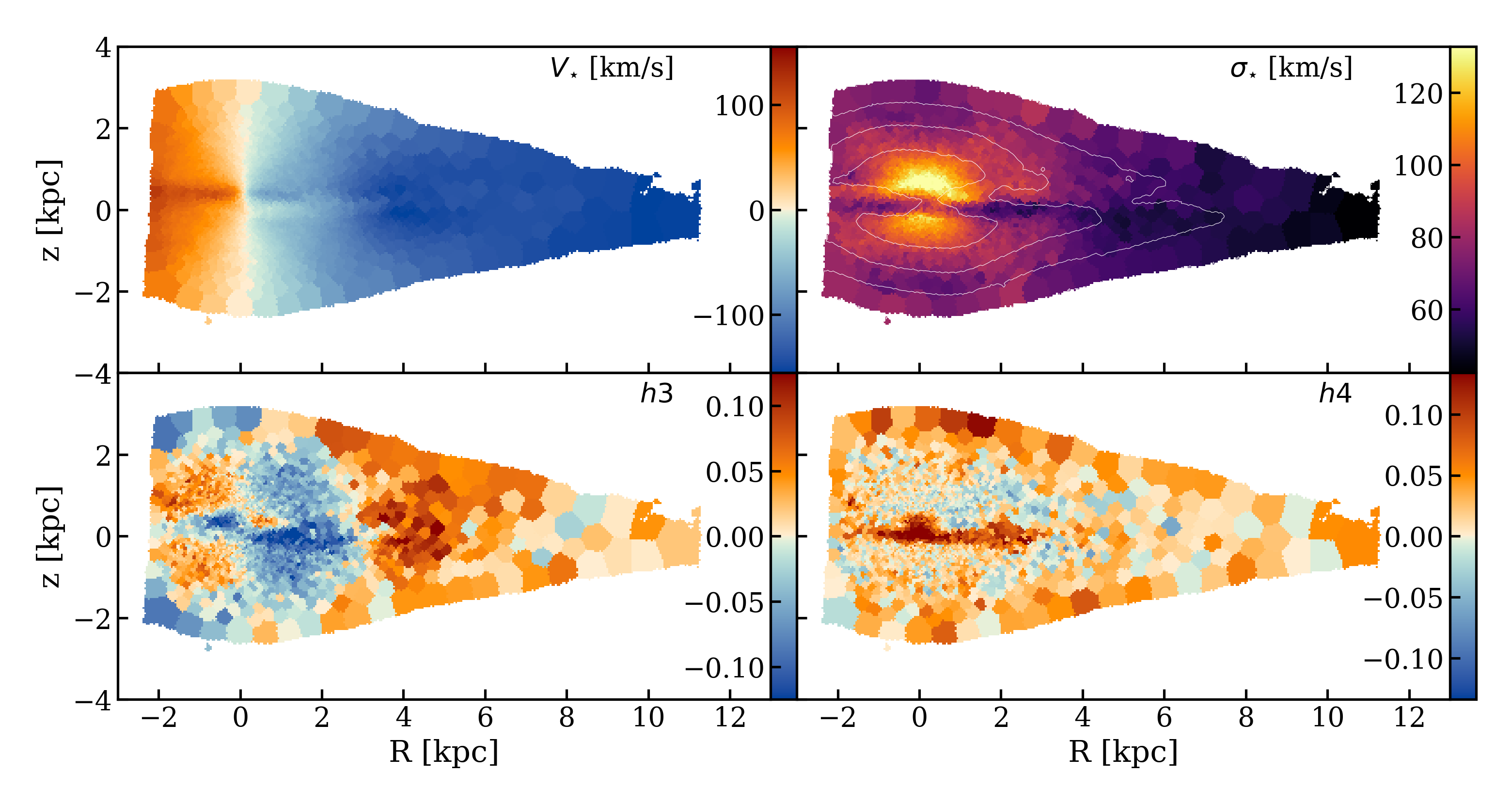} 
\end{subfigure}
 \caption{As for Figure~\ref{Fig:vel_maps1}, but for \textbf{IC 1711} - BP bulge.}
  \label{Fig:vel_maps5}
 \end{figure*}

\begin{figure*}[h]
\centering
\begin{subfigure}[t]{0.97\textwidth}
\centering
\includegraphics[trim={0cm 0cm 0cm 0cm},clip,width=\textwidth]{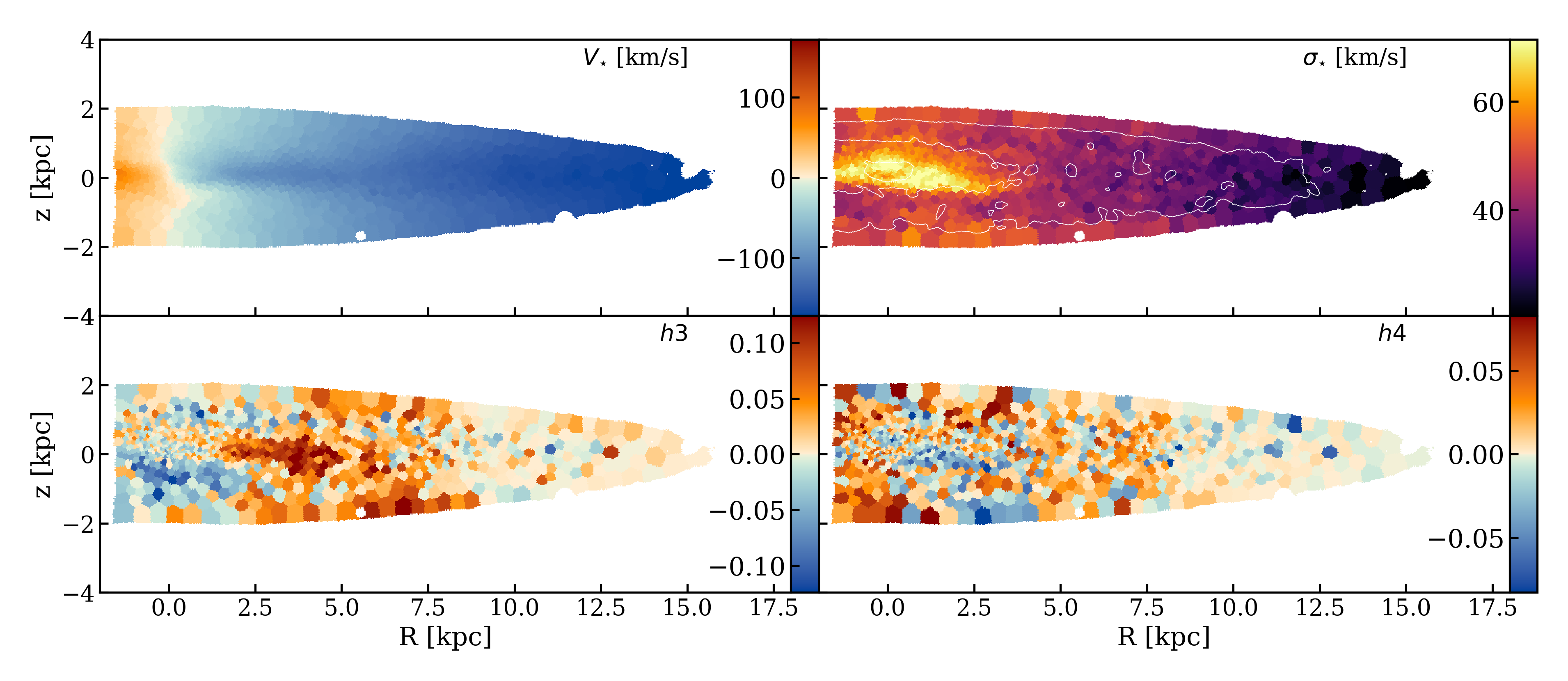} 
\end{subfigure}
 \caption{As for Figure~\ref{Fig:vel_maps1}, but for \textbf{NGC 0360} - no obvious visual BP bulge structure.}
  \label{Fig:vel_maps6}
 \end{figure*}

\begin{figure*}[h]
\centering
\begin{subfigure}[t]{0.97\textwidth}
\centering
\includegraphics[trim={0cm 0cm 0cm 0cm},clip,width=\textwidth]{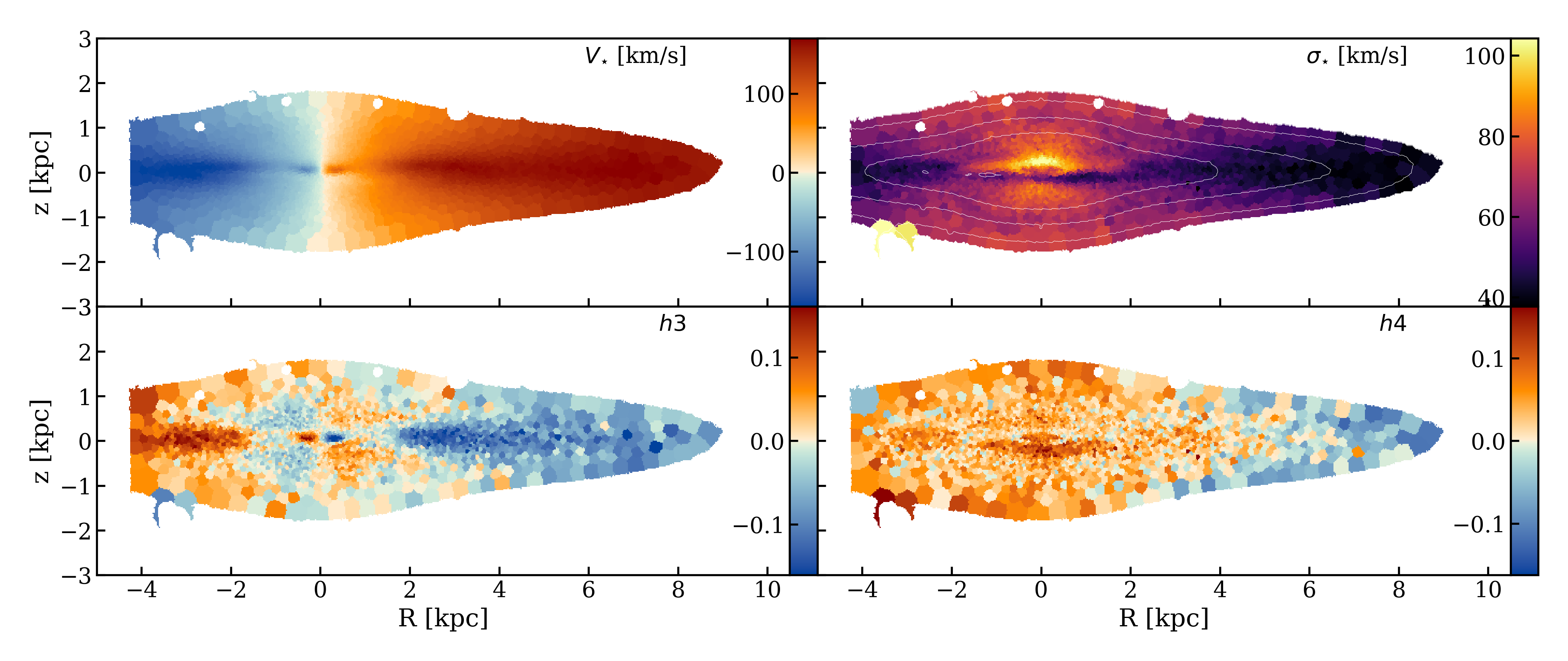} 
\end{subfigure}
 \caption{As for Figure~\ref{Fig:vel_maps1}, but for \textbf{NGC 3957} - BP bulge.}
  \label{Fig:vel_maps7}
 \end{figure*}

\begin{figure*}[h]
\centering
\begin{subfigure}[t]{0.97\textwidth}
\centering
\includegraphics[trim={0cm 0cm 0cm 0cm},clip,width=\textwidth]{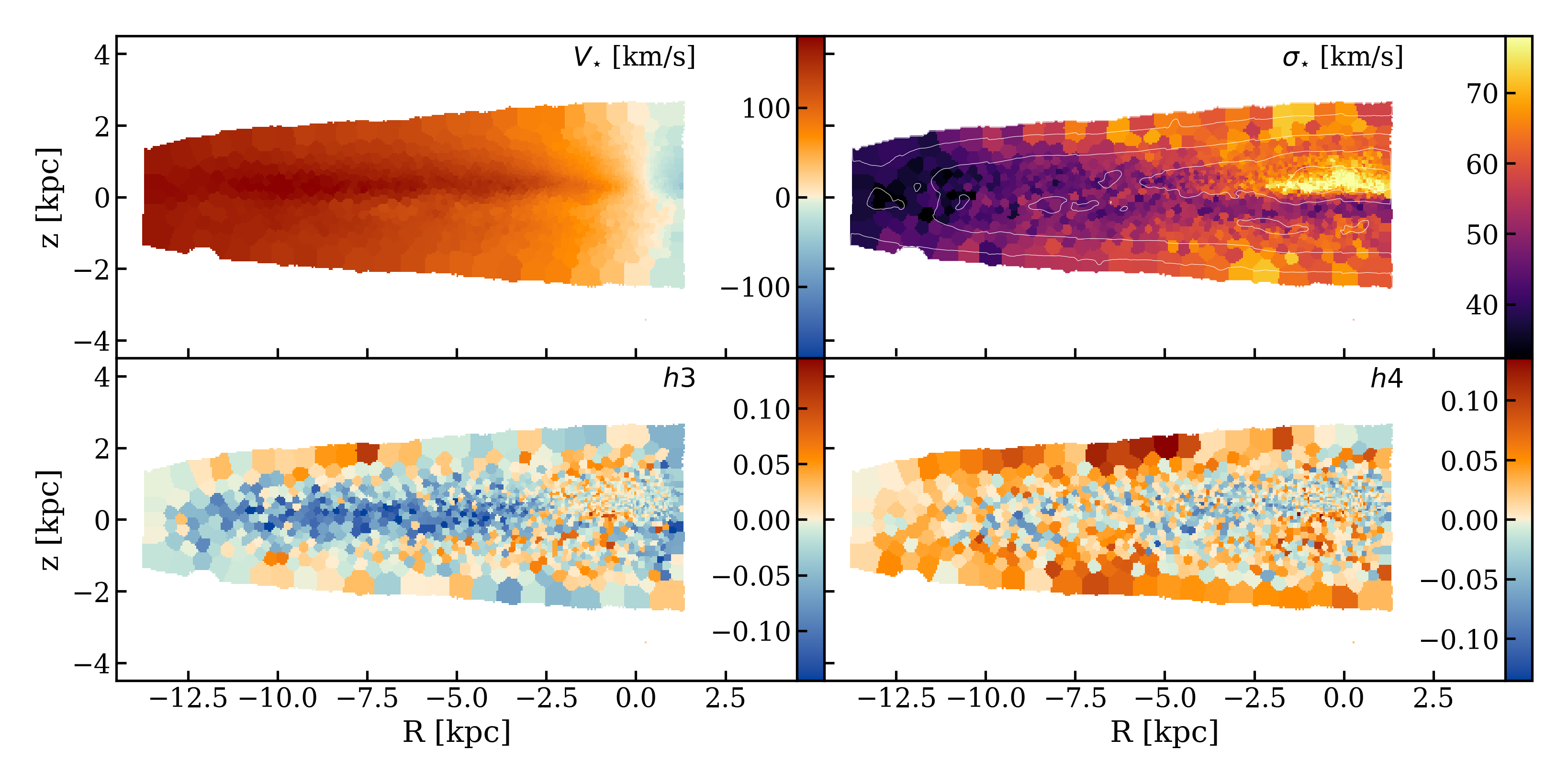} 
\end{subfigure}
 \caption{As for Figure~\ref{Fig:vel_maps1}, but for \textbf{ESO 120-016} - BP bulge.}
  \label{Fig:vel_maps8}
 \end{figure*}

\begin{figure*}[h]
\centering
\begin{subfigure}[t]{0.97\textwidth}
\centering
\includegraphics[trim={0cm 0cm 0cm 0cm},clip,width=\textwidth]{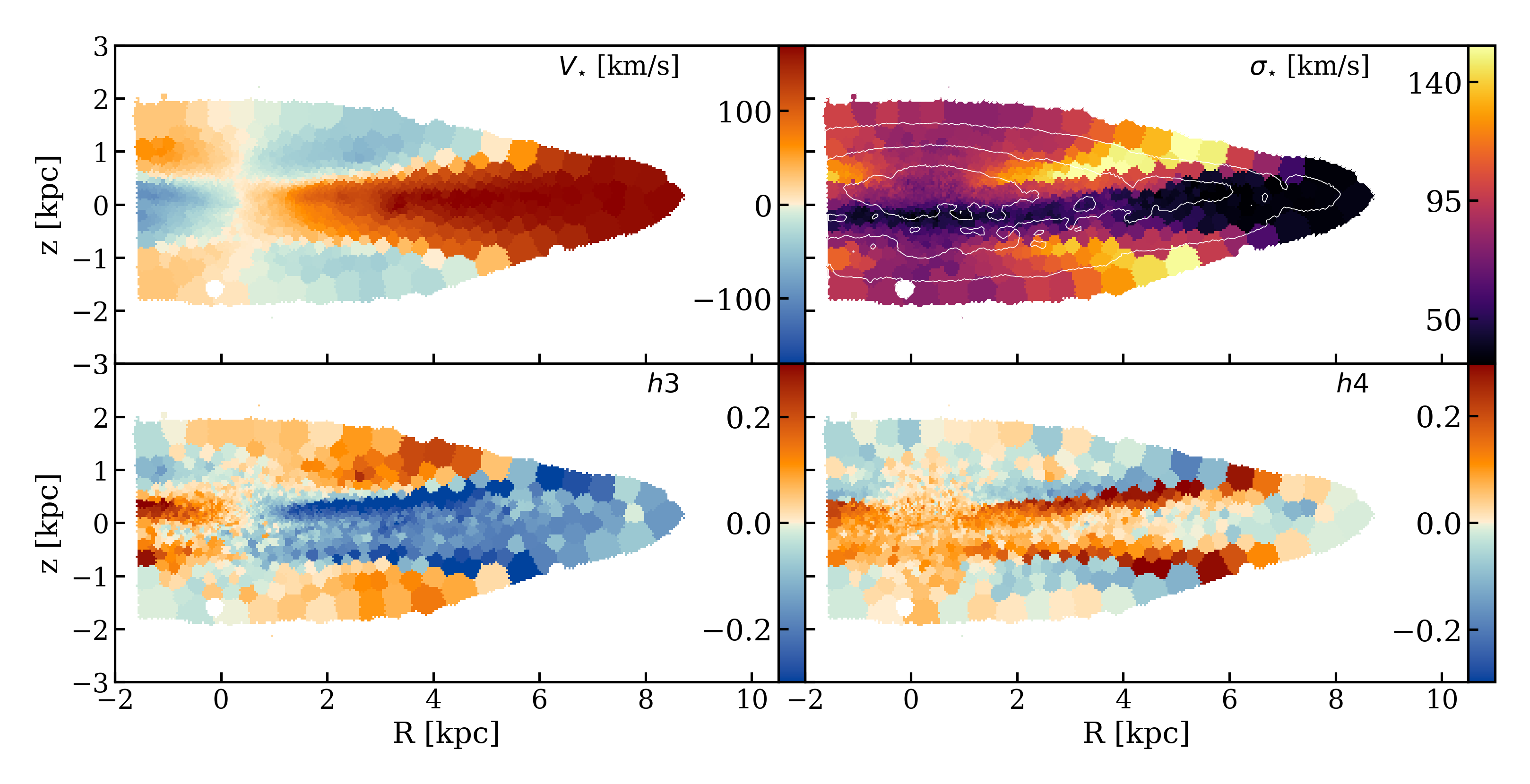} 
\end{subfigure}
 \caption{As for Figure~\ref{Fig:vel_maps1}, but for \textbf{UGC 00903} - no obvious visual BP bulge structure.}
  \label{Fig:vel_maps9}
 \end{figure*}

\begin{figure*}[h]
\centering
\begin{subfigure}[t]{0.97\textwidth}
\centering
\includegraphics[trim={0cm 0cm 0cm 0cm},clip,width=\textwidth]{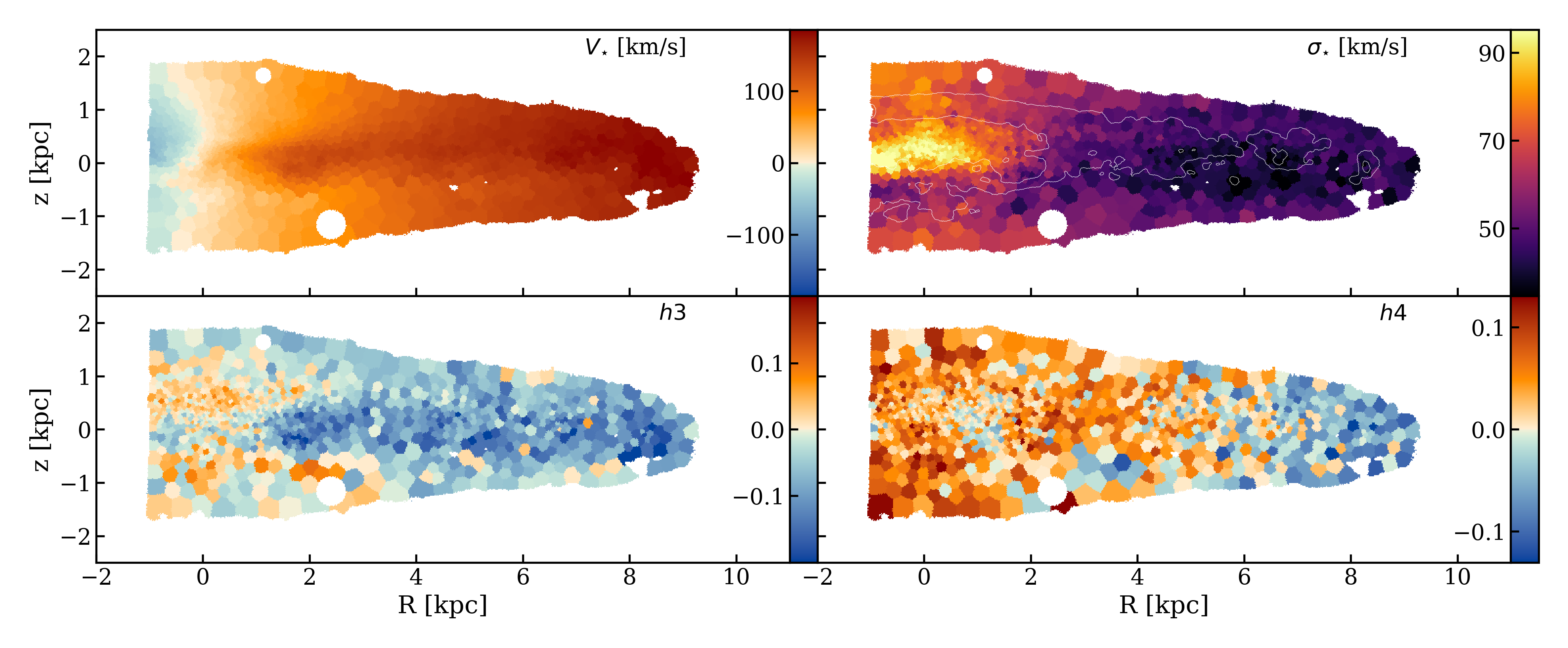} 
\end{subfigure}
 \caption{As for Figure~\ref{Fig:vel_maps1}, but for \textbf{NGC 5775} - no obvious visual BP bulge structure.}
 \label{Fig:vel_maps10}
 \end{figure*}

\begin{figure*}[h]
\centering
\begin{subfigure}[t]{0.97\textwidth}
\centering
\includegraphics[trim={0cm 0cm 0cm 0cm},clip,width=\textwidth]{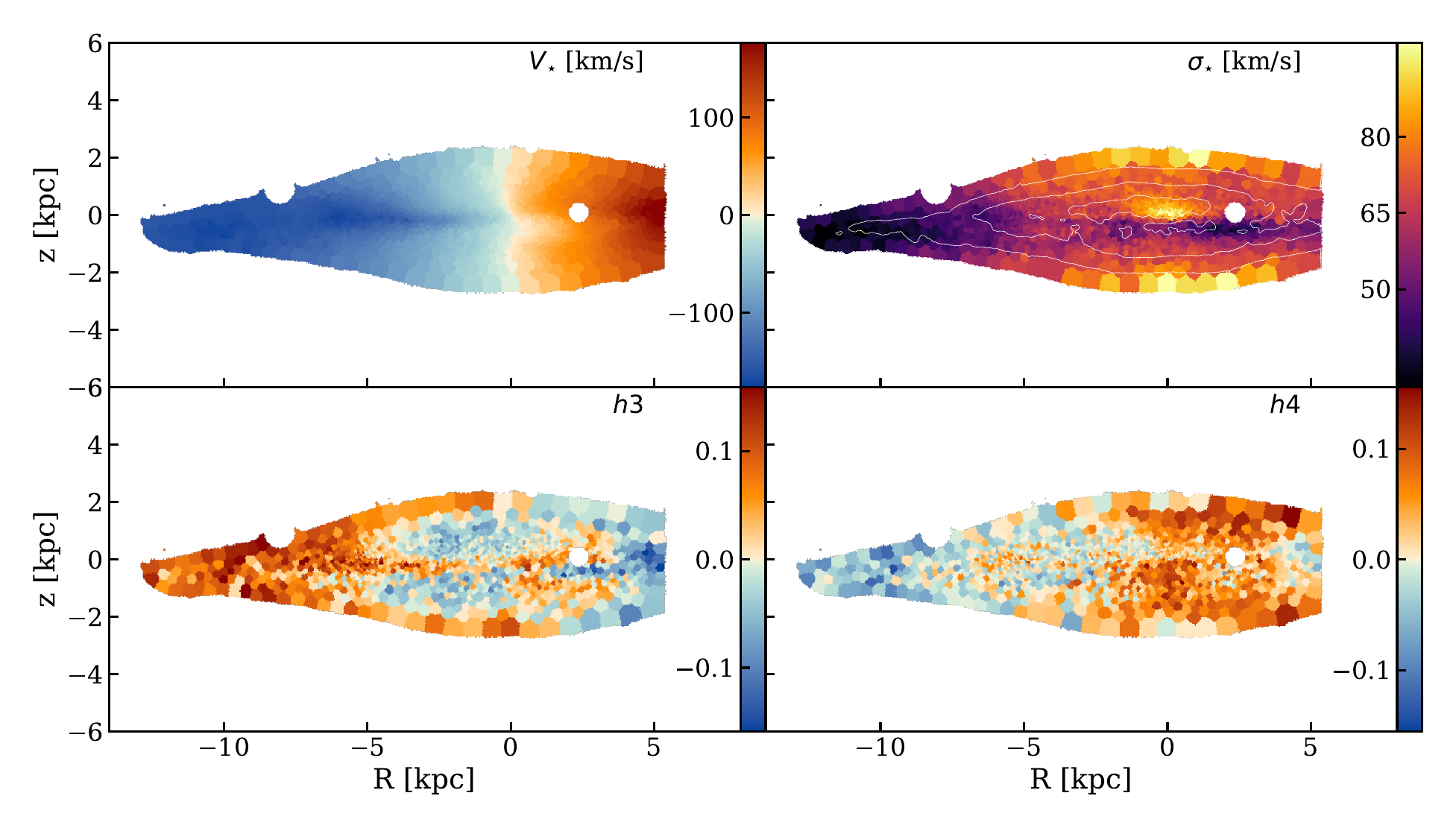} 
\end{subfigure}
 \caption{As for Figure~\ref{Fig:vel_maps1}, but for \textbf{ESO 079-003} - BP bulge.}
  \label{Fig:vel_maps11}
 \end{figure*}

\begin{figure*}[h]
\centering
\begin{subfigure}[t]{0.97\textwidth}
\centering
\includegraphics[trim={0cm 0cm 0cm 0cm},clip,width=\textwidth]{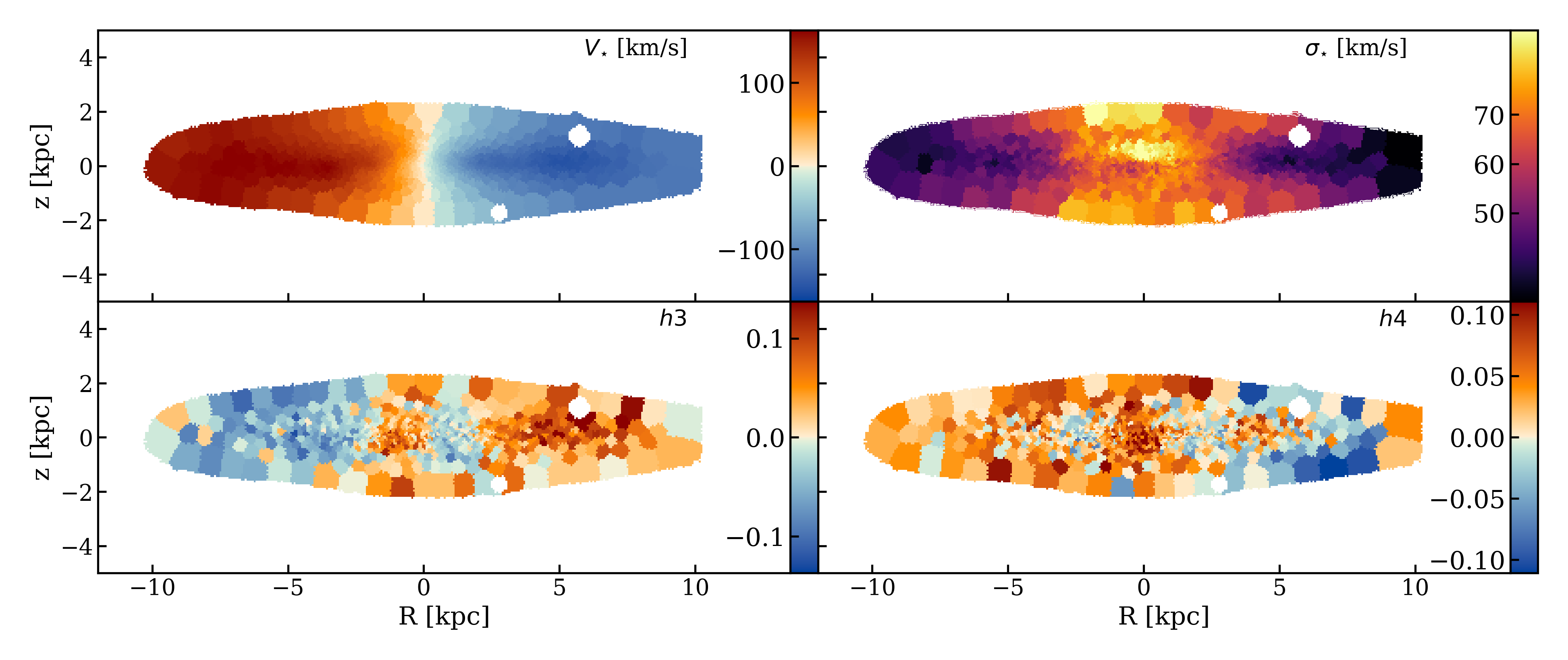} 
\end{subfigure}
\caption{As for Figure~\ref{Fig:vel_maps1}, but for \textbf{ESO 484-036} - BP bulge.}
\label{Fig:vel_maps12}
\end{figure*}

The top left panels of Figures~\ref{Fig:vel_maps1}--\ref{Fig:vel_maps12} display $V_{\star}$ maps, corrected for systemic velocity using a box of width 2 kpc extending perpendicular to the galaxy kinematic major axis, centred on the flux centre of the galaxy, and extending to only the side of the galaxy that is less dust affected (i.e.\ positive values of z in Figures~\ref{Fig:vel_maps1}--\ref{Fig:vel_maps12}). We mask regions with E(B-V)>0.4, and recalculate the median velocity within this region before subtracting it from the maps.
In all cases, we see evidence of well-behaved, regularly-rotating discs, with the notable exception of UGC 00903, which appears to comprise two counter-rotating stellar discs (see van de Sande et al.\ in prep.\ for further insight). In some cases, the dust lane is evident as a region lighter in colour, indicating velocity magnitude lower than its surroundings, located just below the z=0 line. For many galaxies the  $V_{\star}=0$ isovelocity line is not straight; rather, often warps, wedges, or other structures are visible both along the z=0 line, and at higher latitudes. 
For several galaxies (e.g.\ PGC 044931, NGC 0522, NGC 0360, and ESO 079-003), the kinematic minor axis displays a wedge-like structure within $\pm0.5$kpc of the z=0 line. The co-spatial nature of young stars and dust make light-weighted stellar kinematic maps difficult to interpret in highly obscured regions, though we speculate on the commonly observed features in Section~\ref{subsubsec:off-plane}.

\begin{figure*}[h]
\centering
\begin{subfigure}[t]{0.47\textwidth}
\centering
\includegraphics[width=\textwidth]{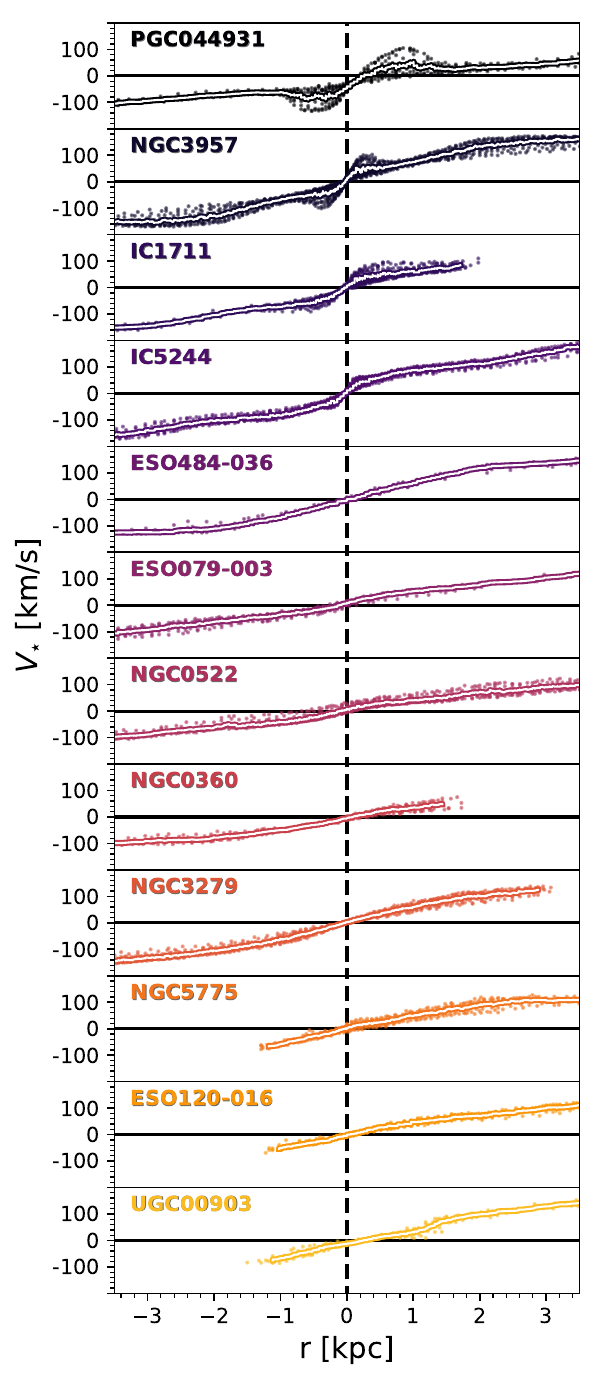} 
\caption{1D line maps of $V_\star$}
\end{subfigure}
\begin{subfigure}[t]{0.465\textwidth}
\centering
\includegraphics[width=\textwidth]{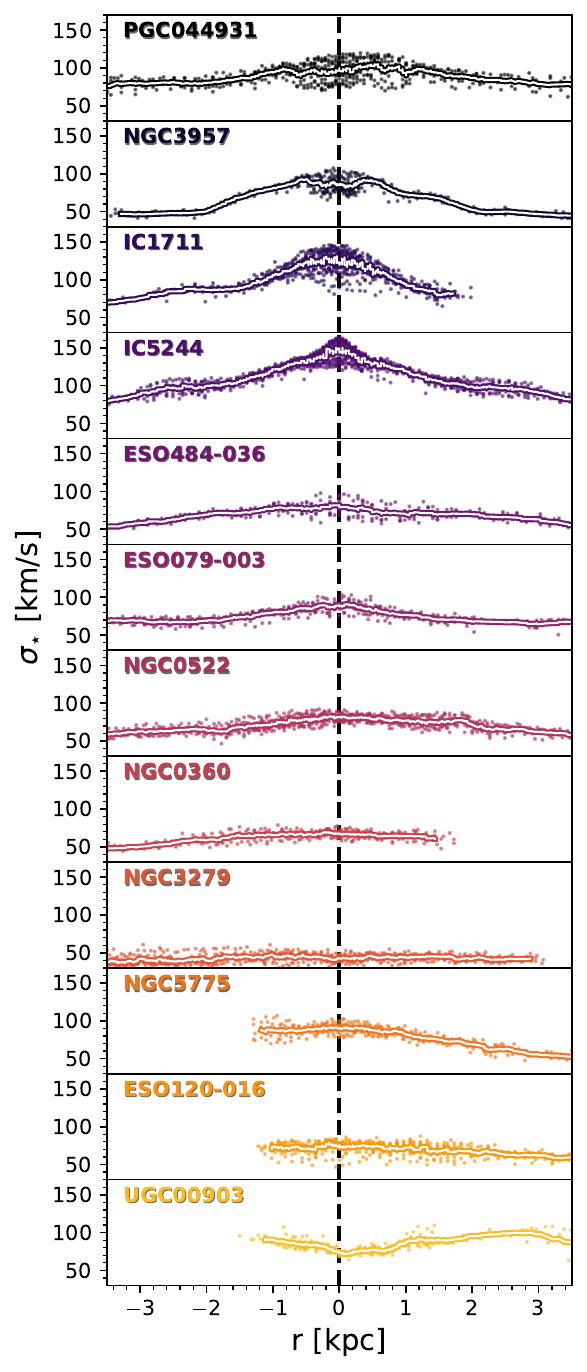} 
\caption{1D line maps of $\sigma_{\star}$}
\end{subfigure}
\caption{1D line profiles, extracted from slits of width the $0 < \rm{z} < 0.7$ kpc. Each coloured point represents a Voronoi bin, and the white line is a running median. To enable easier comparisons, in all cases, the galaxy is rotated such that the receding side is to the right. Typical median errors are $V_{\star,err}\sim2$ km s$^{-1}$ and $\sigma_{\star,err} \sim 5$ km s$^{-1}$}
\label{Fig:linemaps1}
\end{figure*}

\begin{figure*}[h]
\centering
\begin{subfigure}[t]{0.47\textwidth}
\centering
\includegraphics[width=\textwidth]{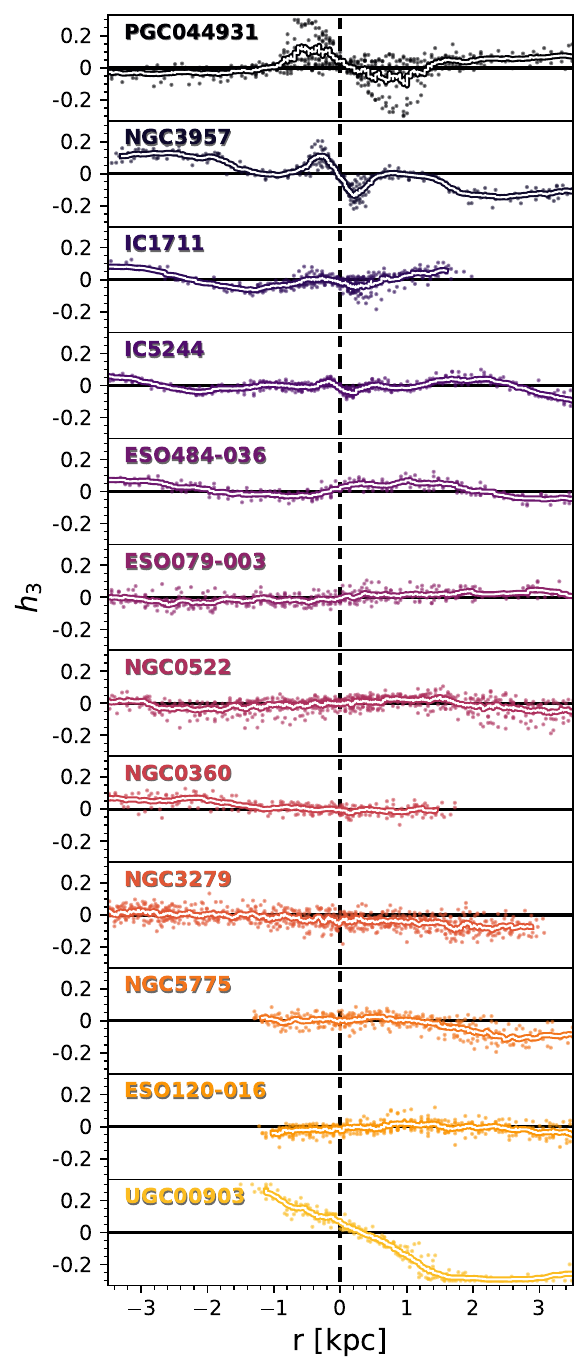} 
\caption{1D line maps of $h_{3}$}
\end{subfigure}
\begin{subfigure}[t]{0.47\textwidth}
\centering
\includegraphics[width=\textwidth]{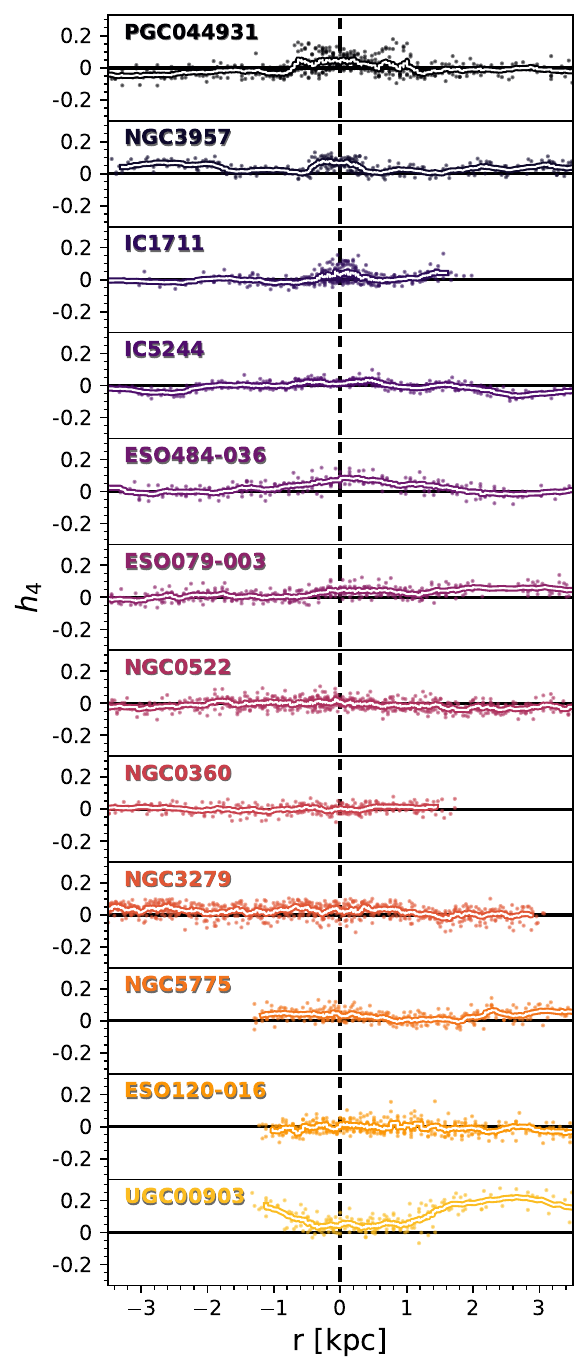} 
\caption{1D line maps of $h_{4}$}
\end{subfigure}
\caption{As in Figure~\ref{Fig:linemaps1}, but 1D line maps of $h_{3}$ (left) and $h_{4}$ (right). Typical median errors are $h_{3,err}=0.016$ and $h_{4,err}=0.017$.}
\label{Fig:linemaps2}
\end{figure*}

For IC 1711, NGC 3957, and PGC 044931, the most striking feature is a small, darker-coloured region within the central kiloparsec that is rotating faster than the surrounding disc, likely a kinematically-decoupled discy component. For comparison to previous literature, in Figure~\ref{Fig:linemaps1}a, we plot the galaxy rotation curves, extracted from slits of width 1 kpc. All galaxies host rapidly rising rotation curves, though the slope of the curve is much steeper in the central regions of PGC 044931, NGC 3957, IC 1711, and IC 5244. The characteristic `double-humped' rotation curve (i.e.\ a rotation curve that first rises rapidly to reach a local maximum, then drops slightly to create a local minimum, and then rises again slowly up to its flat section) described by \citet{bureau2005} is seen in these four galaxies. 

 \subsubsection{Stellar velocity dispersion, $\sigma_{\star}$, maps and profiles}
 
The top right panels of Figures~\ref{Fig:vel_maps1}--\ref{Fig:vel_maps12} show $\sigma_{\star}$ maps. Here, we see a great deal of variation between galaxies, and several features of note. First, we acknowledge the dark purple strips of low $\sigma_{\star}$ crossing most galaxies, coincident with the position of the dust lane, and highlighted by the non-symmetric nature of the flux contours overlaid in white. Given that dust blocks light from the galactic plane regions from reaching the observer, we expect to be viewing just the stars which are intrinsically exterior to the dust lanes in these regions. 

In addition to the dust lanes, we also observe central regions of high-velocity dispersion, transitioning to lower dispersion at greater radii in most of the galaxies, except NGC 3279 and UGC 00903. Central $\sigma_{\star}$ maxima range from 70--130 km s$^{-1}$, with an average of $\sigma_{\star}\sim 90$ km s$^{-1}$. The increase in $\sigma_{\star}$ from the outskirts to the central regions, $\Delta \sigma_{\star}$, is on average $\sim50$ km s$^{-1}$. In NGC 3957 and PGC 044931, this central high $\sigma_{\star}$ region is `croissant-shaped', with a lower dispersion region embedded within the high $\sigma_{\star}$ region at the centre of the galaxy. We expect that these are not the result of dust obscuration, as neither of these galaxies is perfectly edge-on and the effect of the dust is visible below the z=0 line. Additionally, the low $\sigma_{\star}$ regions are offset from the projection of the main dust lane in the 2D maps; co-located with the kinematically-decoupled disc indicator seen in the $V_\star$ maps. 
For IC 5244 and IC 1711, the central region with $\sigma_{\star}>100$ km s$^{-1}$ is extended and nearly circular, with dispersions exceeding 110 km s$^{-1}$.

Correspondingly, the 1D $\sigma_{\star}$ profiles shown in Figure~\ref{Fig:linemaps1}b are also quite varied. While some galaxies host gently peaked line profiles, the strong peaks of IC 1711 and IC 5244 are clear. Interestingly, a large spread in central $\sigma_{\star}$ is seen in the four galaxies with the steepest-rising rotation curves: IC 5244, IC 1711, NGC 3957, and PGC 044931. 

\subsubsection{Skew, $h_{3}$, maps and profiles}

The bottom left panels of Figures~\ref{Fig:vel_maps1}--\ref{Fig:vel_maps12} show the high-order Gauss-Hermite moment $h_{3}$ (or skew). The colour map is such that red indicates a positive skew value (where excess low velocities are present in the LOSVD beyond that of a pure Gaussian), and blue a negative (where excess high velocities are present in the LOSVD). While there is more stochasticity between neighbouring bins than the $V_\star$ and $\sigma_{\star}$ maps, some spatially-coherent features are readily visible. 

We observe small, oval-shaped regions of highly positive and negative $h_{3}$ at the very centre of the galaxy and only along the z=0 line for IC 5244, IC 1711, NGC 3957, and PGC 044931. Compared to their $V_{\star}$ maps, these regions are opposite in sign.  
Additionally, this feature is also seen in the 1D radial profiles of Figure~\ref{Fig:linemaps2}a. A strong, negative gradient in $h_{3}$ as a function of galactic radius is seen in the $h_{3}$ profiles of PGC 044931, NGC 3957, IC 1711, IC 5244, and UGC 00903. For these galaxies, the extrema of the $h_{3}$ coincide radially with the $V_{\star}$ humps pointed out in Section~\ref{sect:velocity}. The two $h_{3}$ extrema present close to the galaxy centre imply large gradient in the central $h_{3}$ radial 1D profile between them, and since the signs of these extrema are opposite to those of $V_{\star}$, the corresponding gradient of the $V_{\star}$ radial profile is of opposite sign and therefore $h_{3}$ and $V_{\star}$ anti-correlate within the radius of the $h_{3}$ extrema. 
We expect these features to be the signature of a dynamically cold disc, and discuss them further in Section~\ref{disc:NDs}.

At larger scales, we see varying h3 behaviour along the z=0 line compared to off the plane for a given R in several galaxies: NGC 3957, NGC 0522, PGC 044931, IC 5244, and perhaps best illustrated in ESO 079-003. We observe $\sim 0.5$ kpc-wide strips where $h_{3}$ and $V_{\star}$ are of opposite sign along the kinematic major axis of these galaxies, discussed in Section~\ref{sec:inclination}. These strips often extend to the radial extent of the MUSE field of view. 
We note that the correlations seen in 1D profiles often extend only to small scales (e.g.\ $\sim$1 kpc for IC 1711), whereas when sign-matching, additional coherent regions of sign matching and mismatching are also obvious at larger scales. For IC 1711, both above and below the z=0 line, these opposing signs in $V_{\star}$ and $h_{3}$ switch to matching signs, and we observe coherent positive (negative) regions of $h_{3}$, corresponding to positive (negative) values of $V_{\star}$, extending 0--3 kpc radially, and 0.5--2 kpc off the z=0 line of the disc. 
We see the largest regions of matching $h_{3}$ and $V_{\star}$ sign in IC 5244, and IC 1711, but also see this feature to varying degrees in all other BP bulge galaxies, though the signal is faint in ESO 120-016 and NGC 0522. 
In the radial outskirts of all galaxies in this sample, we note coherent regions of $h_{3}$ and $V_{\star}$ opposing signs, extending to the edges of the MUSE field of view. 
For galaxies with the most coherent $h_{3}$ structure, moving from the central regions to the galactic outskirts, we observe the signs of $h_{3}$ and $V_{\star}$ to oppose, match, and then oppose again.  

We note that for PGC 044931 (Fig~\ref{Fig:vel_maps3}) and NGC 0360 (Fig~\ref{Fig:vel_maps6}) the $h_{3}$ values in the very outer disc region tend to zero. There are several possible causes for this behaviour, including inappropriate penalisation of the \ppxf solution in the outermost bins, most of which possess a lower S/N than the desired value. Even with a S/N-dependent \ppxf bias determination, we found that outer bins were extremely susceptible to small changes in penalising bias.
The $h_{3}$ behaviour seen is also common when the measured velocity dispersion is well below the instrumental resolution, and also present in the work of e.g.\ \citet{seidel2015} for the SAURON spectrograph, and in mock IFS datacubes by \citet{wang2024}. In the wavelength range investigated in this work, the MUSE spectral resolution is $\sim 40$ km s$^{-1}$ \citep[e.g.][]{bacon2017}. The measured $\sigma_{\star}$ of both PGC 044931 and NGC 0360 in the outskirts of their discs is $\sim20$ km s$^{-1}$. For this work, we are interested mostly in the central regions of the galaxies, where the dispersion is higher, whereas the analysis of the kinematics in the outskirt regions will be presented in future work. 

\subsubsection{Kurtosis, $h_{4}$, maps and profiles}

In Figures~\ref{Fig:vel_maps1}--\ref{Fig:vel_maps12}, we present the line-of-sight kurtosis, or $h_{4}$, maps. 
There are fewer spatially coherent features visible in these maps, possibly the result of observational challenges related to retrieving high-order velocity moments from medium-resolution spectra \citep[e.g.][]{wang2024}.
Nevertheless, we do notice some trends. Given their co-spatial nature, we expect that the large longitudinally-extended regions of high positive $h_{4}$ around the centres of some galaxies that correspond to low $\sigma_{\star}$ and clear deviations from symmetry in the surface brightness contours are caused by the dust lanes. 
In some galaxies, including PGC 044931, NGC 3957, NGC 0522, and ESO 079-003, we see regions of high positive $h_{4}$ extending from the central regions of the galaxy 1--2 kpc off the z=0 line. 
In PGC 044931, IC 1711, and NGC 3957, the small, inner areas of opposing $h_{3}$ and $V_{\star}$ sign are co-spatial with a coherent region of high positive $h_{4}$. The strength of these features with respect the the surrounding disc are highlighted in the 1D profiles of Figure~\ref{Fig:linemaps2}b.

Also of note are the asymmetries in $h_{4}$ in the disc regions on IC 5244 and ESO 484-036. While analysis is ongoing, we expect that these features are likely unphysical and caused by variations in sky subtraction across the frames (and between neighbouring mosaics).

\section{Discussion}
\label{discussion}

Here we attempt to connect the observations of Section~\ref{results_maps} to an understanding of the central kinematic sub-structures in the GECKOS galaxies. Already clear is the diversity of structures in the $V_\star$ maps (visible central kinematically-decoupled rapidly-rotating structures), $\sigma_{\star}$ maps (structure and intensity of sigma peaks at the centre of the galaxy, effects of dust lanes, central `croissant' features), $h_{3}$ (coherent areas of matching/opposing sign with $V_{\star}$), and $h_{4}$ (small, highly-peaked central regions). 
That said, some features are similar throughout the sample: all galaxies possess a rotating disc structure and some degree of $h_{3}$ and $V_\star$ sign mismatch in the outer disc regions of the galaxy. 
In the following, we explore these observations as possible tracers of kinematic sub-structures.

\subsection{The presence of bars and the effect of bar viewing angle}
\label{bar_features}

Several studies have investigated the kinematic properties of bars and BP bulges, both theoretically and observationally, and in one and two dimensions. We focus first on the 1D, midplane results. 

\subsubsection{Midplane features}
From N-body models, \citet{bureau2005} developed a set of characteristic stellar-kinematic bar signatures, and predicted that along a slit aligned with the photometric major axis of an edge-on galaxy one should observe:
\begin{itemize}
    \item A `double-hump' rotation curve, with a stellar mean line-of-sight velocity ($V_\star$) profile exhibiting an initial rise, followed by a hump or plateau, and then a secondary rise, before flattening out. The authors surmise that the hump is caused by (mostly inner) bar $x_{1}$ orbits, the elongated family of orbits that follow the major axis of the bar \citep[e.g.][]{contopoulos1980,athanassoula1992}, adding a distinct component to the LOSVD. 
    
    \item  A flat-topped or weakly peaked radial velocity dispersion ($\sigma_{\star}$) profile, sometimes with a local central minimum. Broad `shoulders' or secondary maxima may be present in strong bars, usually towards the end of the bar, just inside the inner ring. \citet{bureau2005} attributed the central $\sigma_{\star}$ peak to the large variety of orbital shapes encountered in strongly-barred discs, particularly along the major axis. They posited that the $\sigma_{\star}$ `shoulders' are produced by the tips of the $x_{1}$ orbits and the inner 4:1 orbits. In particular, $x_{1}$ orbits with loops around the major axis are expected in strong bars, causing a local increase in the velocity dispersion, strongest for bars seen end-on \citep{athanassoula1992}. The higher energy inner 4:1 orbits have similar loops, which increase $\sigma_{\star}$ for bars seen both side-on and end-on.

    \item A $h_{3}$ profile that is correlated with $V_\star$ over the expected bar length. This key tracer of triaxiality is driven by the elongated bar orbits (e.g. $x_{1}$, 4:1, etc) creating a tail of high-velocity material, required for $V_\star$ and $h_{3}$ to correlate \citep{bureau2005}. 
\end{itemize}

Crucially, \citet{bureau2005} also show that the strength of these kinematic features is correlated with bar viewing angle, such that the strongest signals are seen when the bar is oriented end-on to the viewer.
Through observations of 30 edge-on spiral galaxies, \citet{chung2004} provided an alternative explanation for the `double-hump' rotation curve by taking into account the central minimum behaviour of the $\sigma_{\star}$ profile and $h_{3}$-$V_\star$ profile anti-correlation in this region. From these observations, they suggest that the inner regions of the galaxy are both cold, and also very close to axisymmetric. They speculate that these regions have largely decoupled from the rest of the galaxy (and the bar) and circularised, to form a nuclear disc. 

Our Figures~\ref{Fig:linemaps1} and \ref{Fig:linemaps2} can be directly compared to the \citet{bureau2005} predictions to search for kinematic evidence of bar structure in the GECKOS galaxies. Clear `double hump' rotation curves are observed in PGC 044931, NGC 3957, IC 1711, and IC 5244. 
In all galaxies except NGC 3279 and UGC 00903, we observe either a flat-topped or peaked dispersion profile, but given that we expect that other structures could also give rise to central dispersion peaks, taken alone, this is not an adequate bar predictor. Shoulders are seen in the dispersion profiles of PGC 044931, NGC 3957, IC 1711, and the strongest example in IC 5244, the profile of which is strikingly similar to that that the intermediate bar strength and 45\textdegree\ viewing angle case of \citet{bureau2005}. 

Comparing the $h_3$ 1D profiles of Figure~\ref{Fig:linemaps2}a with $V_{\star}$ in Figure~\ref{Fig:linemaps1}a, some $h_{3}-V_{\star}$ \textit{correlations} are seen in the central few kpc of PGC 044931, NGC 3957, IC 1711, IC 5244, ESO 484-036, and ESO 079-003. Taken with the trends seen in the $V_{\star}$ and $\sigma_{\star}$ profiles, we can reliably report that these galaxies host strong evidence of bar structure from their stellar kinematics along the z=0 line.

Finally, we discuss the effects of dust on the 1D line profiles. Prominent dust lanes are seen in every galaxy in this sample. When this dust absorbs or scatters stellar photons, it does the same to kinematical information contained within these photons. 
\citet{baes2003} present model galaxy rotation curves and $\sigma_{\star}$ profiles both with and without the application of radiative transfer models incorporating absorption and scattering by interstellar dust grains. They find that for highly inclined galaxies, rotation curves and central $\sigma_{\star}$ peaks are significantly flattened, leading to significant underestimation of the true rotational velocity of the inner regions \citep[e.g.][]{davies1990, bosma1992, catinella2006}. For this reason, we do not further quantify the shapes of the rotation curves observed beyond the presence of `double hump' features, but do note that \citet{baes2003} report that even a slight deviation from exactly edge-on inclinations dramatically reduces dust effects.

\subsubsection{Off-plane features}
\label{subsubsec:off-plane}

\citet{iannuzzi2015} extended previous long-slit work by providing 2D, projected line-of-sight kinematics for a set of barred/BP bulge dynamical models featuring star formation, one of which also hosted a spherical bulge. 
The velocity maps presented in that work, whilst appearing qualitatively similar to those seen here, differ in the uniform nature of the kinematic minor axis. In all cases, the kinematic minor axis was aligned perpendicular to the photometric major axis. In the GECKOS galaxies, we often see central `wedge-shaped' deviations, as illustrated in e.g.\ NGC 0522. These deviations are generally coincident with the dust lane, as shown by the overlaid flux contours on the $\sigma_{\star}$ maps, and so it is reasonable to attribute them to dust. The kinematic minor axis is straight in the model predictions as regardless of bar viewing angle, all star particles on bar orbits will be accounted for, regardless of their position on the near or the far side of the disc. Dust breaks the bi-symmetry of a bar (and indeed any disc), as light from stars on the near side of the bar will contribute far more than those from the far side obscured by dust. This net radial flow manifests itself as the `wedge' shape of the kinematic axis of GECKOS galaxies PGC 044931, NGC 0522, ESO 079-003, all of which are barred, but also NGC 0360 which we have not classified as barred. Further analysis of these structures in conjunction with complementary ALMA data will give us additional information on the location of the near side of the bar, and is planned for future work.

Relatedly, we examine the lighter-coloured regions of the $V_{\star}$ maps, indicative of a velocity magnitude lower than the surroundings. These regions are co-spatial with the dust lane, as seen from the surface brightness contours overlaid on the $\sigma_{\star}$ maps. As illustrated in \citet{davies1990}, the component of rotational velocity recovered from a highly inclined galaxy with an optically thick inner region is lower than expected. Coupled with observational results showing that dust-free rotation curve measures such as HI and CO generally produce steeper inner rotation curves compared to optical measures \citep[e.g.][]{bosma1992}, we assume that the regions of lower velocity magnitude are the direct result of dust attenuation.

Finally, we mention also the off-plane deviations from a straight $V_{\star}=0$ isovelocity line in PGC 044931, NGC 0360, ESO 120-016, NGC 5775, ESO 079-003, and ESO 484-036. These deviations are clearest on the lower sides of the velocity maps, given that we calculated the systemic velocities from the top half. Given that the lower half of the $V_{\star}$ maps contains the dust lane (if the galaxy is not perfectly edge-on), we expect this side to be more dust-obscured than the top side. We hypothesise that it is the greater degree of dust obscuration that causes these deviations from a perfectly straight $V_{\star}=0$ isovelocity line.

In addition to confirming 1D predictions and observations, \citet{iannuzzi2015} found that the 2D line-of-sight mean velocity maps do not provide a robust predictor of the presence of BP bulges as the degree of cylindrical rotation (i.e. mean stellar velocity independent of height above the disc) varies. In their work, \citet{iannuzzi2015} reported a tendency for the degree of cylindrical rotation to weaken when simulations are seen from side-on to progressively more end-on bar angles, though the importance of this effect varies from case to case without a clear trend in BP bulge strength. A variation in BP bulge rotational properties is also reported by \citet{athanassoula2002}, who attribute this behaviour to different disc and halo contributions to the total circular velocity curve in their various models. 
\citet{williams2011}'s observations support this idea with observations of five edge-on discs hosting BPs that show a variety of projected off-plane velocity behaviours. However, \citet{molaeinezhad2016} developed a new technique to measure the degree of cylindrical rotation and found a wide range of cylindrical rotation, but on average, BP bulges displayed higher values for this parameter than non-BP bulges. We do not look further into cylindrical rotation indicators in this work, though note that the GECKOS sample will provide an excellent opportunity to understand the off-plane velocity behaviour of BP bulges.   

When considering $\sigma_{\star}$ maps, \citet{iannuzzi2015} reported a gradual increase of the central velocity dispersion as the bar viewing angle changes from side-on to end-on, corresponding to the
line-of-sight being increasingly along the long axis of the $x_{1}$ orbits, leading to higher velocities and thus higher dispersions \citep[also][]{athanassoula1992, bureau1999}.
IC 1711 and IC 5244 are the galaxies in the GECKOS sample with the highest peak $\sigma_{\star}$ values (remembering that all galaxies should be approximately Milky Way stellar mass). This, coupled with the shape of the high $\sigma_{\star}$ regions in these galaxies supports a hypothesis that these two galaxies host bars that are close to end-on orientation.
\citet{iannuzzi2015} concluded that full 2D spatial coverage is not required to describe adequately the $\sigma_{\star}$ behaviour; the major-axis information is sufficient to describe the global $\sigma_{\star}$ variation. 

Next, we look at  to high-order velocity moments, which can encode additional information on galaxy sub-structures. \citet{iannuzzi2015} noted that, contrary to $V_\star$ and $\sigma_\star$, $h_{3}$ and $h_{4}$ require full 2D maps to fully capture the richness of the features expected, as confirmed by \citet{li2018}.
\citet{iannuzzi2015} added the identification of strictly peanut-related signatures in rough spatial correspondence with the projected edges of the structure, offset from the major axis:

\begin{itemize}
\item Elongated `wings' of large values $h_{3}$. Again, the bar viewing angle is important for both the amplitude and morphology of $h_{3}$ maps. \citet{iannuzzi2015} and \citet{li2018} analyse separate N-body models of barred systems hosting a BP bulge at several different bar viewing angles and conclude that while bar-driven signatures in $h_{3}$ are barely visible when the bar is side-on to the observer, considerable changes occur as the bar is rotated, such that large regions of $h_{3}$ maxima and minima develop roughly in the regions corresponding to the peanut, increasing in size such that they are largest at intermediate bar viewing angles (30-60\textdegree), and magnitude  such that they are strongest when the bar is viewed end-on. 

\item An `X'-shaped region of deep $h_{4}$ minima, becoming deeper and better spatially defined as bar viewing angle moves from side-on to end-on.
\end{itemize}

Both $h_{3}$ and $h_{4}$ features become stronger at larger BP bulge strengths (for a given bar viewing angle) in edge-on galaxies. 

Taking these predictions to the GECKOS sample, we see coherent areas of $h_{3}$ of same sign as $V_\star$ at the centres of all BP bulge galaxies (though to varying degrees).   
We see an excellent example of the predicted wing structure in IC 5244 and IC 1711. These areas of coherent $h_{3}$ sign originate in the central galactic regions and extend up to 2 kpc off the disc, and up to 5 kpc radially. We expect that these regions correspond to the area of influence of the bar, with the radial width related to the bar viewing angle. 

Along the midplane, \citet{bureau2005} observe central $h_{4}$ minima, a peak at the ends of the BP, and then secondary minima at radii beyond the bar. These features are strongest when the bar is seen end-on, and weaken as the bar is rotated towards a side-on position. Given the correlation between $h_{4}$ and $\sigma_{\star}$ behaviour, they found that these described features are likely due to the inner $x_{1}$ orbits.
The GECKOS $h_{4}$ maps do not match the \citet{bureau2005} nor the \citet{iannuzzi2015} predictions; we do not witness any deep $h_{4}$ minima in any BP bulge galaxies. Rather, apart from structure relating to dust lanes, we note three distinct features. 
In some galaxies, rather than areas of deep $h_{4}$ minima, we notice large-scale regions of $h_{4}$ \textit{maxima} emanating off the disc to a height of at least 2 kpc in the centre and extending 2-4 kpc radially in a bi-conical manner; PGC 044931 and NGC 0522 present good examples of this behaviour.
In the case of ESO 484-036 and IC 5244 a wide-scale $h_{4}$ gradient is seen spanning the entire disc. Finally, the other coherent structure that we see is small highly positive $h_{4}$ regions at the very centres of PGC 044931, NGC 3957, and IC 1711.

There could be many reasons for the lack of similarity with simulation predictions, ranging from physical to instrumental. 
In the case of the disc-wide $h_{4}$ gradient seen in ESO 484-036 and IC 5244, it is possible that our kurtosis measures are being affected by a spatially-variable LSF or residuals from an imperfect sky subtraction or telluric correction. Further investigation into possible instrumental effects is ongoing.
The large-scale regions of $h_{4}$ maxima and small-scale regions are more difficult to explain, though given the co-spatial nature of the small-scale feature with other nuclear disc indicators (discussed further in Section~\ref{disc:NDs}), we expect that a nuclear disc is what is causing this feature. 

\citet{wang2024} examined true and recovered $h_{4}$ in an edge-on Milky Way mock (unbarred) galaxy. The mock was created from integrated spectra using SSP models and the mock stellar catalogue from \textsc{E-Galaxia} (Sharma et al., in prep.) and then \ppxf was applied to recover stellar kinematic moments.
Their analysis indicated that, at a MUSE spectral resolution, \ppxf could recover similar structures in the true $h_{3}$ distribution of particles in \textsc{E-Galaxia}. However, structures in the true $h_{4}$ distribution were not detectable in the \ppxf outputs. This suggests that $h_{4}$ is inherently more difficult to measure accurately than $h_{3}$ using full spectral fitting methods.

\citet{iannuzzi2015} applied Voronoi binning to their particle data before fitting the LOSVD with a Gauss-Hermite to obtain their kinematics; an equivalent strategy to \citet{wang2024}. 
We believe that the reason that these simulation studies found strong $h_{4}$ features that are not seen in the GECKOS data is most likely due to observational limitations, i.e., the process of transferring particle statistics to spectra, and then extracting the $h_{4}$ from weak spectral features. Due to this process, all of the `real' $h_{4}$ features become more shallow or disappear. We refer the reader to Sections 3.3 and 4.1 of \citet{wang2024} for a more thorough discussion on this point.

\subsubsection{Inclination effects}
\label{sec:inclination}
\citet{iannuzzi2015} investigate the effects of galaxy inclination on high order kinematic maps. When a galaxy is inclined to be perfectly edge-on to the viewer, a radially-extended ($2\lesssim\rm{R[kpc]}\lesssim7$ ) region of $h_{3}$ maxima corresponding to a sign match with $V_{\star}$ occurs along the z=0 line. In the 80\textdegree\ case, these regions move off the z=0 line to z$\sim$2 kpc, extending upwards of 5 kpc both above and below the z=0 line. The $h_{3}$ maxima along the z=0 line changes sign, though remains of high magnitude.

\citet{bureau2005} report similar behaviour along the midplane, attributing the switch in sign from a $h_{3}$-$V_{\star}$ correlation to an anti-correlation to the projected signature of the outer disc disappearing, which in turn increases the observed mean velocities, and shifts the asymmetric tail of material in the LOSVDs from high to low velocities. We observe qualitatively similar behaviour to the \citet{iannuzzi2015} 80\textdegree\ inclination case in NGC 3957, NGC 0522, PGC 044931, IC 5244, and ESO 079-003, all of which exhibit $h_{3}$ maxima corresponding to a sign match with $V_{\star}$ located as coherent regions both above and below the z=0 line. 
Given the slight inclinations apparent from the unsharp masked optical images and Spitzer maps, we expect that the majority of the GECKOS galaxies possess inclinations between 80--90\textdegree. Work to accurately characterise the GECKOS sample inclinations via detailed orbital modelling techniques are underway, however we speculate that a slight deviation from perfectly edge-on geometry is the cause of these regions of $h_{3}$-$V_{\star}$ sign match moving off the z=0 line.

\subsubsection{Bar viewing angle}
$h_{3}$ maps vary considerably in both amplitude and morphology with bar viewing angle, according to both the \citet{iannuzzi2015} and \citet{li2018} model predictions. 
The physical extent and maximum value of the observed $h_{3}$ `blobs' appears least conspicuous in the side-on case, and increases in both size and magnitude (whilst still maintaining the appearance of extended blobs), as the bar is rotated toward the viewer. The $h_{3}$ blobs are most prominent when the bar is seen end-on. In Figures~\ref{Fig:vel_maps1}-\ref{Fig:vel_maps12}, we see the strongest degree of coherent $h_{3}$ structure (correlating in sign with regions of $V_{\star}$) in the central regions of IC 5244 and IC 1711, adding weight to the idea that these galaxies possess end-on bars. We see the lowest degree of $h_{3}$ structure in ESO 120-016 and NGC 0522. While it is possible that ESO 120-016 is indeed not a BP bulge galaxy, NGC 0522 possesses a definite X shape in the DECaLS and Spitzer imaging. For this reason, we expect that NGC 0522 must possess a bar that is close to side-on orientation. For all other galaxies, we infer from the $h_{3}$ maps that the bar viewing angle is somewhat intermediate between side-on and end-on. 

\citet{chung2004} note that the kinematic signatures of bar presence often contrast almost perfectly with visual morphology, such that the strongest box/peanut shapes often show very little indication of bar structure in their kinematic maps, and those with strong $h_{3}$-$V_\star$ correlations often display the rounded morphology of end-on bars (incidentally often mistaken for classical bulges). Our findings are mostly in line with this observation, though we note the case of PGC 044931: this galaxy possesses both strong kinematic and morphological indicators of a bar. `Strength' is a difficult term to define in terms of BPs, and it may be that BPs do not grow linearly in strength with time, nor appear the same for a given bar viewing angle from galaxy to galaxy. 

While we cannot constrain the bar viewing angle through kinematic measurements alone (though this will be investigated further with cold gas in future work), we can infer that IC 5244 and IC 1711 are both close to end-on bars. The kinematic signatures of NGC 0522 are very faint, and the corresponding BP bulge is very strong in imaging, leading us to conclude that this bar is likely close to side-on. For the other BPs in the sample, we expect that the bars are at intermediate viewing angles. 

Deep CO observations could provide additional information to help understand bar frequency in BP bulge galaxies \citep[e.g.][]{alatalo2013, topal2016}. Molecular gas is colder than ionised, both dynamically and in temperature, so it reacts more strongly to a bar potential than warmer components. All non-axisymmetric signatures are therefore more prominent. Additionally, molecular gas observations, being at mm wavelengths, are unaffected by dust, and so the (near) edge-on orientation of the GECKOS sample is inconsequential.
The more highly-spectrally-resolved cold gas maps could be examined for the telltale `X' shape in position-velocity diagrams \citep[e.g.][]{kuijken1995, bureau1999}, a feature not readily distinguished in ionised gas profiles at MUSE's spectral resolution. 

\subsection{Evidence for nuclear discs}
\label{disc:NDs}

Following from the predictions of \citet{bureau2005}, an additional predictor of bar structure along the kinematic major axis of edge-on galaxies was provided by \citet{chung2004}, who illustrate the features mentioned in Section~\ref{bar_features} in their Figure 9, and observed a strong \textit{anti}-correlation between $h_{3}$ and $V_\star$ profiles in the innermost region of BP bulge galaxies, which they attributed to the presence of a cold, rapidly-rotating nuclear disc built by a bar. This anti-correlation switches back to a correlation outside of the nuclear disc region but still within the bar region, in line with the \citet{bureau2005} prediction.

Evidence for the presence of nuclear discs is observed in the $V_{\star}$, $\sigma_{\star}$, $h_{3}$, and $h_{4}$ maps of IC 1711, IC 5244, NGC 3957, and PGC 044931. Whilst all show modest increases in rotational velocity in small regions at their very centres, NGC 3957 and PGC 044931 also show a `croissant'-shaped feature in their velocity dispersion maps, with a central depression of lower $\sigma_{\star}$. All four galaxies also show $h_{3}$-$V_{\star}$ sign mismatches in these central regions, co-located with regions of high positive $h_{4}$.

These features are perhaps better shown in the 1D line profiles of Figures~\ref{Fig:linemaps1} and ~\ref{Fig:linemaps2}, where the four galaxies in question lie at the top of these figures. The central minima in the dispersion profiles of PGC 044931, NGC 3957, and IC 1711 are features that are sometimes referred to as `sigma drops' \citep[e.g.][]{emsellem2001}.
Observationally, this feature has been frequently seen in long-slit spectra \citep[e.g.][]{emsellem2001, marquez2003, chung2004, comeron2008, mendez-abreu2014} and IFS data \citep[e.g.][]{crocker2011, lin2017, pinna2019a, shimizu2019}. We expect that these $\sigma$ drops are due to the presence of a dynamically cold and kinematically-decoupled central stellar disc originating from gas inflow. 
Interestingly, upon closer inspection of the $\sigma_{\star}$ radial profiles of PGC 044931, NGC 3957, and IC 1711, we see that their profiles bifurcate in the central regions, with one population of stars of low $\sigma_{\star}$ (the nuclear disc stars), and one population with higher $\sigma_{\star}$. In fact, in IC 5244 we see only a $\sigma_{\star}$ peak, with perhaps only a tiny hint of a drop. We expect that the peaks are due to bar orbits, as discussed in Section~\ref{bar_features}.

The N-body models of \citet{bureau2005} do not include nuclear discs, but do reproduce this central $\sigma_{\star}$ minima, attributing it to the orbital structure of strongly barred discs. More recent N-body and SPH simulations that include stars, gas, and star formation show that young stars born in the nuclear regions from dynamically cold gas have a velocity dispersion lower than the older stellar population \citep{wozniak2003, michel-dansac2004, wozniak2007, cole2014, portaluri2017}, and hence attribute the sigma drop feature to a dynamically cold, nuclear disc. 
Contextually, the presence of a nuclear disc in a barred galaxy makes sense: we expect the bar to funnel gas towards the central regions of the galaxy \citep[e.g.][]{kim2024, verwilghen2024}, and often observe higher molecular gas surface densities in central regions as a result \citep[e.g.][]{fisher2013,yu2022}. Through the bar structure and associated torques, gas loses angular momentum, allowing it to sink to the centre of the galaxy, creating star-forming structures including rings, filled discs, or nuclear spirals that in turn build a nuclear disc \citep[e.g.][]{seo2019,lin2020}.

The $h_{3}$ and $h_{4}$ maps also encode information on the presence of nuclear discs at the centres of some of the GECKOS BP bulge galaxies. There is a small region of high positive and negative $h_{3}$ at the centres of IC 5244, IC 1711, NGC 3957, and PGC 044931, all of which correspond to an opposing sign in the same regions of the $V_{\star}$ maps.  
Presumably, this feature is related to the central $h_{3}$-$V_{\star}$ anti-correlation seen by \citet{chung2004}, who explain this observation via a region that appears to have largely decoupled from the rest of the galaxy (and the bar) and circularized, forming a dense and (quasi-) axisymmetric central stellar disc. This feature was also observed in most of the TIMER galaxies \citep{gadotti2020}, NGC 1381 \citep{pinna2019a}, NGC 5746 \citep{martig2021}, and in NGC 4643 \citep{erwin2021}.
We also observe coherent regions of $h_{4}$ maxima co-spatial with the central $h_{3}$-$V_{\star}$ sign mismatch and $\sigma$ drop regions. We expect that we are observing the projected equivalent of the regions of $h_{4}$ maxima seen in most TIMER survey galaxies \citet{gadotti2020} \citep[also e.g. NGC 4643][]{erwin2021}. These are explained by \citet{gadotti2020} as areas in which there is a superposition of multiple kinematic components - in this case, a low-dispersion nuclear disc component on top of a relatively higher-sigma underlying disc component.

Observationally, \citet{seidel2015} investigated the high-order moments of the LOSVD of 16 barred galaxies using the SAURON spectrograph. They reported a $h_{3}$--$V_\star$ anti-correlation within 0.1 of the bar radius in around 50\% of cases. For some of their galaxies, however, the $h_{3}$ and $h_{4}$ results are hard to interpret because $\sigma_{\star}$ falls well below the instrumental resolution and $h_{3}$ and $h_{4}$ drop to zero. Kinematically separating the nuclear disc component from the rest of the disc is difficult to do without excellent spatial resolution. 

From Figure~\ref{Fig:linemaps2}a, it appears that the magnitude of the peak $h_{3}$ value is correlated with the physical distance between the $h_{3}$ minima and maxima for the four galaxies hosting nuclear disc signatures.
The $h_{3}$ extrema correspond to changes in the sign of the gradient of $h_{3}$ with radius, which also corresponds to the turnover in peak $V_{\star}/\sigma$ values used to measure nuclear disc size for the TIMER galaxy sample, and termed $r_{\rm{k}}$ \citep{gadotti2020}. We therefore measure the size of the nuclear discs detected in the four GECKOS galaxies as half of the physical distance between $h_{3}$ extrema, terming this value $r_{h3}$. We note that we could also have measured the sizes from the transitions in $h_{4}$, or the bumps in the rotation curves, but found that the signal was clearest in $h_{3}$. We note here that these techniques of measuring nuclear disc size will not return the full extent of the nuclear disc, just the (light-weighted) region where they dominate over background stellar populations. In all likelihood, they will be larger than these kinematically-defined sizes.
There are also other techniques for measuring the size of the disc, for example measuring the radius for which $h_{3}$ reaches zero again after its peak (which \citet{chung2004} attributed to a dense and (quasi-)axisymmetric central stellar disc). However, to maximise comparison with recent observational works such as \citet{gadotti2020}, we choose to stick to the above definition of $r_{h3}$.
We list the nuclear disc sizes and peak $h_{3}$ values for the four GECKOS galaxies that exhibit signs of nuclear disc presence in Table~\ref{table:ND}, noting that the sizes obtained are in line with those measured for the TIMER survey by \citet{gadotti2020} and \citet{desa-freitas2023}. 

PGC 044931 possesses the strongest peak in $h_{3}$, and the largest $r_{h3}$ value, which we expect to be the brightest (and possibly largest) nuclear disc. Interestingly, while NGC 3957 displays the second greatest $h_{3}$ peak, it is IC 1711 with the larger $r_{h3}$. If we interpret greater negative and positive $h_{3}$ to be excess material at approaching and receding velocities in the LOSVD (beyond that of a pure Gaussian), respectively, then we can assume that there are either more bright stars (or many more faint stars) in the nuclear disc of NGC 3957, or the difference in rotation speed is greater between the nuclear disc and main disc of NGC 3957 than IC 1711.
Alternatively, we might be observing geometrical differences in nuclear disc orientations. Given that $x_{2}$ orbits are often somewhat elongated, and aligned perpendicular to the direction of elongation of the bar \citep{contopoulos1980}, the measured projected radius of the nuclear disc may vary as a function of bar viewing angle. In this manner, the combination of peak $h_{3}$ and $r_{h3}$ may be able to be used to infer bar viewing angles. Simulations should be employed to further investigate this idea.

\begin{table}
\caption{Nuclear disc properties in the GECKOS sample.}            
\label{table:ND}      
\centering                          
\begin{tabular}{l c c}       
\hline\hline                
Galaxy & $r_{h3}$ [kpc]& Max $h_{3}$ \\    
\hline      
PGC 044931 & 0.71 & 0.179 \\
NGC 3957 & 0.28 & 0.120 \\  
IC 1711 & 0.37 & 0.052  \\
IC 5244 & 0.18 & 0.035 \\
 
\hline                                   
\end{tabular}
\end{table}

Given the boxy-peanut structure present in 8/12 galaxies analysed, and assuming that BP bulge structure is synonymous with the presence of a bar, why do we not observe nuclear stellar discs in all eight galaxies? It may be that not all Milky Way-mass barred disc galaxies harbour nuclear discs, but if they do the answer may come down to dust: the more perfectly edge-on and star-forming galaxies could certainly contain enough dust that at such a viewing angle it is impossible to peer into the centres of these galaxies at MUSE wavelengths. Spatial resolution may also come into play. The closest galaxy in this sample of BP bulges is NGC 3957, at $z=0.0054$ ($D=24.8$ Mpc), corresponding to a spatial scale of $\sim0.11$ kpc arcsec$^{-1}$. With a typical seeing of $\sim0.8^{\prime\prime}$, this corresponds to a spatial resolution of $\sim0.1$ kpc. The most distant galaxy is ESO 484-036, at $z=0.017$ ($D = 68.7$ Mpc), corresponding to a spatial scale of 0.35 kpc arcsec$^{-1}$. Reported literature nuclear disc radii range from 0.1 to 1 kpc \citep[e.g.][]{gadotti2020,desa-freitas2023}. This means that for the most distant galaxies, we are typically searching for structures that are only a few resolution elements across; these nuclear discs may be at the limit of being spatially resolved.

\subsection{Other kinematic sub-structures}
\label{what_bulge}
We have so far discussed kinematic indicators for the presence of bars, BP bulges, and nuclear discs. There are several other galactic sub-structures described in previous literature, including classical bulges, and disc structure(s).

\subsubsection{Classical bulges}
\label{sec:classicalbulge}
A common predictor for the presence of a visually compact spheroidal structure at the centre of a galaxy \citep[often referred to as a classical bulge, and expected to be dynamically hotter than surrounding components e.g.][]{kormendy2004} is a high central value for $\sigma_{\star}$. \citet{mendez-abreu2014}, for example, defined any galaxy with $\sigma_{\star}>200~\rm{km}~{s}^{-1}$ to possess a classical bulge. 
Based on an extensive literature review, \citet{fisher2016} updated this value to $\sigma_{\star}>130~\rm{km}~{s}^{-1}$. 
Other works combine photometric and spectroscopic indicators, for example \citet{neumann2017} measure the inner slope of the radial velocity
dispersion profile within the photometrically-defined bulge region.
Several GECKOS galaxies possess a peak in their dispersion profile in their central regions, and for both IC 5244 and IC 1711, this peak value is $> 130$ km s$^{-1}$. Following the literature, we would then attribute these peaks as evidence of classical bulges, however, there are also alternate explanations.

\citet{iannuzzi2015} compare the 2D stellar kinematic maps of simulated bulgeless and BP bulge galaxies. They found that for all bar viewing angles, $\sigma_{\star}$ increases significantly when a BP bulge is present, which they attribute to the larger number of orbital shapes populating the structure. Interestingly, their `composite' case of a BP + classical bulge resulted in characteristic kinematic signatures being considerably weakened - i.e. when a classical bulge's contribution was added to the LOSVD, the BP bulge-related signatures were damped. 

Examining the end-on bar case of \citet{bureau2005} we see that high $\sigma_{\star}$ could also be explained by an end-on (or nearly end-on) bar. The reason for this is that in the end-on case, the $x_{1}$ orbits are elongated along the line-of-sight, resulting in higher line-of-sight velocities and also dispersions. This, and the overall variety of orbits in barred galaxies lead to an increase in central velocity dispersion, particularly when the bar is seen end-on. This case of an end-on (or near end-on) bar masquerading as a classical bulge is important; if other galactic structures can imitate their morphological and kinematic features, then the concept, ubiquity, and utility of classical bulges is brought into question.

Returning to the cases of IC 1711 and IC 5244, despite their high $\sigma_{\star}$ peaks, we conclude that the BP bulge morphology, clear kinematic nuclear disc signatures, and additional high-order kinematic information point to end-on bars being the likely cause of the $\sigma_{\star}$ behaviour in these galaxies. The observations can be fully explained by the presence of an end-on bar, without the need to invoke significant contribution from a classical bulge. 
It will be informative to examine the full GECKOS sample for such features, though we note that statistical photometric studies find low fractions of classical bulges at Milky Way stellar masses \citep[e.g.][]{fisher2011}; it may be that the GECKOS sample is simply below the stellar mass range in which we expect to find such features. Indeed, the Milky Way itself possesses cylindrical rotation \citep{shen2010} and structural and kinematic properties \citep{bland-hawthorn2016} inconsistent with the presence of a classical bulge. 

\subsubsection{Kinematic disc structures}

In Section~\ref{sec:classicalbulge}, we explained $\sigma_{\star}$ central peak behaviour in BP bulge galaxies through the new stellar orbits introduced by the BP and bar structures themselves. 
However, there are two galaxies in the current sample that possess a peak in central stellar velocity dispersion, but no other visual or kinematic indicators of a bar. NGC 5775 (Figure~\ref{Fig:vel_maps10}) and NGC 0360  (Figure~\ref{Fig:vel_maps6}) both possess increased $\sigma_{\star}$ in their centres, though smaller than the threshold proposed for classical bulges by \citet{fisher2016}.

\citet{wang2024} present the analytic chemo-dynamical model of the Milky Way of \citet{sharma2021} that includes a prescription of the evolution of [$\alpha$/Fe] with age and [Fe/H], and a new set of relations describing the velocity dispersion of stars.
They produce mock IFS cubes and run them through \ppxf in the same manner as an observation. Their Figure 6 shows that once Voronoi-binned and flux-weighted, $\sigma_{\star}$ is almost always overestimated compared to the input mock cube. The structure of their $\sigma_{\star}$ maps that were obtained through \ppxf is very similar to that seen for NGC 5775 and NGC 0360.
It seems that an analytic,  axisymmetric chemo-dynamical model could replicate the observed features of NGC 5775 and NGC 0360 without the need to invoke further central structure \citep[see also][]{guerou2016}. Dynamical modelling  \citep[e.g.][]{poci2019,tahmasebzadeh2024} will help in this space.

The remaining galaxies that are yet to be discussed are UGC 00903, and NGC 3279. The former hosts a complex stellar kinematic structure, thought to be due to the presence of a counter-rotating stellar disc. Such structures are seen statistically in IFS survey samples in e.g.\ \citet{bevacqua2022}, though they find that they are rare (1.3 per cent).
NGC 3279 is similarly intriguing, in that it hosts a disc with almost uniformly low $\sigma_{\star}$ across the whole galaxy. We expect this galaxy to comprise chiefly a single, dynamically-cold disc. Understanding its evolutionary path will be illuminating.

With this, we have explained the kinematic behaviour of 11/12 GECKOS galaxies (excluding UGC 00903, whose kinematics are considerably more complex and require detailed modelling) by a combination of bars, BP bulges, nuclear discs, and simple disc structure(s). We see no physically-motivated reason to invoke classical bulges in this sample, a conclusion also drawn by \citet{bittner2020} for the TIMER sample. While certainly not statistical nor representative, the GECKOS sample does span a range of bulge morphologies and bulge-to-total ratios derived from two-component decompositions. The full sample in combination with dynamical modelling will provide excellent insight into the frequency but also necessity (or lack thereof) of compact ($\sim$100s of pc) spheroidal bulges in the local Universe.

\begin{table*}
\caption{Summary of central sub-structures classifications and kinematic properties}          
\label{table:1}      
\centering                          
\begin{tabular}{l c c c c c}       
\hline\hline                 
Galaxy & BP bulge & Kinematic evidence & $h_{3}$ correlated   & Central $\sigma_{\star}$ & Bar viewing angle \\    
    & from imaging & for ND? & with $V_\star$ & behaviour  & if barred \\
\hline                        
   NGC 3957 & Yes & Yes & Yes & Central `croissant' on high $\sigma_{\star}$  & Intermediate \\      
   PGC 044931 & Yes & Yes & Yes & Central `croissant' on high $\sigma_{\star}$ & Intermediate \\
   IC 5244 & Yes & Yes & Yes & Large and circular & End-on bar \\
   NGC 0522 & Yes & No & Yes - weak & Elongated `tophat' & Side-on bar \\
   IC 1711 & Yes & Yes & Yes & Large and circular & End-on bar \\ 
   ESO 484-036 & Yes & No & Yes & Slightly elongated along z=0 line & Intermediate\\
   ESO 079-003 & Yes & No & Yes & Slightly elongated along z=0 line & Intermediate\\
   ESO 120-016 & Maybe & No & Maybe & Slightly elongated along z=0 line & Intermediate (if barred) \\
   NGC 0360 & No & No & -- & Centrally-peaked & Unbarred \\
   UGC 00903 & No & No & -- & No central increase & Unbarred\\
   NGC 3279 & No & No & -- & No central increase & Unbarred \\
   NGC 5775 & No & No & -- &  Slightly elongated along z=0 line & Unbarred \\
 
\hline                                 
\end{tabular}
\end{table*}

\section{Summary and conclusions}

To understand the diversity of kinematic sub-structures in Milky Way-mass galaxies, we examine the stellar kinematics of the first 12 targets observed as part of the GECKOS Survey of nearby edge-on galaxies. 
After first visually examining optical and mid-IR images to identify BP bulge structures, we examine both the 1D radial profiles and 2D maps of line-of-sight mean velocity ($V_{\star}$), velocity dispersion ($\sigma_{\star}$), and high-order skew ($h_{3}$) and kurtosis ($h_{4}$). Eight of the 12 galaxies possess boxy and/or peanut-shaped central structures, present in either $r$-band optical or 3.6-$\mu$m Spitzer imaging.

There is a diversity of kinematic structures in the centres of the GECKOS galaxies, yet some trends emerge that we compare to literature predictions in both 1D and 2D:
\begin{itemize}
    \item All galaxies show $h_{3}-V_{\star}$ sign mismatch in outer disc regions consistent with a (quasi-)axisymmetric, rotating disc of stars.
    \item All galaxies that show evidence for a clear boxy-peanut shaped structure in images also possess kinematic indicators of a bar. In particular, all of these galaxies show extended regions of $h_{3}$ sign matching with $V_\star$ permeating up to 2 kpc off the disc z=0 line. These features are not present in the four non-BP bulge galaxies.
    \item Four BP bulge galaxies also host strong kinematic signatures of nuclear discs, including a `double hump' rotation curve, central $\sigma_{\star}$ minima, $h_{3}-V_{\star}$ sign mismatch in the central regions, and strong $h_{4}$ maxima co-spatial with the expected location of the nuclear disc itself.
\end{itemize}

Comparing to the 2D maps produced by N-body simulations of \citet{iannuzzi2015}, there is an array of central velocity dispersions, and varying degrees of coherent regions of $h_{3}$-$V_\star$ sign match across the bar-dominated regions. In IC 5244, IC 1711, NGC 3957, and PGC 044931, there is strong evidence for the existence of nuclear discs, in the form of central $\sigma_{\star}$ depressions (NGC 3957 and PGC 044931), and central $h_{3}$-$V_\star$ sign mismatch (in all four galaxies). 
Dust obscuration and resolution effects can significantly hinder the recovery of the stellar kinematics in the central regions of edge-on galaxies, so we cannot rule out the possibility that a nuclear disc is present in all of the BP bulge galaxies; we may simply not be able to detect them. 

Combining the $\sigma_{\star}$, $h_{3}$, and visual morphology of each object, we can constrain the bar viewing angle. For IC 5244 and IC 1711, the large central velocity dispersions are evidence for the many orbits of an end-on bar contributing to the LOSVDs. Conversely, NGC 0522 possesses one of the strongest BP bulge structures, but very weak bar signatures in its stellar kinematics, suggesting that the bar is oriented almost side-on to the viewer. Future work combining the stellar kinematic indicators presented here with ALMA cold-gas measurements will provide more insight into bar viewing angles.

We discuss our findings in the context of other central structures. We see no kinematic evidence for classical bulges that cannot also be explained by the presence of a bar, or simple axisymmetric disc structure(s). We conclude that no galaxy with BP bulge structure (nor indeed any galaxy in this sample) displays signs of a classical bulge \cite[see also][]{bittner2020}.

We conclude that the variety of kinematic sub-structures found in these twelve GECKOS galaxies promotes a modern interpretation of central galaxy structure - one in which the latest chemodynamical models of the Milky Way may be used as a template to understand disc galaxy evolution.

\begin{acknowledgements}
      Author contribution statement: This project was devised by AFM, with analysis completed by AFM, and some scripts provided by JvdS. Comments were provided by the whole author list. GECKOS data reduction is led by JvdS and AFM, with assistance from EE, LC, BMC, TM, and BC.
    \ngist\ development is led by AFM and JvdS, with contributions and testing from TB, AW, ZW, LSL, LC, FP, MM, DG, EE, and CdSF.\\
      
      Based on observations made with ESO Telescopes at the La Silla Paranal Observatory under programme ID 110.24AS. We wish to thank the ESO staff, and in particular the staff at Paranal Observatory, for carrying out the GECKOS observations.\\

      The authors wish to thank the anonymous referee, whose careful reading and interpretation of this manuscript resulted in comments that improved its quality.
      The authors also wish to thank Adrian Bittner for useful conversations, and his blessing to continue the development of the \gist~ pipeline. We also thank Lodovico Coccato for useful conversations relating to MUSE data reduction and analysis, Alex Vazdekis for his permission to distribute the MILES models with the \ngist\ package, Livia Casagrande whose Honours work provided useful insights, and Michele Cappellari for useful conversations related to \ppxf usage.\\
      
      This research has made use of the NASA/IPAC Extragalactic Database (NED; https://ned.ipac.caltech.edu/) operated by the Jet Propulsion Laboratory, California Institute of Technology, under contract with the National Aeronautics and Space Administration.
      This research has made use of the NASA/IPAC Infrared Science Archive, which is funded by the National Aeronautics and Space Administration and operated by the California Institute of Technology.
      This research used the Canadian Advanced Network For Astronomy Research (CANFAR) operated in partnership by the Canadian Astronomy Data Centre and The Digital Research Alliance of Canada with support from the National Research Council of Canada the Canadian Space Agency, CANARIE and the Canadian Foundation for Innovation.

    Part of this research was conducted by the Australian Research Council Centre of Excellence for All Sky Astrophysics in 3 Dimensions (ASTRO 3D), through project number CE170100013.\\

    AFM gratefully acknowledges the sponsorship provided by the European Southern Observatory through a research fellowship.
    DAG is supported by STFC grant ST/X001075/1.

    TAD acknowledges support from the UK Science and Technology Facilities Council through grants ST/S00033X/1 and ST/W000830/1.
      FP acknowledges support from the Agencia Estatal de Investigaci\'on del Ministerio de Ciencia e Innovaci\'on (MCIN/AEI/ 10.13039/501100011033) under grant (PID2021-128131NB-I00) and the European Regional Development Fund (ERDF) "A way of making Europe”. FP acknowledges support also from the Horizon Europe research and innovation programme under the Marie Skłodowska-Curie grant “TraNSLate” No 101108180.

      FF is supported by a UKRI Future Leaders Fellowship (grant no. MR/X033740/1).

      AP acknowledges support from the Hintze Family Charitable Foundation. 
      
    LSL acknowledges CAPES for the support through grant 88887.637633/2021-00 and CNPQ through grant 200469/2022-3
    
      LMV acknowledges support by the German Academic  
    Scholarship Foundation (Studienstiftung des deutschen Volkes) and the Marianne-Plehn-Program of the Elite Network of Bavaria.

    LC and ABW acknowledge support from the Australian Research Council Discovery Project funding scheme (DP210100337)

    J.F-B acknowledges support from the PID2022-140869NB-I00 grant from the Spanish Ministry of Science and Innovation.
    
     MM acknowledges support from the UK Science and Technology Facilities Council through grant ST/Y002490/1.
\end{acknowledgements}

\bibliographystyle{aa}
\bibliography{references_short}{}

\begin{appendix}
\label{label:appendix}

\section{The \ngist\ pipeline}
\label{ngist_appendix}
Modern IFS datasets require modern data analysis pipeline solutions.
The new Galaxy IFU Spectroscopy Tool (\ngist) pipeline is a continuation of the \gist~ pipeline of \citet{bittner2019}, though led by a new team of developers. Borne out of the need for a robust but flexible analysis pipeline for an influx of MUSE galaxy data, the decision was made to build upon existing publicly available software, rather than starting again from scratch. 

The need for further development arose from both the increased computational and scientific requirements of recent IFS surveys. \ngist\ pushes beyond current limits in terms of memory requirements and deals better with longer optical wavelength ranges and sky residuals that are particularly problematic at redder wavelengths ($\lambda>7000\AA$). Building on the foundations of the \gist\ pipeline, \ngist\ also incorporates elements of pipelines developed for recent IFS surveys, including the SAMI stellar kinematics pipeline \citep{vandesande2017}, and the PHANGS data analysis pipeline \citep[DAP;][]{emsellem2022}, which in turn is rooted in portions of the MaNGA DAP \citep[][]{belfiore2019, westfall2019}. \ngist\ represents a significant upgrade in both scientific output and computational performance over \gist.

\subsection*{Overview}
Starting with a reduced IFS datacube, and after running preparatory steps, \ngist\ runs up to six science modules, creating science-ready value-added products including stellar kinematics, continuum-subtracted cubes, emission line measures, and stellar population parameters via both full spectral fitting and line strength index measures. Any subset of these science modules can be run at a given time, and in any order. Each module possesses associated routines that use different methods to compute the desired result. For example, the emission line measurement module can be run using one of three routines: one employing \textsc{gandalf} \citep{sarzi2006}, one employing a modified version of \textsc{gandalf} developed by the MAGPI team (Battisti et al., in prep), and a third employing a \textsc{pPXF}-based routine. Some modules currently contain only one routine. A user can easily add their own module or routine to the \ngist\ framework, making it flexible and adaptable for modern and future datasets.  

\ngist\ commenced from the final release of the \gist~ pipeline, v3.1.0\footnote{We note that the final release of \gist~ requires \textsc{python $<$ v3.6}, which is increasingly deprecated and difficult to install on modern machines via package managers such as \textsc{conda}. At the time of writing, \ngist\ has been tested up to \textsc{python v3.11}.}, and has continued to be developed by adding new functionality, and updates that are appropriate for increasingly complex IFS datasets. \ngist\ is publicly available via a GitHub repository\footnote{\href{https://github.com/geckos-survey/ngist}{https://github.com/geckos-survey/ngist}}, with documentation available\footnote{\href{https://geckos-survey.github.io/gist-documentation/}{https://geckos-survey.github.io/gist-documentation/}}. It has been thoroughly tested for MUSE data, but can be employed with any galaxy IFS dataset. Below, we detail the \ngist\ philosophy, workflow, and major changes introduced since the last version of \gist. For a full description of the original \gist\ pipeline, we refer the reader to \citet{bittner2019}.

\subsection*{\ngist\ pipeline philosophy}
The \ngist\ pipeline aims to maintain the core principles of the \gist~ pipeline, namely 
\begin{itemize}
    \item \textbf{Convenience}: an all-in-one framework that takes an input fully reduced IFS data cube and outputs consistent 2D maps of value added products.
    \item \textbf{Extensive functionality}: using popular industry-standard tools and routines but allow for their modification to the user's specification. 
    \item \textbf{Flexibility \& Modularity}: The ability to customise pipeline outputs to the user's specifications. The ability to run only the modules that the user wishes.
\end{itemize}

\section*{Performance upgrades}
Several significant performance-based enhancements have been made to the original \gist\ code base with the main goals of optimizing memory usage and parallelization when the input data cube is large ($\gtrsim 5$GB). The important code optimizations are as follows:

\begin{itemize}
    \item {\bf Optimized Voronoi Binning:}
    The spatial binning module was modified to use \textsc{scipy}'s \texttt{spatial.cKDTree} method when assigning pixels to each bin. This modification improves the speed of the code by leveraging nearest-neighbor searches and significantly reduces RAM usage by avoiding unnecessary pairwise distance calculations.
    \item {\bf Switched Parallelization to Joblib:} \ngist\ switches from the \textsc{multiprocess python} package to \textsc{joblib} for parallelization, improving overall performance. \textsc{joblib} is designed specifically for parallel computing with large data sets, offering better memory management and more efficient serialization of large objects. Additionally, \textsc{joblib} handles shared memory more effectively, reducing the overhead associated with data transfer between processes.
    \item {\bf Transitioned to HDF5 for Pipeline Data Tables:}
    Previously, the pipeline reformatted and saved both binned and unbinned spectra to multi-dimensional FITS tables by loading the entire dataset into memory. The consequence of this choice is that the pipeline cannot be run when the size of these data tables approaches that of the available memory. \ngist\ transitions to HDF5 tables, which support very large out-of-memory operations and n-dimensional arrays with metadata, similar to FITS. Writing and reading HDF5 tables is now implemented across all relevant modules, significantly improving memory management.
\end{itemize}

\section*{Module Summary}
\begin{figure*}
   \centering
   \includegraphics[trim={1cm 2cm 1cm 2cm},clip, width=0.98\textwidth]{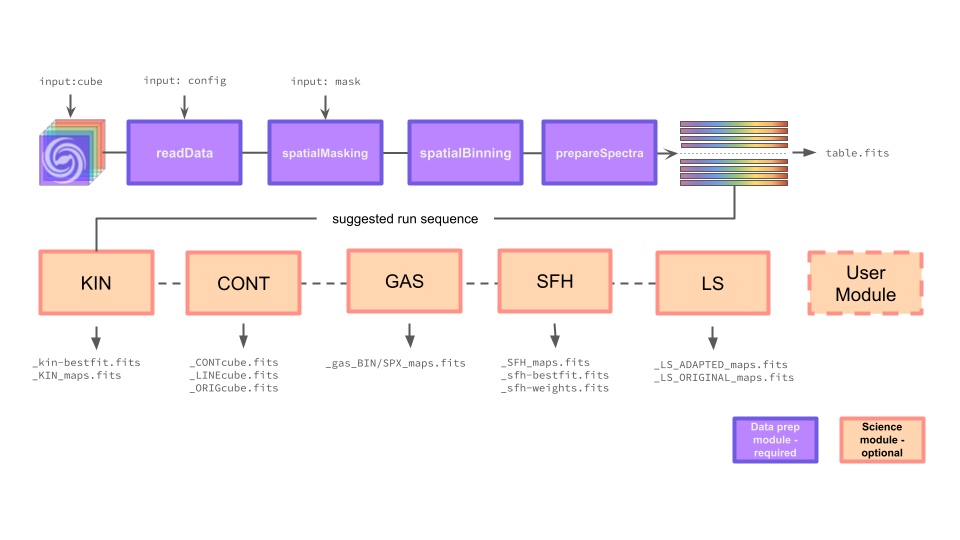}
   \caption{Flowchart of the \ngist\ workflow. Purple boxes denote mandatory preparatory modules, and orange boxes denote the optional science modules. The suggested running sequence of the science modules is shown, though we note that the modules may be run in any order. Selected science-ready output .fits files are shown, and we refer the reader to the \href{https://geckos-survey.github.io/gist-documentation/}{\ngist\ documentation} for a full list of inputs and outputs.}
              \label{Fig:ngist_flowchart}%
    \end{figure*}

Below, we briefly summarise the nine modules that make up \ngist\ (pictorially represented in Figure~\ref{Fig:ngist_flowchart}), and detail any major changes that have taken place since \gist. We note that as \ngist\ is currently being actively developed, new additions may take place at any time. For this reason, we refer the reader to the GitHub repository and documentation for the most up-to-date version and changes. 

\subsubsection*{Read Data (\texttt{READ\_DATA})}
\textbf{Purpose}: Read in the input datacube, and return the relevant information as a \textsc{python} dictionary structure. \\

\noindent{\textbf{Updates since \gist}: }

$\bullet$ Both \gist~ and \ngist\ rely on an input configuration file, within which resides all user-defined options for each module. The \gist~ version of this file had all variable names on one line, and the values on another, which made editing the correct variable difficult. For \ngist, we have re-written this input configuration file as a YAML, clearly separating out each module, and with one input option per line. 

$\bullet$ Added the ability to read in a variance spectrum from the cube if available. This was added to provide the option of non-constant noise for the science routines. 
If no variance spectra are available, then the variance can be estimated via a DER\_SNR algorithm \citep{stoehr2008}, or a constant noise vector assumed. 

$\bullet$ Allowances made for IFS data of different formats, including if the data are in the first HDU, and support for \texttt{CDELTi} keywords.

$\bullet$ Added the ability to perform a Milky Way foreground dust extinction to the input spectra using a \citet{schlegel1998} correction if an E(B-V) value is provided. This E(B-V) value can be easily calculated using, for example, the \texttt{SFDQuery} routine of the \textsc{python} \texttt{dustmaps} package. 

\subsubsection*{Spatial Masking (\texttt{SPATIAL\_MASKING})}
\textbf{Purpose:} Mask spatial regions in an IFS cube. Masks all spaxels below an input minimum signal-to-noise ratio. If specified, will also apply a user-defined mask.

\noindent{\textbf{Updates from \gist:} None.}

\subsubsection*{Spatial Binning (\texttt{SPATIAL\_BINNING})}
\textbf{Purpose:} Spatially bin the data to obtain an approximately constant signal-to-noise ratio throughout the field of view. Currently, the only implemented binning routine is the Voronoi technique of \citet{cappellari2003}, though a user-defined binning scheme can be easily implemented.

\noindent{\textbf{Updates from \gist:}} 

$\bullet$ \texttt{\_SPATIAL\_BINNING\_maps.fits} now output, a file that contains maps of the binning scheme, S/N per spaxel, S/N per bin, flux per spaxel, and the number of spaxels per bin.

$\bullet$ \texttt{KIN} and \texttt{SFH} modules now calculate the post-fit signal-to-noise ratio from the residuals and output these in the saved fits files. We use the residual of the data minus the best-fit model to calculate the standard deviation, which is then compared to the mean of the original noise spectrum. The difference between the two values is used to scale the original noise spectrum. The post-fit SNR is defined as the median of the flux divided by the new noise spectrum.

\subsubsection*{Prepare Spectra (\texttt{PREPARE\_SPECTRA})}
\textbf{Purpose:} Prepare spectra for analysis by binning them, and then log-rebinning them spectrally. Currently, there is only one routine to perform these tasks, \texttt{default.py}

\noindent{\textbf{Updates from \gist:} None.}

\subsubsection*{Stellar Kinematics (\texttt{KIN})}

\begin{figure*}
\centering
\includegraphics[width=0.33\textwidth,trim={0.75cm 0cm 0.5cm 0cm},clip]{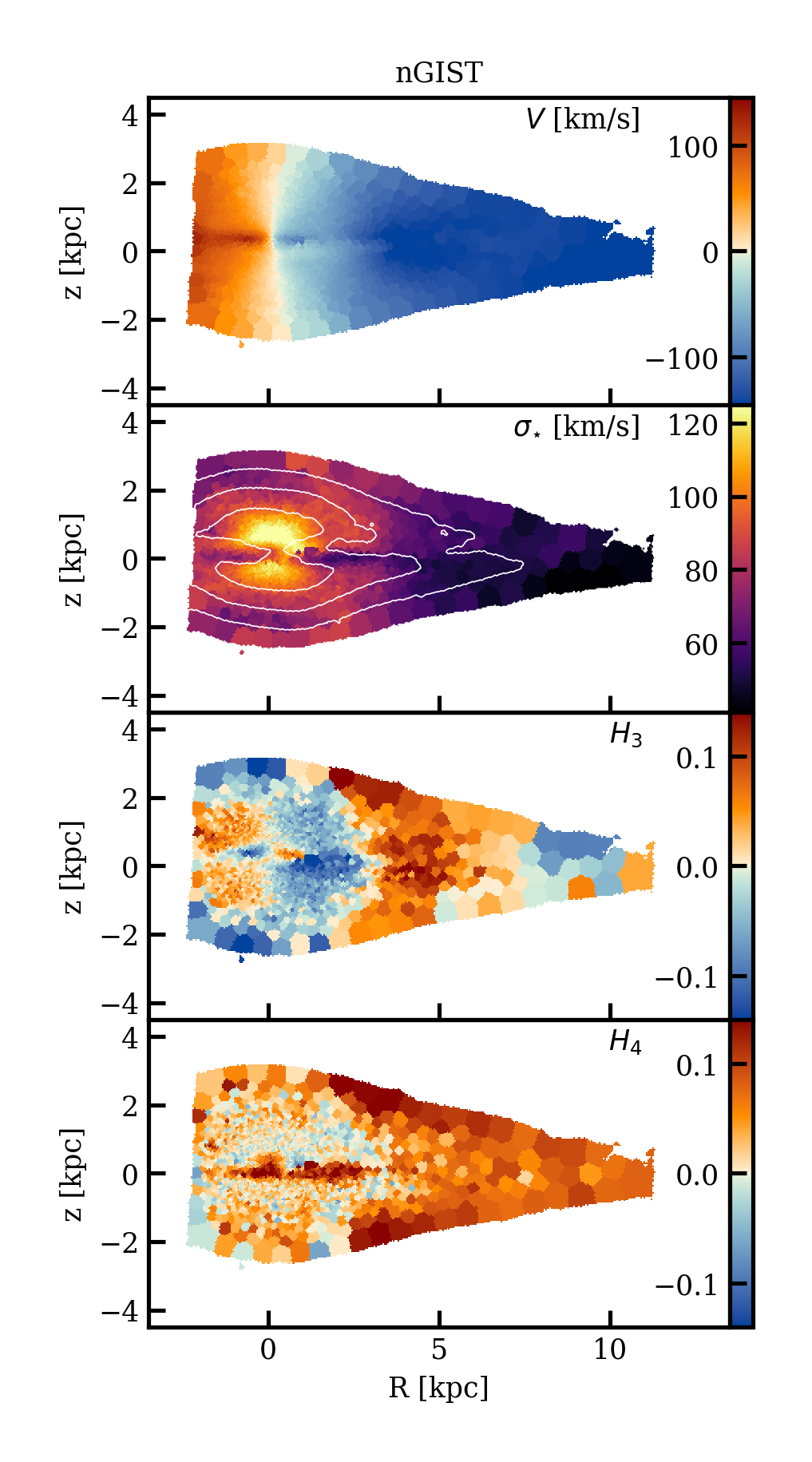}
\includegraphics[width=0.33\textwidth,trim={1.25cm 0cm 0cm 0cm},clip]{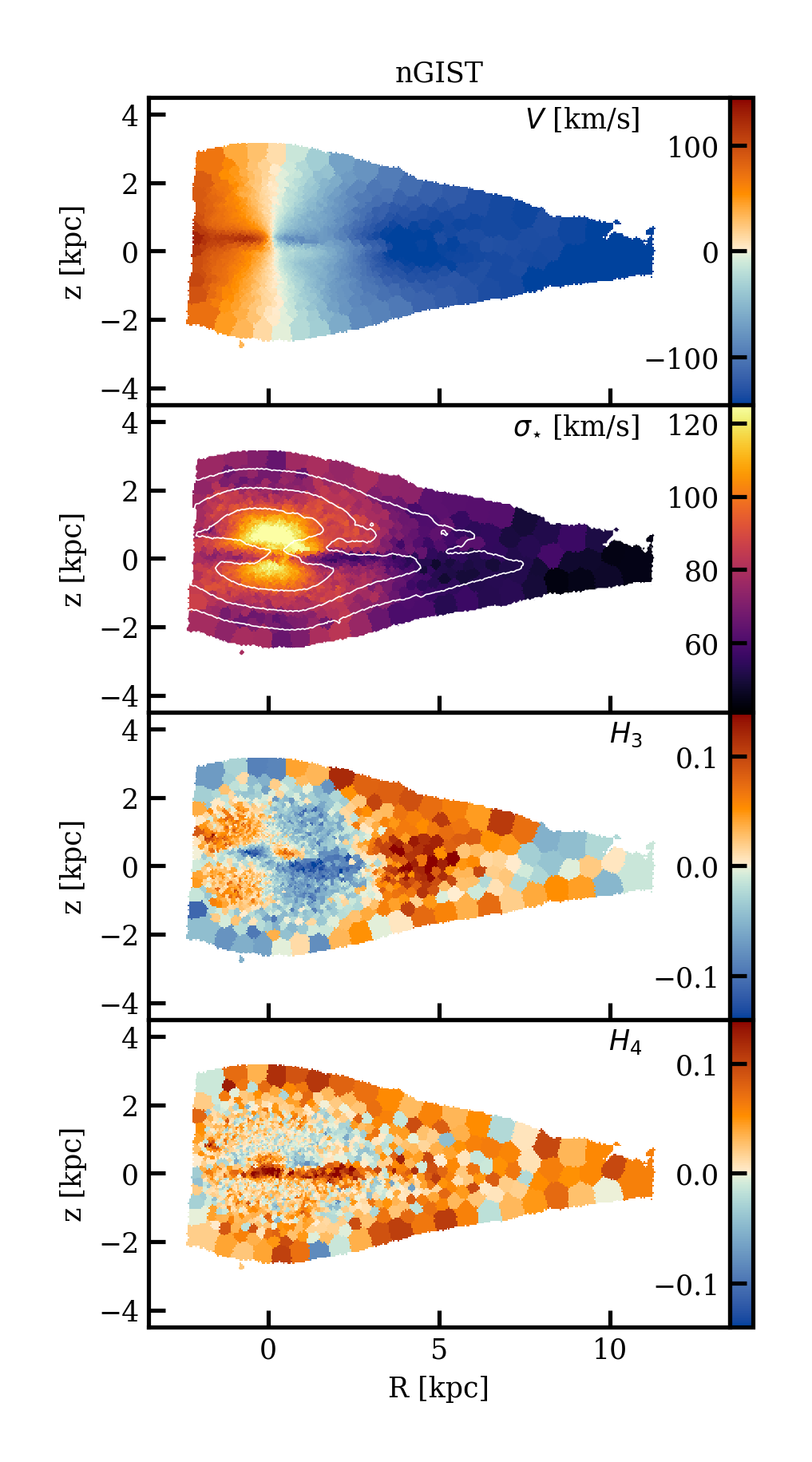}
\includegraphics[width=0.33\textwidth,trim={1.25cm 0cm 0cm 0cm},clip]{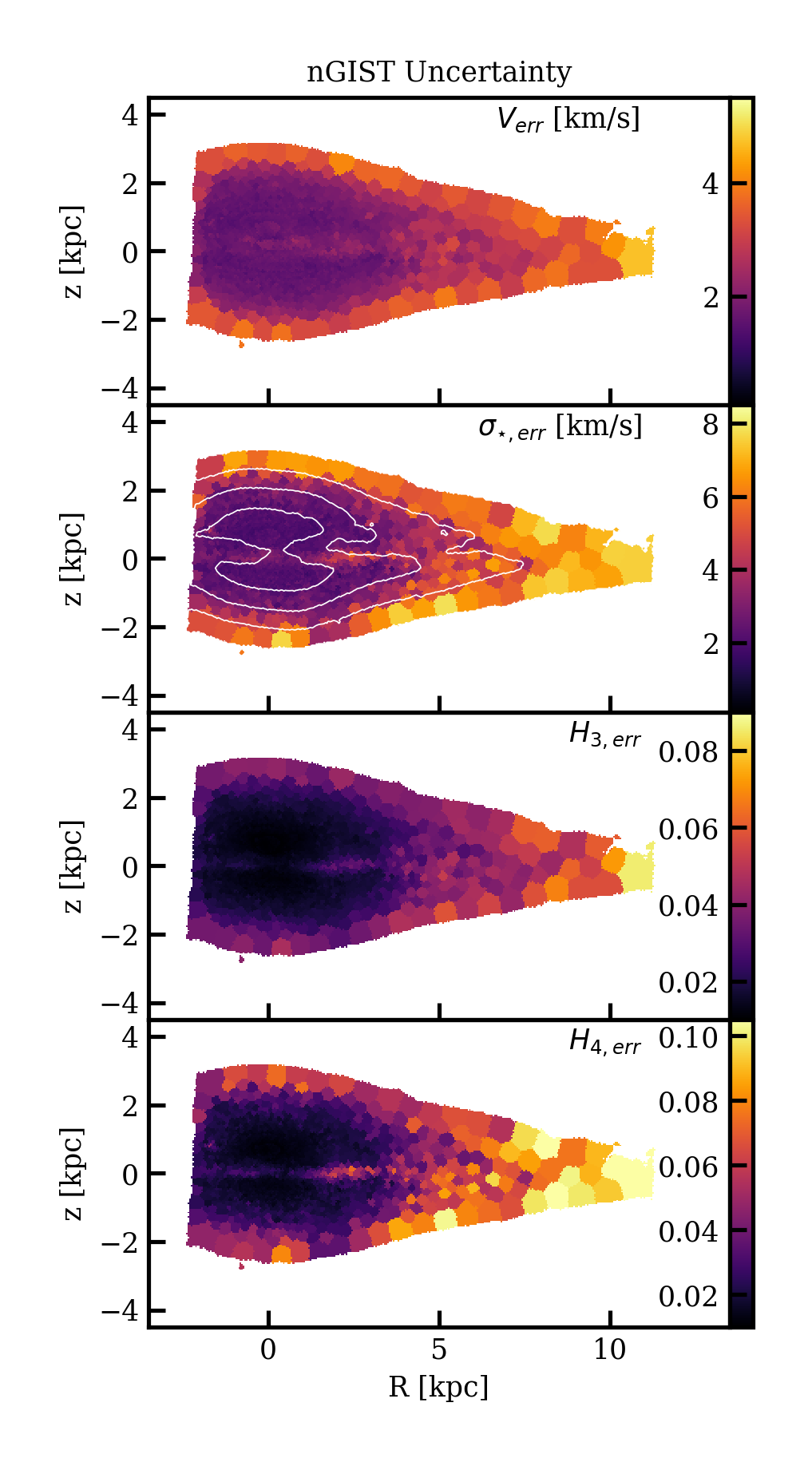}
\caption{Comparison of derived stellar kinematic moment maps between \gist\ and \ngist\ for galaxy IC 1711. The left column shows \gist\ output for (top to bottom) $V_{\star}$, $\sigma_{\star}$ (with surface brightness contours overlaid), $h_{3}$ and $h_{4}$. The middle column shows the same, but from \ngist\, and the right-most column shows the formal error derived by the \ppxf fit of \ngist. For both the \gist\ and \ngist\ runs, cubes were binned to S/N=100 (using the binning maps created by \ngist), and all other input parameters were kept identical.} 
\label{Fig:ogist_ngist_kinmaps}
\end{figure*}

\begin{figure*}
\centering
\includegraphics[width=1.0\textwidth,trim={0.5cm 0cm 0.5cm 0cm},clip]{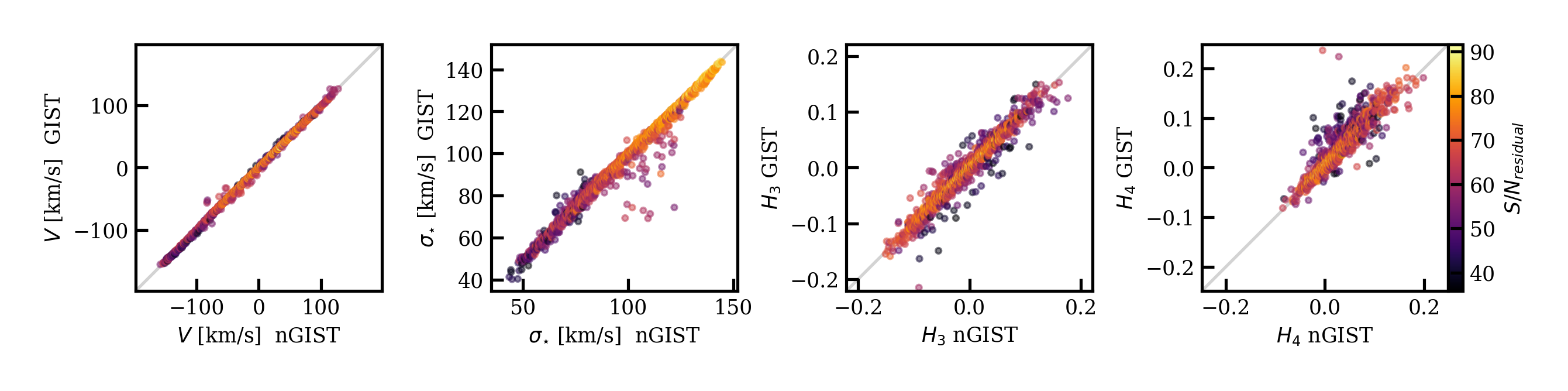}
\caption{A direct, bin-by-bin comparison of derived stellar kinematic moments from \gist\ (y-axis) and \ngist\ (x-axis) for GECKOS galaxy IC 1711. Every point is a Voronoi bin (as shown in Figure~\ref{Fig:ogist_ngist_kinmaps}), and points are colour-coded by the S/N determined from the residual of the data minus the best-fit model.} 
\label{Fig:ogist_ngist_scatter_plots_kin}
\end{figure*}


\begin{figure*}[!ht]
\centering

\includegraphics[width=0.80\textwidth,trim={0 1.6cm 0 0.75cm},clip]{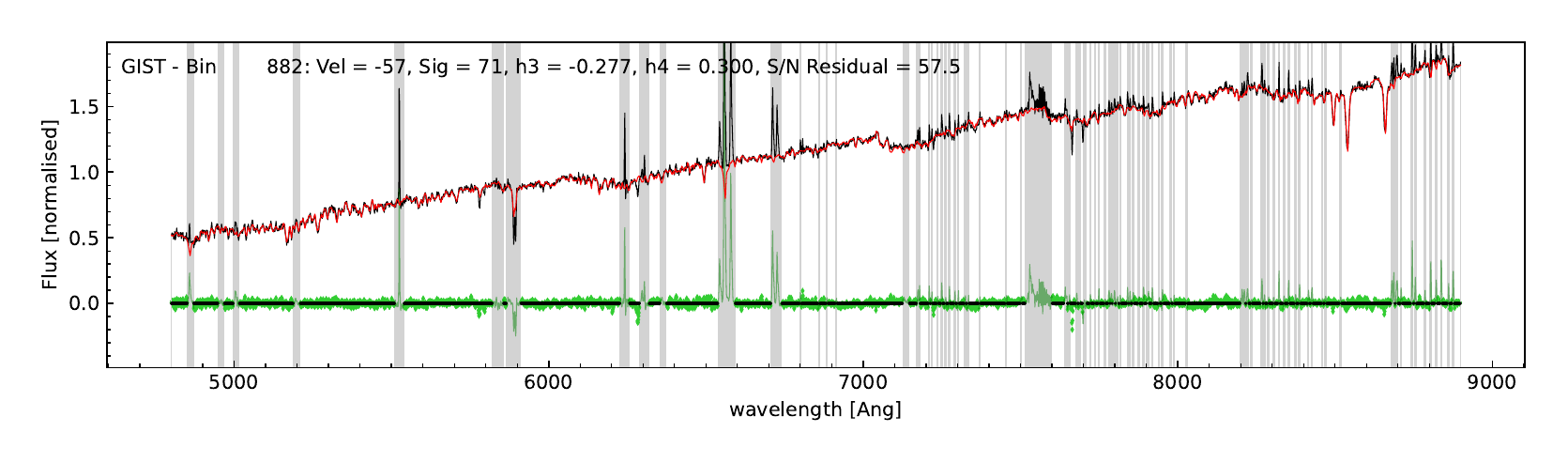}

\includegraphics[width=0.80\textwidth,trim={0 1.6cm 0 0.75cm},clip]{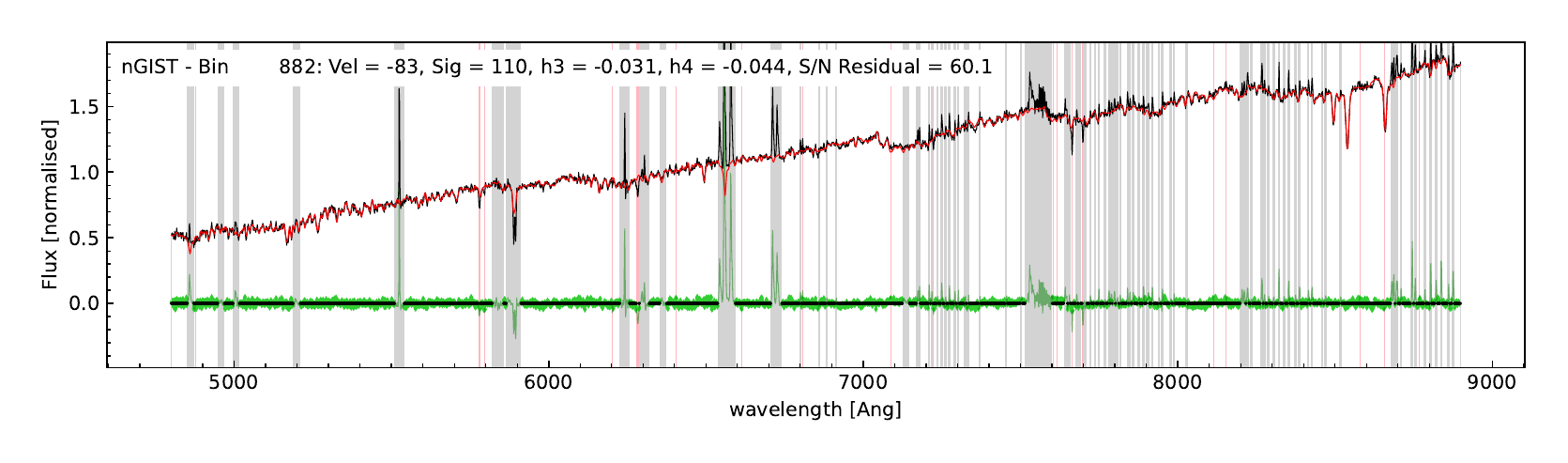}

\vspace{0.25cm}
\includegraphics[width=0.80\textwidth,trim={0 1.6cm 0 0.75cm},clip]{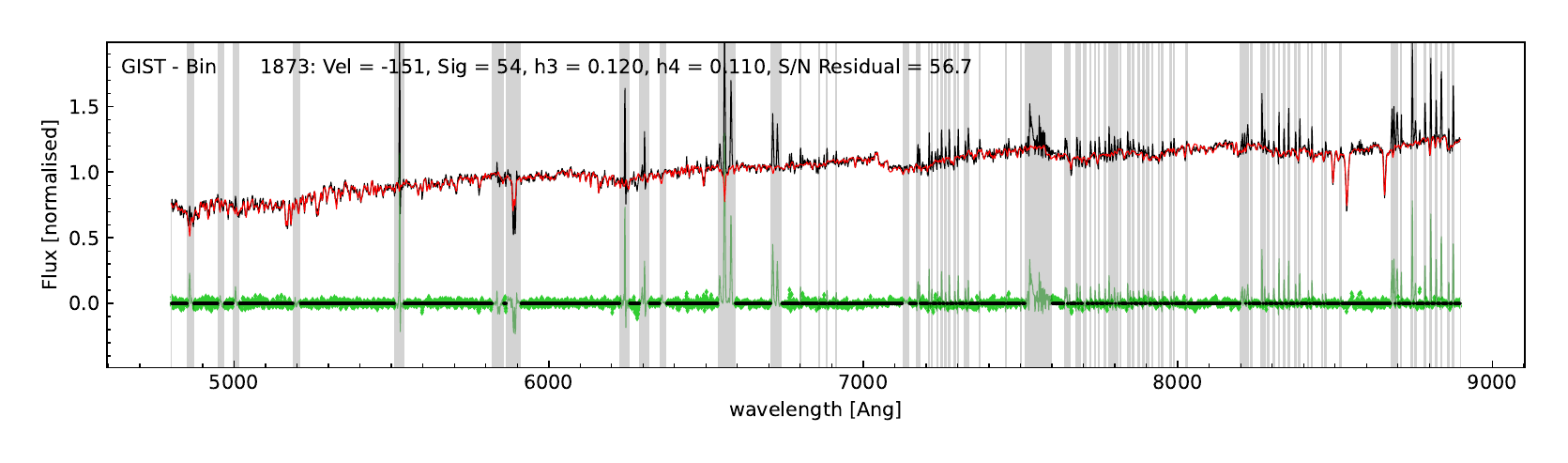}

\includegraphics[width=0.80\textwidth,trim={0 1.6cm 0 0.75cm},clip]{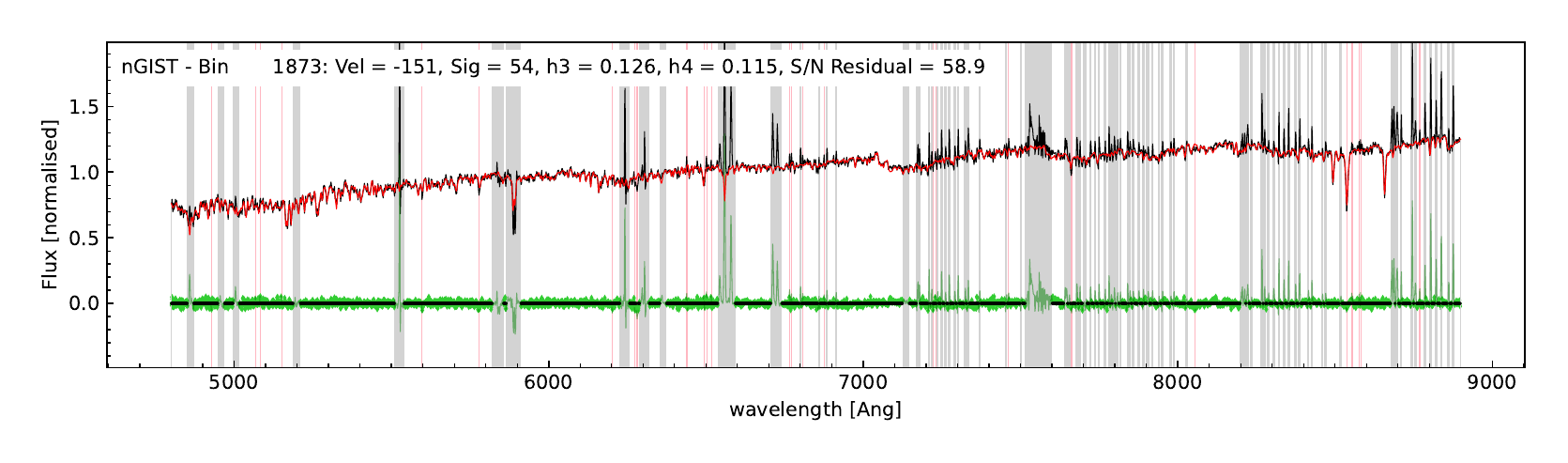}

\vspace{0.25cm}
\includegraphics[width=0.80\textwidth,trim={0 1.6cm 0 0.75cm},clip]{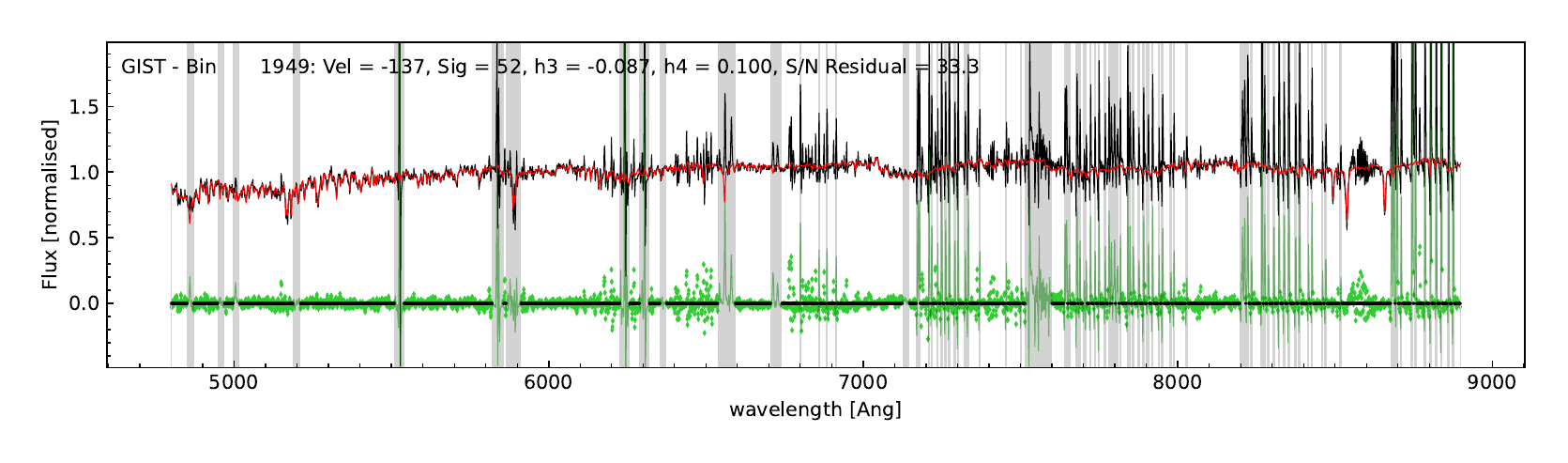}

\includegraphics[width=0.80\textwidth,trim={0 0 0 0.5cm},clip]{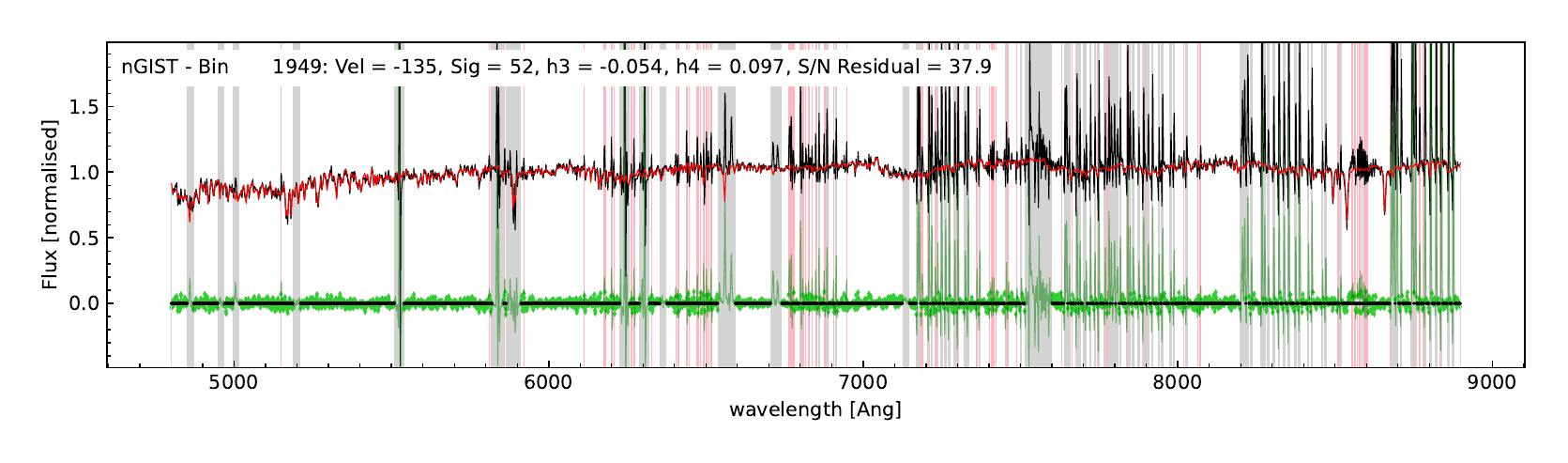}

\caption{Comparison of the masking and subsequent \ppxf best-fit for three Voronoi bins of the \gist\ (top) and \ngist\ (bottom) fits shown in Figures~\ref{Fig:ogist_ngist_kinmaps} and \ref{Fig:ogist_ngist_scatter_plots_kin}. In each case, the original spectrum of the bin is shown as a black line. The emission and sky line masks provided to \ppxf are shown as the shaded grey regions. \ngist\ additionally performs further cleaning of the spectra, and these additional clipped regions are shown as shaded red regions. The best fit to the stellar continuum is shown in red, and residuals in green. The derived stellar kinematic parameters for each spectrum are also listed.} 
\label{Fig:ogist_ngist_spectra}
\end{figure*}

\textbf{Purpose:} Measure the stellar kinematics of the observed spectra. Currently, there is only one routine implemented, that uses the \textsc{pPXF} method of \citet{cappellari2004}, \citet{cappellari2017}, and \citet{cappellari2023}. 

\noindent{\textbf{Updates from \gist:} A considerable rewrite to the \textsc{pPXF} routine has occurred, including: }

$\bullet$ Removed the ability to obtain spatially-resolved $\lambda_{r}$ measures and print maps, as this is an integrated quantity. \\

$\bullet$ Added the option for the user to specify the BIAS value (see \textsc{pPXF} documentation) input to \textsc{pPXF}. If not stated, then the bias is calculated following \citet{cappellari2004} as
\begin{equation}
bias = 0.7 \times \sqrt{500/N_{\lambda pixels}},
\end{equation}

\noindent{where $N_{\lambda pixels}$ is an integer vector containing the indices of the good pixels within the input spectrum (i.e. those to be used in the fit). }

$\bullet$ Normalised the input spectrum as suggested by \citet{cappellari2017}to avoid numerical issues. The output best-fit model is then scaled back to match the original flux of the galaxy spectrum.

$\bullet$ Added the ability to turn the \textsc{pPXF} CLEAN function on or off using the keyword `DOCLEAN'. Useful in finding a proper estimate of the noise. \\

\noindent $\bullet$ Included the option to use a multi-step process to determine the stellar kinematics following the method from \citet{vandesande2017}, which can be used when the absolute value of the noise spectrum is uncertain and requires scaling.
In this method, to increase the fitting speed for individual Voronoi bins, \textsc{pPXF} is run once on a combined mean spectrum of all Voronoi bins to get a single optimal template. For each Voronoi bin, \textsc{pPXF} is then fitted three times:\\

\noindent 1) Using a constant noise spectrum. A new estimate of the noise is derived from the residual of the data minus the best-fit model, using a bi-weight estimate of the scale (standard deviation). The approach is described in \citet{hoaglin1983}.

\noindent 2) Using the scaled noise spectrum from step 1), \textsc{pPXF} is fitted with the single optimal template, this time employing the CLEAN option to clip 3$\sigma$ outliers from the comparison of the best-fit model to the data. 

\noindent 3) In the final step, we use the updated spectral mask from step 2, the scaled noise spectrum, and the full set of templates.

In Figure~\ref{Fig:ogist_ngist_kinmaps}, we show the comparison of the stellar kinematic maps from \gist\ and \ngist\ for GECKOS iDR1 galaxy IC 1711, with a target $S/N=100$ and minimum spaxel $S/N=5$. For this test, we adopted the v11.0 EMILES Base models with the BaSTI scaled-solar isochrones, that can be used in both \gist\ and \ngist. For these fits, we removed the `unsafe' SSPs with metallicities [M/H] = -2.27, -1.79, and +0.40. The wavelength range was set to 4800--8900\AA, we assumed a constant noise spectrum, and fixed the \textsc{pPXF} bias value to 0.14.

There is a good agreement in Figure~\ref{Fig:ogist_ngist_kinmaps} for the results from \gist~(left column) and \ngist\ (middle column) for all kinematic parameters, though there are noticeable differences in the outer bins where the contribution from the sky lines and telluric absorption features become more dominant. In particular, $h_3$ and $h_4$  reach more extreme values in the \gist\ maps. A one-to-one comparison of the results are presented in Figure \ref{Fig:ogist_ngist_scatter_plots_kin}, where the data are colour-coded by the S/N calculated from the residual of the data minus the best-fit. Here $\sigma_{\star}$ can be seen to exhibit some scatter even at moderate values of S/N$_{residual}$.

Even in the central regions, we find subtle but significant differences, best seen in Figure \ref{Fig:ogist_ngist_spectra}. Here, we present the \gist\ (upper) and \ngist\ (lower)  spectra from a central bin (\#882; x,y = 1,0 [kpc]; top two spectra, highest S/N), a mid-disk bin (\#1873; x,y = 5, 0.5 [kpc]; middle two spectra), and an outer bin (\#1949; x,y = 9, -0.75 [kpc]; lower two spectra, lowest S/N). For each spectrum in Figure~\ref{Fig:ogist_ngist_spectra}, the original, Voronoi-binned spectrum is shown in black, the \ppxf-derived best fit spectrum in red, and the residuals of the fit in green. For both \gist\ and \ngist\ runs, the emission and sky line masks employed in the fit are shown as grey-shaded regions.

\ngist\ includes an additional spectral cleaning step, and the regions masked as a result of this step are shown as shaded pink regions. While the sky line masks are extensive (based on the UVES sky emission atlas, where every sky line of flux $>5\times10^{-16}$ erg s$^{-1}$ \AA$^{-1}$ cm$^{-2}$ arcsec$^{-2}$ was masked), some additional masks are always applied by the \ngist\ cleaning step. The number of masked spectral pixels as a result of the cleaning step increases from central to outer galaxy regions where sky lines begin to dominate. There are clear regions in the outer bin spectrum that benefit from the additional masking, particularly a region around $\lambda \sim 8700\AA$, which the Calcium triplet sits on top of at the redshift of this galaxy. We expect this to be the reason behind the scatter in $\sigma_{\star}$ seen in Figure \ref{Fig:ogist_ngist_scatter_plots_kin} for a small number of bins. These bins are located around the edges of the galaxy map, where $\sigma_{\star,err}$ is at its highest.

The derived kinematic parameters are listed in the top left corner of each spectrum. While there are clear differences between all derived kinematic moments in each case, we refer the reader to Figure~\ref{Fig:ogist_ngist_scatter_plots_kin} to see these trends in a statistical sense. Generally, the biggest differences are seen for the high-order kinematic moments, in regions where the sky and telluric residuals are significant, and at lower S/N. We find that the added cleaning step of \ngist\ removes spurious outliers, adding stability and reliability as compared to \gist.

\subsubsection*{Continuum and continuum-subtracted cube creation (\texttt{CONT})}
\textbf{Purpose:} Create continuum-only (i.e. emission line-subtracted), and continuum-subtracted (i.e. line only) cubes. 

\textbf{Updates from \gist:} This is a new module, borne out of the need for these data products within the GECKOS team. This module is closely modelled on the stellar kinematics module, and employs \textsc{pPXF} to model and fit the stellar continuum. 
The best-fit spectrum in each Voronoi bin is rescaled to the individual spaxel level using the ratio SIGNAL$_{bin}$ / SIGNAL$_{spaxel}$, and then subtracted. The continuum-subtracted spectra are then rebinned from the log-rebinned wavelength channels back to the native, linear resolution channels using a cubic spline interpolation. Spaxels without a continuum fit (where the $S/N<S/N_{min}$) are propagated from the original datacube. This technique leaves a baseline jump between fit and unfit regions, which is handled by later processing.

This best fit for each bin is then saved as the continuum data cube. The continuum cube is then subtracted from the raw binned cube to leave just the emission lines, which are saved as the line-only cube.

\subsubsection*{Emission Lines (\texttt{GAS})}

\begin{figure*}
\centering
\includegraphics[width=0.33\textwidth,trim={0.75cm 0cm 0.5cm 0cm},clip]{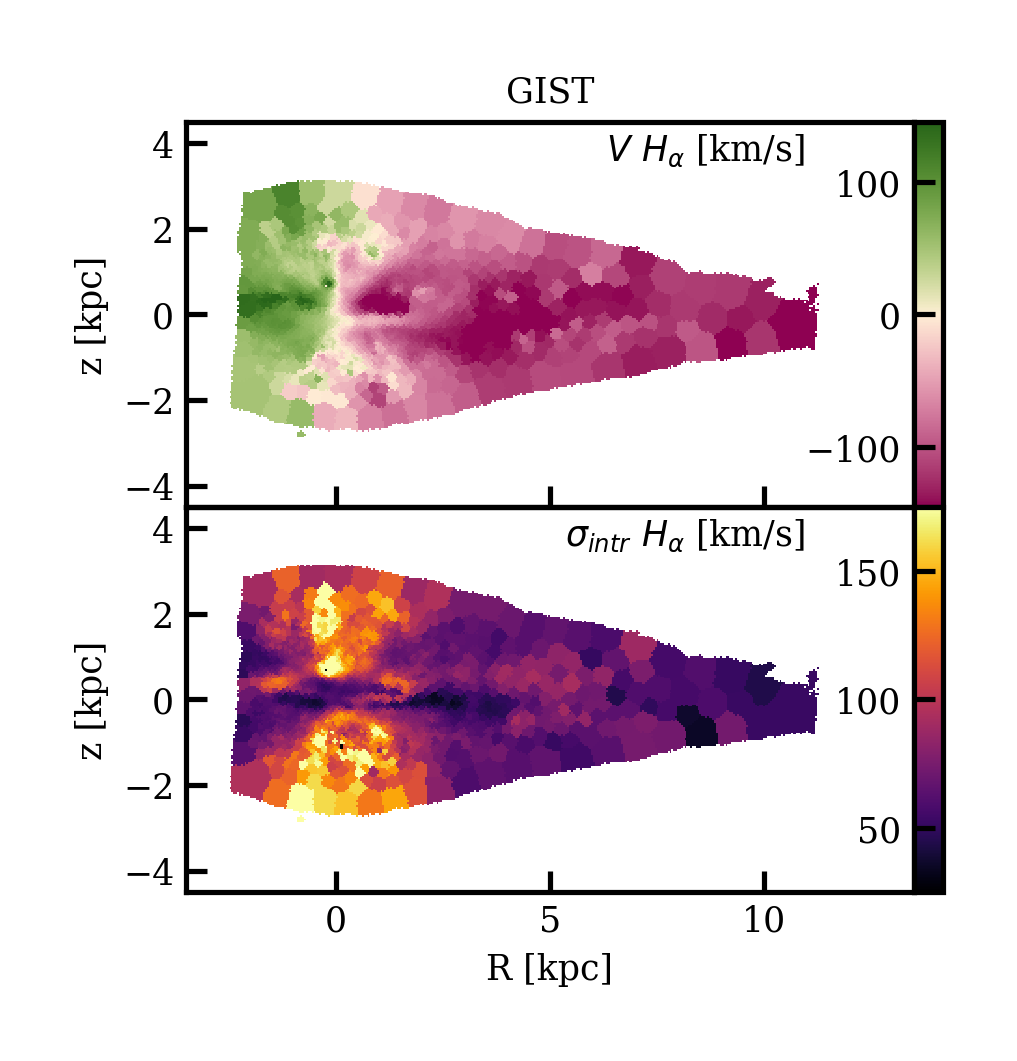}
\includegraphics[width=0.33\textwidth,trim={1.25cm 0cm 0cm 0cm},clip]{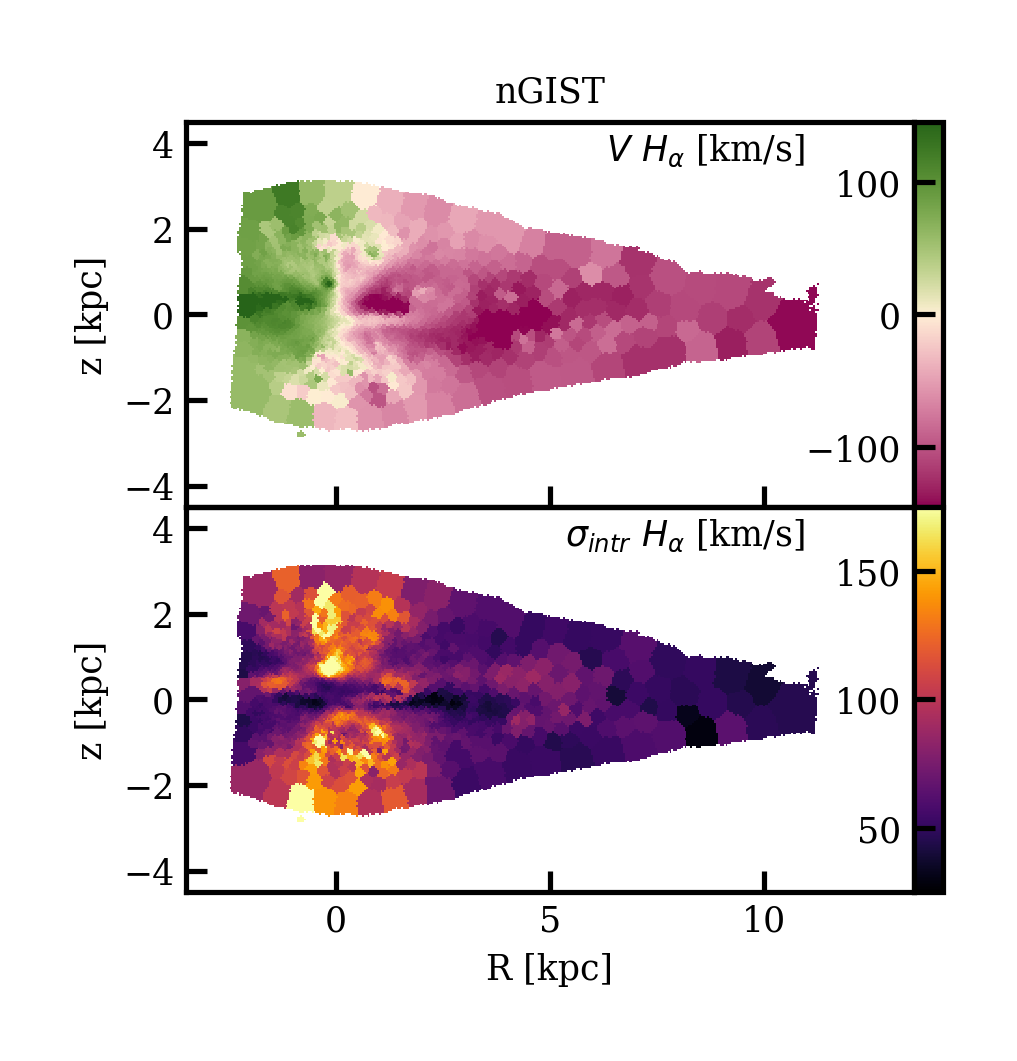}
\includegraphics[width=0.33\textwidth,trim={1.25cm 0cm 0cm 0cm},clip]{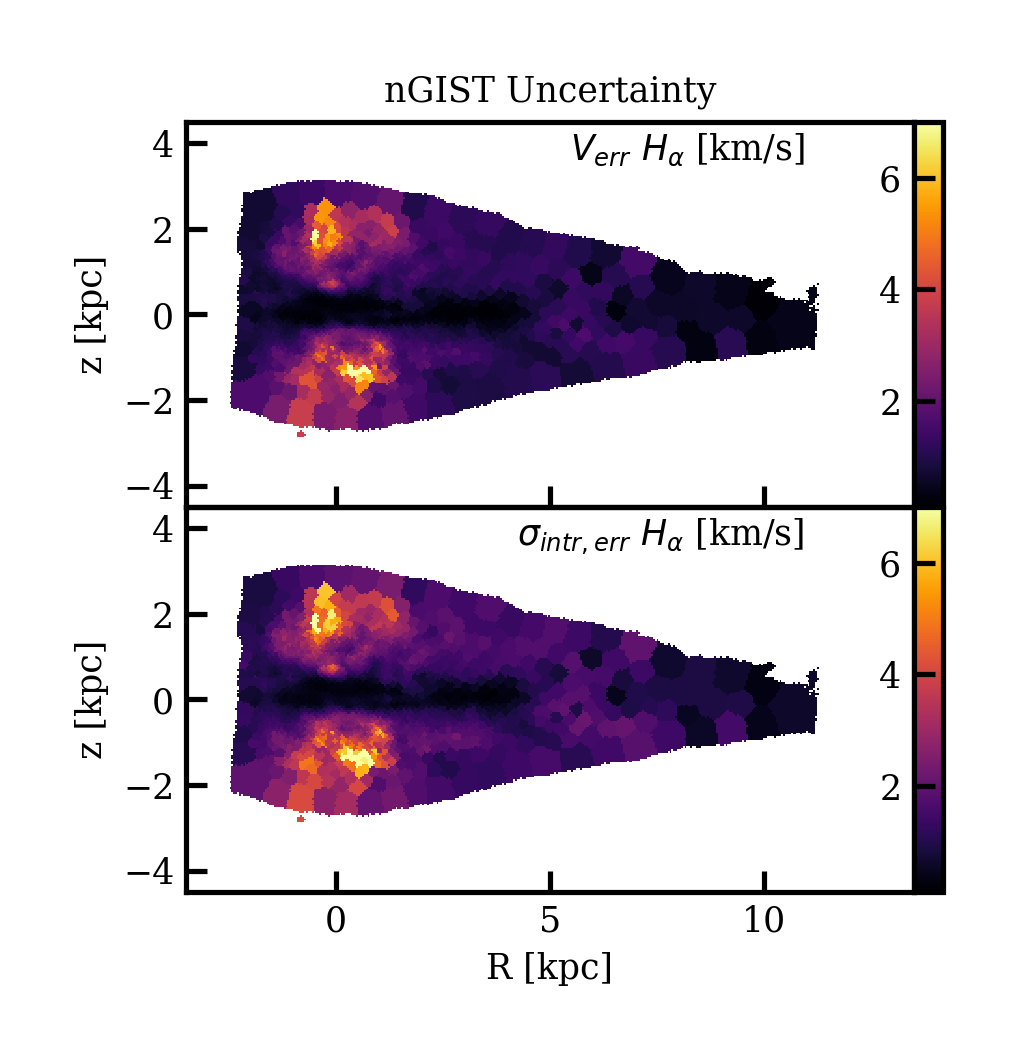}
\caption{Comparison of derived stellar kinematic moment maps between \gist\ and \ngist\ for galaxy IC 1711. The left column shows \gist\ output for $V_{\rm{gas}}$ (top) and $\sigma_{\rm{gas}}$ (bottom). The middle column shows the same, but from \ngist\, and the right-most column shows the formal error derived by the \ppxf fit of \ngist. For both the \gist\ and \ngist\ runs, cubes were binned to S/N=100 (using the binning maps created by \ngist), and all other input parameters were kept identical.} 
\label{Fig:ogist_ngist_gasmaps}
\end{figure*}

\begin{figure*}
\centering
\includegraphics[width=0.6\textwidth,trim={0.5cm 0cm 0.5cm 0cm},clip]{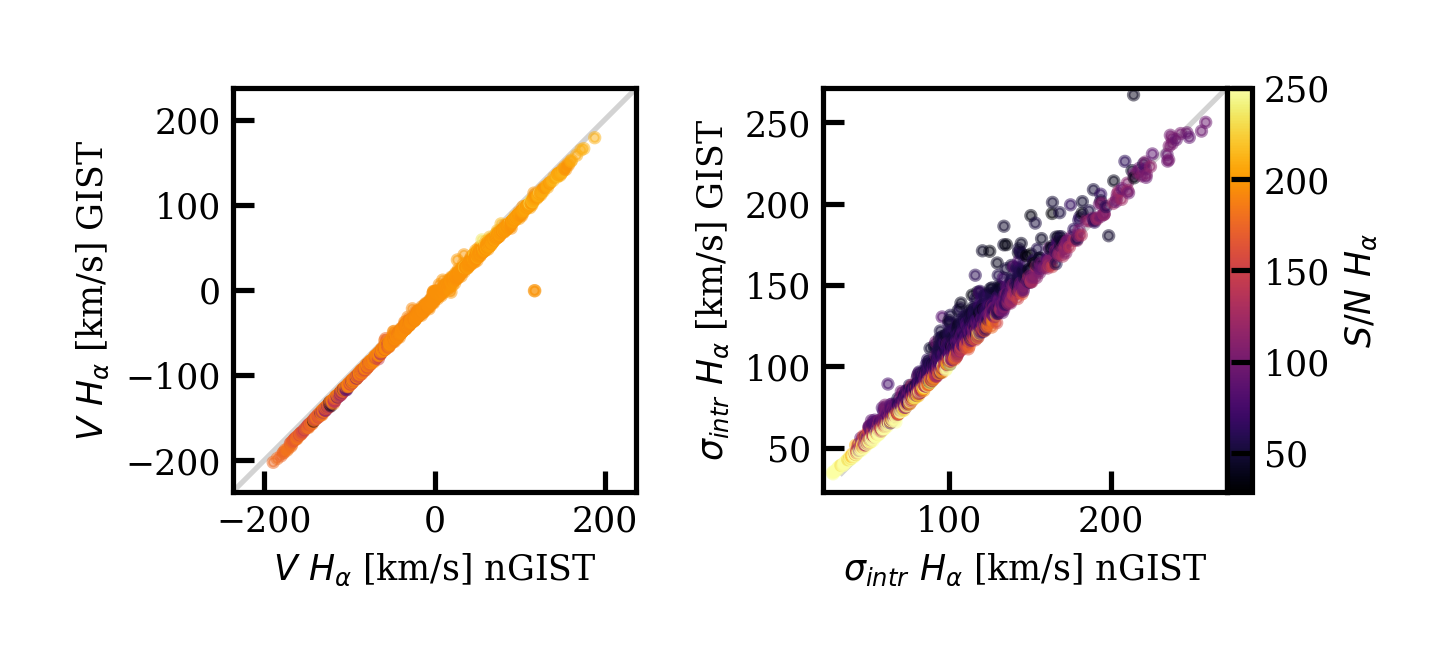}
\caption{A direct, bin-by-bin comparison of derived H$\alpha$ gas kinematics from \gist\ (y-axis) and \ngist\ (x-axis) for GECKOS galaxy IC 1711. Every point is a Voronoi bin (as shown in Figure~\ref{Fig:ogist_ngist_gasmaps}), and points are colour-coded by the H$\alpha$ $S/N$.} 
\label{Fig:ogist_ngist_scatter_plots_gas}
\end{figure*}

\textbf{Purpose:} Measure the flux, velocity, and velocity dispersion of all defined emission lines. 

\textbf{Updates from \gist:} \gist\ had one routine for emission line measurement, employing \textsc{gandalf} \citep{sarzi2006}. 
We retain this routine, and add two other routines: one of which is a modification to the \textsc{gandalf} code by the MAGPI team (Battisti et al., in prep), and the other which is \ppxf-based. The MAGPI \textsc{gandalf} modification includes a chi-squared minimization of the local minima by rerunning spaxels that crash (or don't converge with MPFIT) by adjusting the initial parameter guesses, and then employing a Monte Carlo approach to estimate the flux uncertainties.

To tie the emission line routine in with the \ppxf-based routines available for the \texttt{KIN} and \texttt{SFH} modules, and given a significant speedup, we also developed a \ppxf-based  \texttt{GAS} routine, based on the PHANGS DAP. The input is the raw binned spectrum, and the output of the stellar kinematics module can be used to fix the stellar $V_\star$ and $\sigma_{\star}$ if desired. As with the \textsc{gandalf}-based routine, fits can either be performed on a bin level (matching the binning of the previous modules), or a spaxel-by-spaxel basis. \textsc{pPXF} produces a set of templates, including stellar templates and Gaussian emission lines based on an input line list. There is an option to tie lines together (e.g. the Balmer series), such that their gas velocity will be the same. \textsc{pPXF} simultaneously fits the stellar and gas templates. Gas V and $\sigma$ are determined, as is the line flux in units of $\times 10^{-20}~\rm{erg}~\rm{s}^{-1}~\rm{cm}^{-2}~\AA^{-1}$.\\

Figure~\ref{Fig:ogist_ngist_gasmaps} compares the H$\alpha$ velocity and dispersion derived from the \gist\ \textsc{gandalf}-based routine (left) to that of \ngist's \ppxf-based routine (right). We note that we do not compare emission line fluxes between the two codes as the \gist\ emission line module was relatively untested in this respect. Similar to Figure~\ref{Fig:ogist_ngist_scatter_plots_kin}, Figure~\ref{Fig:ogist_ngist_scatter_plots_gas} compares the binned H$\alpha$ velocity and dispersion values for the \gist\ \textsc{gandalf}-based routine (y-axis) to that of \ngist's \ppxf-based routine (x-axis). There is one clear outlier in the \ngist\ vs. \gist\ $V_{H\alpha}$ plot, where the reported \gist\ $V_{H\alpha}=0$. This behaviour is in line with that seen by the MAGPI team (Battisti et al., in prep).

\subsubsection*{Stellar populations and star formation histories (\texttt{SFH})}

\begin{figure*}
\centering
\includegraphics[width=0.36\textwidth,trim={0.75cm 0cm 0.5cm 0cm},clip]{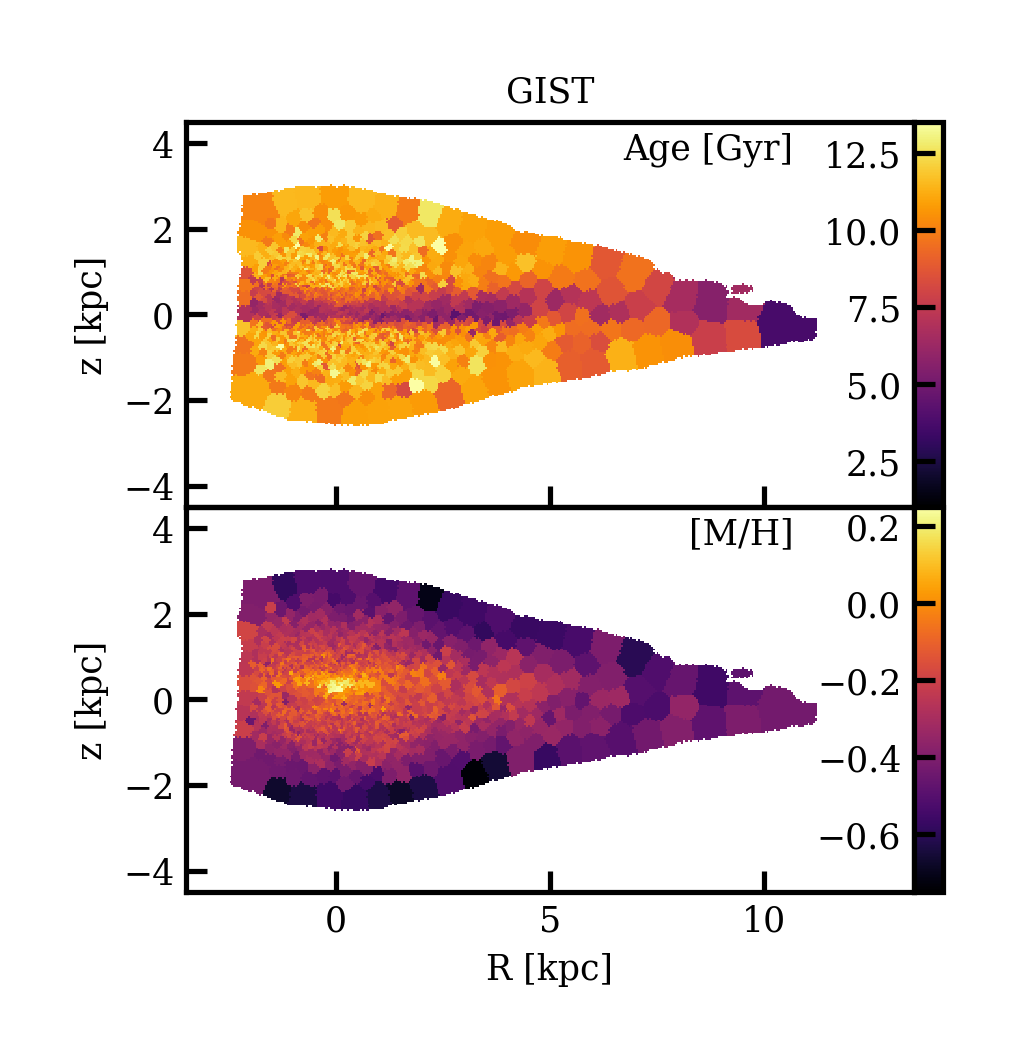}
\includegraphics[width=0.36\textwidth,trim={1.25cm 0cm 0cm 0cm},clip]{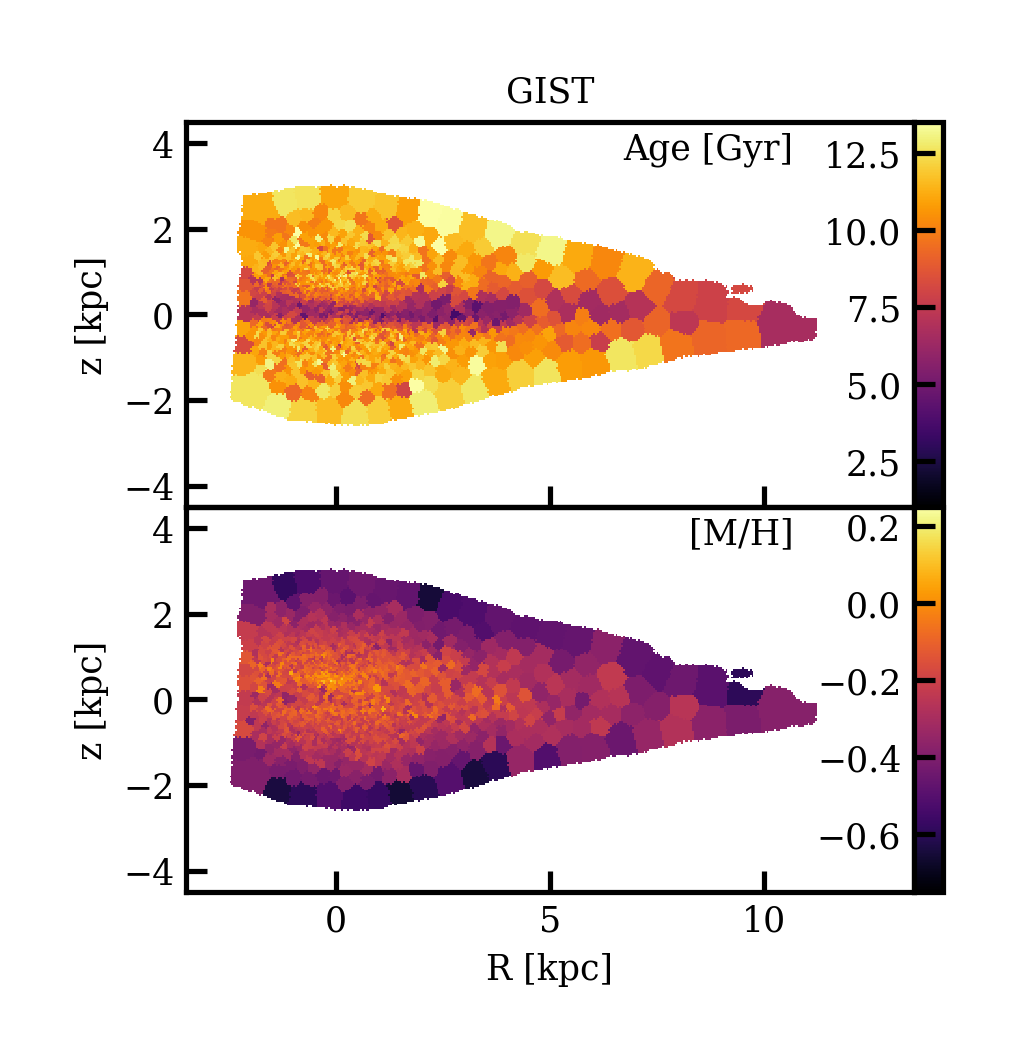}
\caption{Comparison of derived mean stellar population properties between \gist\ and \ngist\ for galaxy IC 1711. The left column shows \gist\ output for Age (top) and metallicity [M/H] (bottom), whereas the right column shows the results from \ngist. For both the \gist\ and \ngist\ runs, cubes were binned to S/N=100 (using the binning maps created by \ngist), and all other input parameters were kept identical.} 
\label{Fig:ogist_ngist_sfhmaps}
\end{figure*}

\begin{figure*}
\centering
\includegraphics[width=0.6\textwidth,trim={0.5cm 0cm 0.5cm 0cm},clip]{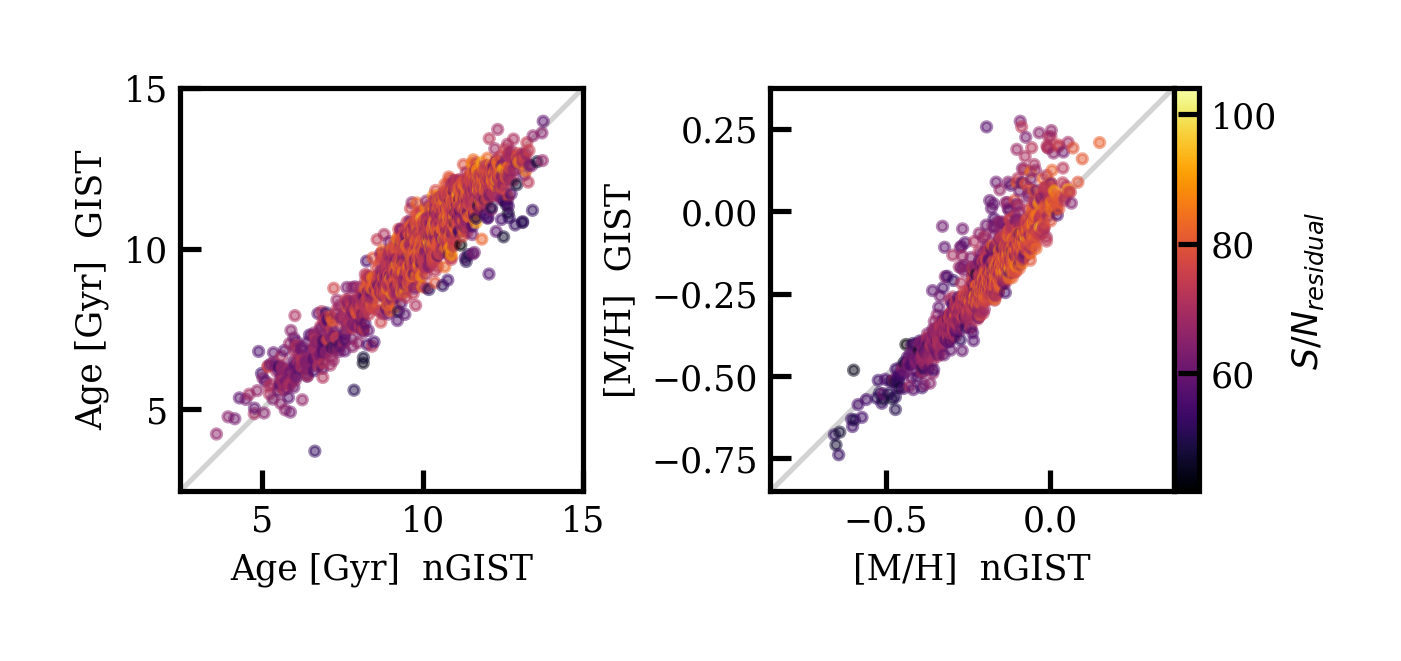}
\caption{A direct, bin-by-bin comparison of derived mean stellar population properties from \gist\ (y-axis) and \ngist\ (x-axis) for GECKOS galaxy IC 1711. Every point is a Voronoi bin (as shown in Figure~\ref{Fig:ogist_ngist_sfhmaps}), and points are colour-coded by the $S/N$ determined from the residual of the data minus the best-fit model.} 
\label{Fig:ogist_ngist_scatter_plots_sfh}
\end{figure*}

\textbf{Purpose:} Derive non-parametric star formation histories for the observed spectra. Light or mass-weighted averages of stellar age, metallicity, and $\alpha$-enhancement are also returned. 

\textbf{Changes from \gist:} As with the stellar kinematics module, significant changes have been made to the \textsc{pPXF} routine, mostly involving the implementation to a near identical three-step process to better estimate the noise in the input spectrum.

$\bullet$ Added continuum normalisation before fit to avoid numerical issues.

$\bullet$ Added the ability to turn the \textsc{pPXF} CLEAN function on or off using the keyword \texttt{DOCLEAN}. Similar to the KIN module, this is useful in obtaining a proper estimate of the noise. 

$\bullet$ Stellar dust absorption E(B-V) maps are now also output, calculated assuming a \citet{calzetti2000} extinction curve by default, and with no polynomials used in the fit. 

$\bullet$ Added the option for using Monte-Carlo (MC) realizations as an alternative to regularisation to obtain stellar population parameters \citep[see for example, Figure 1 of][for a comparison between the two techniques]{cappellari2023}. This is an alternative to \textsc{pPXF} regularization,  which has been found to sometimes introduce strong bias in the star formation history \citep[e.g.][]{wang2024} We follow the procedures outlined by \citet{pessa2023} for the PHANGS survey: in each iteration, the input spectrum is perturbed by adding random Gaussian noise with a mean of zero and standard deviation equal to the spectral flux error. The stellar population parameters are calculated from the median of the MC realisations, and uncertainties are then calculated by taking half the difference between 16$^{\rm{th}}$ and 84$^{\rm{th}}$ percentiles. There are alternative methods for performing the MC realisations, such as using randomly flipped residuals as error or adding noise to the best-fit spectrum, which could also be available for future updates. 

We compare the mean stellar population age and metallicity from \gist\ and \ngist\ in Figures \ref{Fig:ogist_ngist_sfhmaps} and \ref{Fig:ogist_ngist_scatter_plots_sfh}. Similar to \citet{pessa2023}, for this comparison we restrict the wavelength region to 4800-7000\AA, and adopt a multiplicative polynomial of order 12. To avoid discrepancies in the derived stellar populations resultant from differences in the stellar kinematic solutions, we adopt the Voronoi binning and stellar kinematics solution from the  \ngist\ \texttt{SPATIAL\_BINNING} and \texttt{KIN} modules for both.  
We apply minimal regularisation to the resultant mean stellar age and metallicity maps. The new cleaning algorithm in \ngist\ has a significant impact on the derived age and metallicity, most noticeably at large radii and scale heights, where the sky line residuals and telluric correction are most pronounced. We also detect a higher metallicity region in the centre of the \gist\ metallicity map when compared to \ngist\, seen in Figure \ref{Fig:ogist_ngist_sfhmaps}. 
While at first glance, the higher central metallicities of \gist\ may be more in line with expectations of inside-out disc galaxy growth, we find that the higher metallicity returned is actually due to a poorer fit. Specific faint lines (for example KI at 7664.9\AA\ and 7690.0\AA) are particularly prominent in the central regions of IC 1711 (and indeed can be seen in the central spectra at the top of Figure~\ref{Fig:ogist_ngist_spectra}), yet not present in the SSP models used. For the \gist\ pipeline, without automatic masking, these lines bias the best-fit \textsc{pPXF} solution toward higher metallicity, while nGIST's cleaning algorithm identifies them as significant outliers and masks them, resulting in a lower metallicity estimate. 
A full test of the impact of clipping outliers and the dependence of wavelength range on the mean stellar population parameters is beyond the scope of this paper, but will be explored in future work.

\subsubsection*{Line strengths (\texttt{LS})}

\textbf{Purpose:} Measure the line strength indices from the best-fit continuum spectra and derive stellar population parameters in this manner.

\textbf{Updates from \gist:} None.

\subsubsection*{Mapviewer}

\textbf{Purpose:} \gist~ also included a quick-look data visualisation tool called Mapviewer. Mapviewer takes the results of a \gist~ or \ngist\ run, and displays them in an interactive manner. The user can click on an individual spaxel of an IFS map, and the spectrum, and kinematic, emission line, and stellar population fits are displayed. Any map created by a module can also be displayed, as can the 1D star formation history, and a grid of weights assigned to templates in age/metallicity/alpha space. 

\textbf{Updates from \gist:} 

$\bullet$ Small cosmetic changes have been made to the Mapviewer interface, providing updated default colour maps and the ability to log-scale emission line fluxes. \\

$\bullet$ Velocity maps are no longer corrected for the systemic velocity by subtracting the median velocity of the entire map, as this only works when the observations are symmetric and centred on the galaxy centre.

\subsubsection*{General changes}

$\bullet$ Added the ability to use multiple stellar template libraries in the one \ngist\ call. This will be useful if you for example want to use stellar templates for the stellar kinematics module, and SSPs for the emission lines and stellar populations modules. Currently the MILES, EMILES, and SMILES template sets, the IndoUS template set \citet{valdes2004}, the Xshooter stellar library \citet{verro2022} and the \citet{walcher2009} template sets are supported. 

$\bullet$ Output: In addition to the output bin table files, we have also created maps fits files for the KIN, GAS, SFH, and LS modules. These maps files are more user-friendly than the original bin table format, as they can be opened and immediately plotted.

\section{The effect of wavelength range for fitting stellar kinematics in edge-on galaxies}
\label{wavelength_dependence}

\begin{figure*} [!ht]
\centering
\begin{subfigure}[t]{0.3\textwidth}
\includegraphics[width=\textwidth]{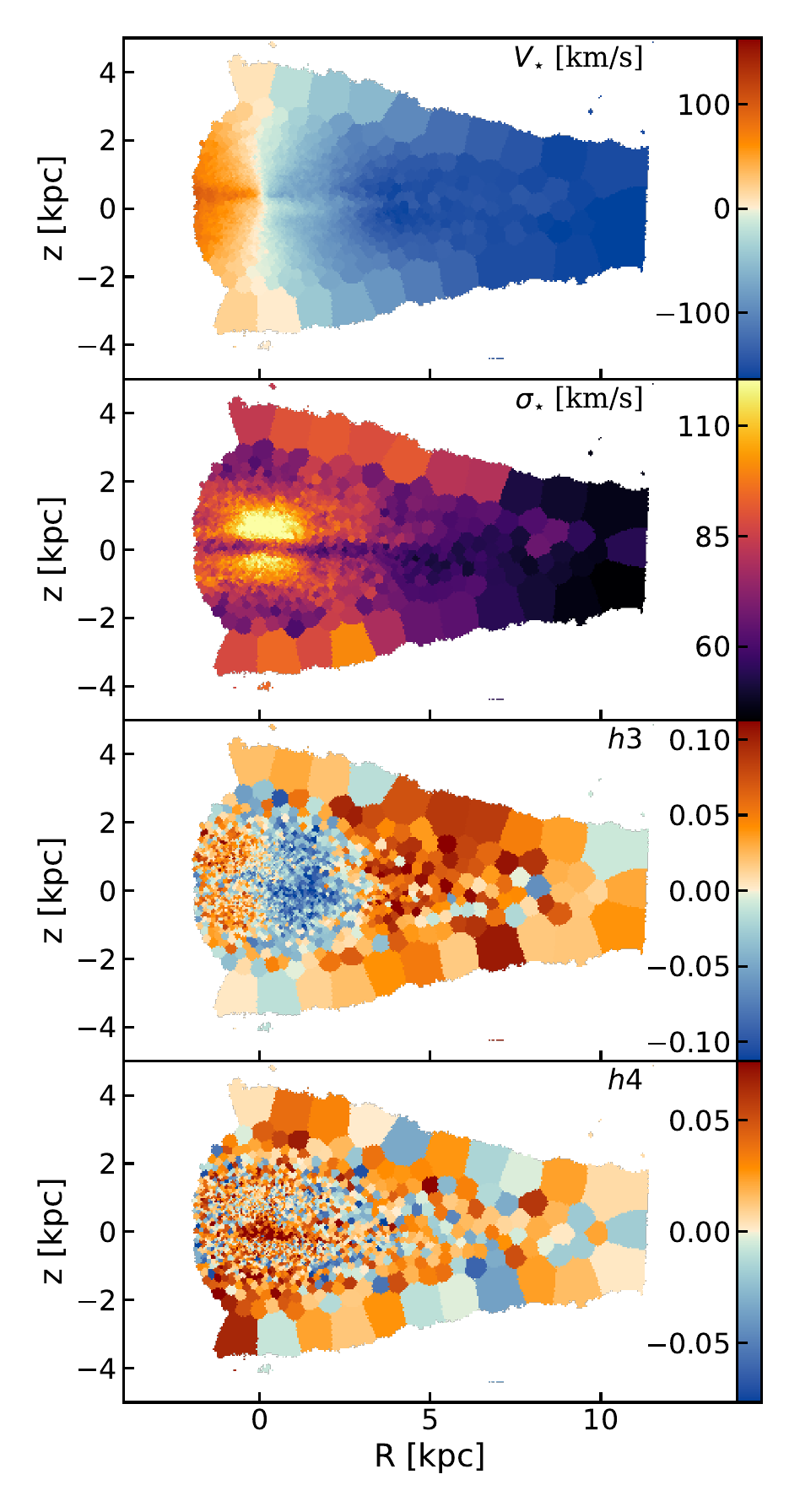} 
\caption{$\lambda$ range: H$\beta$and Mgb (4750--5500\AA)}
\end{subfigure}
\begin{subfigure}[t]{0.3\textwidth}
\includegraphics[width=\textwidth]{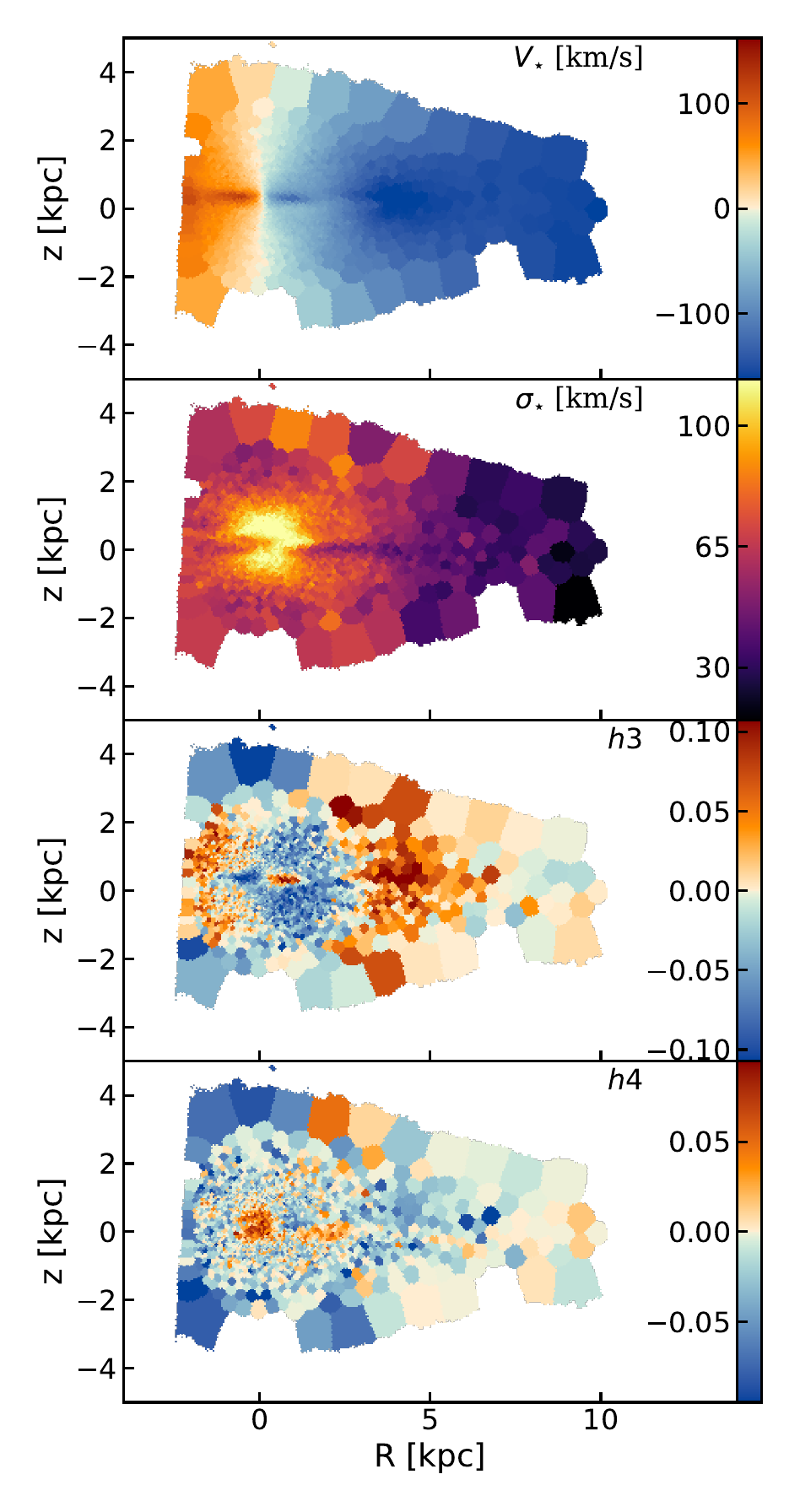} 
\caption{$\lambda$ range:  CaT (8458--8935\AA)} 
\end{subfigure}
\begin{subfigure}[t]{0.3\textwidth}
\includegraphics[width=\textwidth]{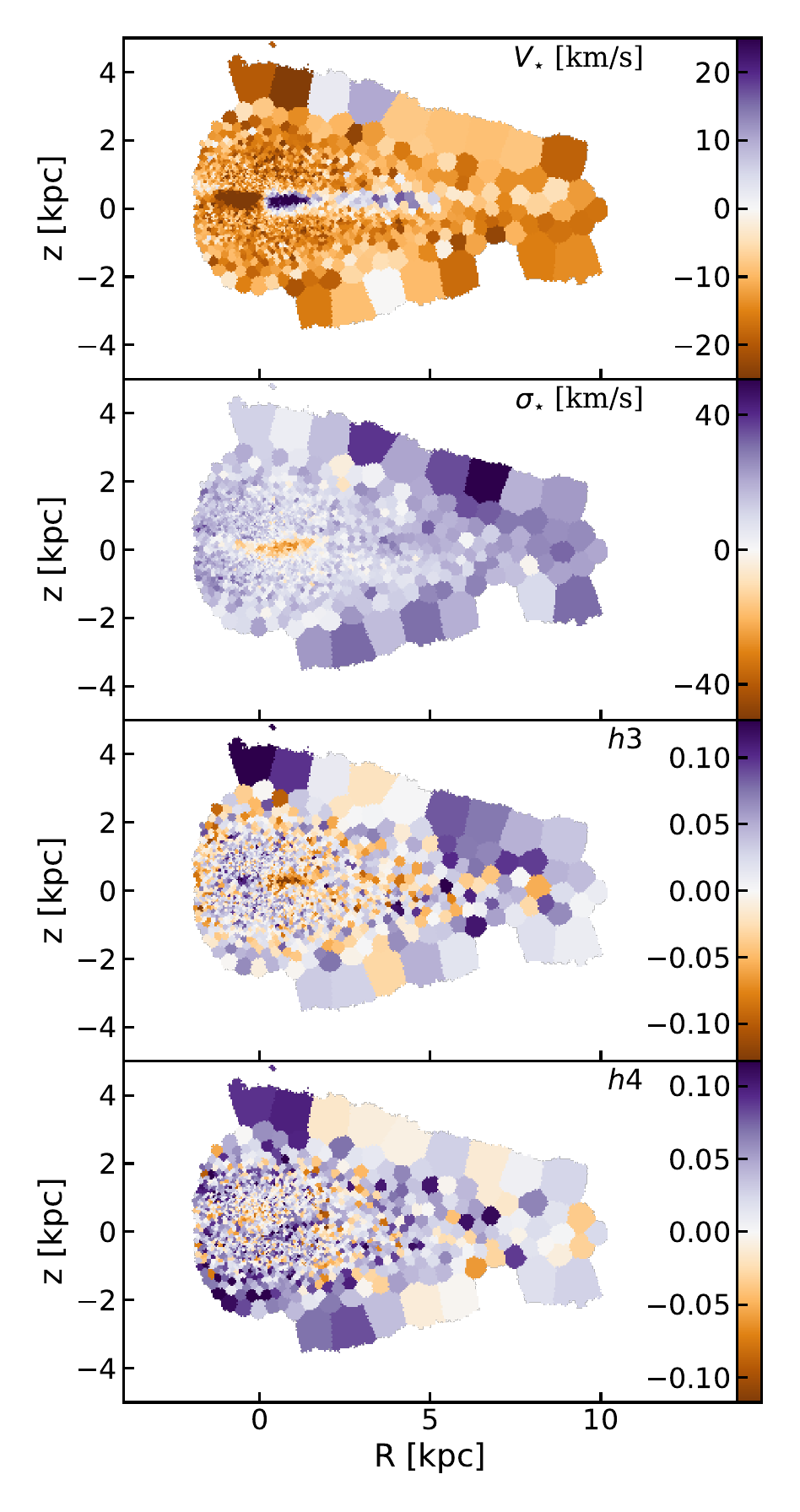} 
\caption{Difference between the two maps H$\beta$ and Mgb - CaT} 
\end{subfigure}

 \caption{Stellar velocity (top row), velocity dispersion (2nd row), $h_{3}$ (third row), and $h_{4}$ (fourth row) for GECKOS galaxy IC1711 calculated over the wavelength ranges of H$\beta$ and Mgb (4750--5500 \AA; first column), and the CaT (8458--8935\AA, middle column) binned to SN=100. The right column shows the difference between the two. The most obvious feature is seen at the centres of the $V_\star$ and $h_{3}$ maps; evidence for a nuclear disc is seen in the CaT maps, but is absent (and presumably hidden by the dust lane) in the H$\beta$ maps.}
 \label{Fig:wavelength_ranges}
\end{figure*}

Given that the amount of light from young vs. old stars and the degree of dust attenuation varies as a function of wavelength in the optical range, the spectral range used for the kinematic fit is important. \citet{mcdermid2002} showed that early-type galaxies in the SAURON/ATLAS 3D sample fitting stellar kinematics around the H$\beta$ line (in the blue region of the visible spectrum) yielded information on the kinematics of young stellar discs, whilst the red CaT region provided information on the older stellar disc.  While some works have leveraged this observation to infer information on the distribution of stellar populations in relatively face-on galaxies \citep[e.g.][]{rosado-belza2020}, in the edge-on case, any variations in derived kinematics (especially in the thin disc regions) are likely due to dust. 

In Figure~\ref{Fig:wavelength_ranges}, we present the cautionary tale of derived stellar kinematics for the GECKOS galaxy IC 1711 using a wavelength range around the H$\beta$-Mg b region (4750--5500 \AA, left column), the Calcium II triplet (CaT; 8458-8935 \AA, middle column), and the difference in derived stellar kinematics between the two, keeping all other input parameters (including the spatial binning) the same (right column). The differences between the three runs are perhaps most evident in the high-order moment maps. While a strong $h_{3}$-$V_\star$ correlation is seen in the inner regions of the disc for all three runs, the small, very central nuclear disc region in which the $h_{3}$ signal is inverted (resulting in a $h_{3}$-$V_\star$ \textit{anti-}correlation) is only seen in the CaT and full wavelength range maps. We infer that the longer wavelengths are capable of penetrating further into the galaxy disc; the blue wavelengths are attenuated such that we receive diminished kinematic signatures from the galaxy central regions. The result of using stellar kinematic tracers in the blue region for edge-on galaxies (or at least not including enough of the red region) is such that central features of the galaxy may be missed. Indeed, for the above analysis in this paper, we would report IC 1711 as not possessing a nuclear disc. \\

\section{Available Spitzer imaging}
\label{spitzer_imaging}

\begin{figure*}
\centering
\begin{subfigure}[t]{0.33\textwidth}
\centering
\includegraphics[width=\textwidth]{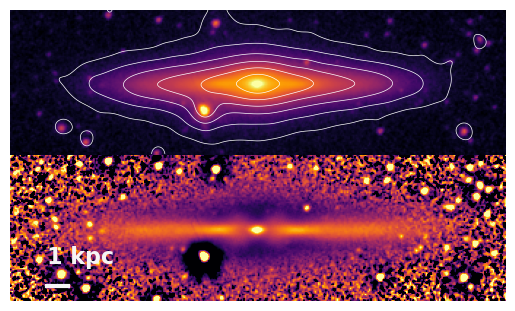} 
\caption{NGC 3957}
\end{subfigure}
\begin{subfigure}[t]{0.33\textwidth}
\centering
\includegraphics[width=\textwidth]{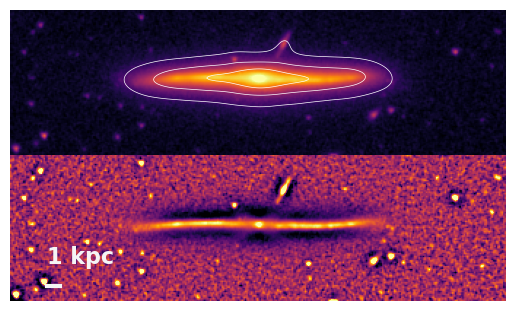} 
\caption{NGC 0522}
\end{subfigure}
\begin{subfigure}[t]{0.33\textwidth}
\centering
\includegraphics[width=\textwidth]{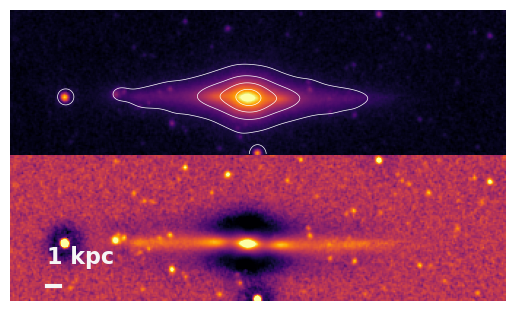} 
\caption{IC 1711}

\end{subfigure}
\begin{subfigure}[t]{0.33\textwidth}
\centering
\includegraphics[width=\textwidth]{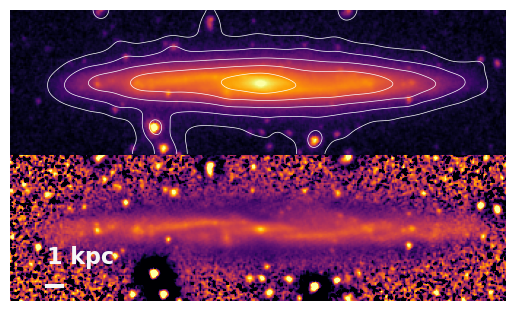} 
\caption{NGC 0360} 
\end{subfigure}
\begin{subfigure}[t]{0.33\textwidth}
\centering
\includegraphics[width=\textwidth]{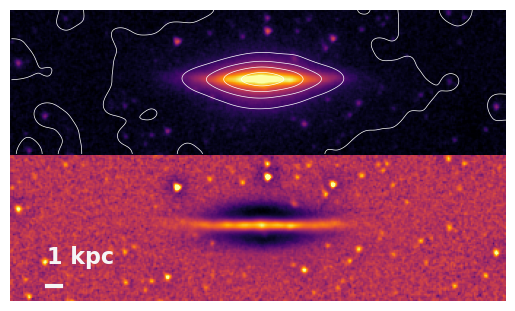} 
\caption{UGC 00903}
\end{subfigure}
\begin{subfigure}[t]{0.33\textwidth}
\centering
\includegraphics[width=\textwidth]{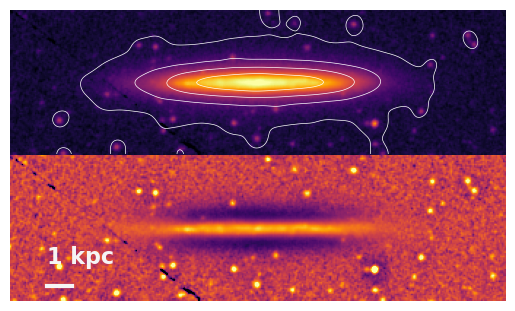} 
\caption{NGC 3279} 
\end{subfigure}

\begin{subfigure}[t]{0.33\textwidth}
\centering
\includegraphics[width=\textwidth]{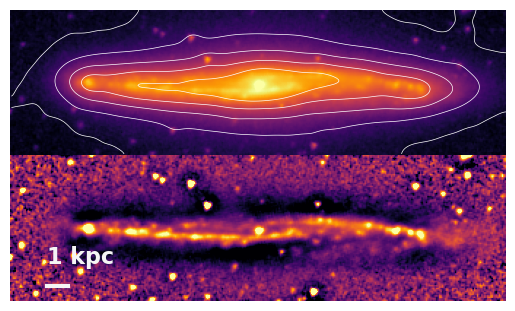} 
\caption{NGC 5775}
\end{subfigure}

 \caption{Spitzer 3.6$\mu$m imaging of available GECKOS iDR1 galaxies with an inverse hyperbolic sine ($\arcsinh$) scaling and surface brightness contours overlaid (top), and unsharp-masked images (bottom). The orientation of galaxies here, as throughout, is the same as in Fig~\ref{Fig:cube_images}.}
 \label{Fig:spitzer}
\end{figure*}

We present Spitzer 3.6-$\mu$m imaging for seven galaxies from the S4G survey \citep{sheth2010} in Figure~\ref{Fig:spitzer}, combined with unsharp-masked images.

\end{appendix}

\end{document}